\documentclass[%
reprint,
superscriptaddress,
nofootinbib,
 amsmath,amssymb,
 aps,
 prl,
]{revtex4-2}

\usepackage{graphicx}
\usepackage{dcolumn}
\usepackage{bm}
\usepackage{xcolor}

\usepackage{xr-hyper}

\usepackage[colorlinks=true
,urlcolor=blue
,anchorcolor=blue
,citecolor=blue
,filecolor=blue
,linkcolor=red
,menucolor=blue
,linktocpage=true
,pdfproducer=medialab
,pdfa=true
]{hyperref}

\DeclareMathOperator*{\argmin}{arg\,min}

\definecolor{orange}{rgb}{0.8,0.35,0}
\newcommand{\orange}[1]{#1}

\newcommand{\mysec}[1]{\paragraph{\textbf{#1}.\textemdash\hspace{-0.25cm}}}
\newcommand{\mysubsec}[1]{\paragraph{#1.\textemdash\hspace{-0.25cm}}}

\usepackage{etoolbox}
\newtoggle{PRL}
\newtoggle{showDeletions}

\togglefalse{PRL}  

\togglefalse{showDeletions}

\iftoggle{showDeletions}
{
    \newcommand{\arxiv}[1]{\iftoggle{PRL}{ \orange{#1}}{ \orange{#1}}}
    \newcommand{\PRLarxiv}[2]{\iftoggle{PRL}{ \red{#1}}{ \orange{#2}}}
}
{
    \newcommand{\arxiv}[1]{\iftoggle{PRL}{}{#1}}
    \newcommand{\PRLarxiv}[2]{\iftoggle{PRL}{#1}{#2}}
}

\begin{document}

\preprint{Preprint number}

\title{The GCE in a New Light: Disentangling the $\gamma$-ray Sky with Bayesian Graph Convolutional Neural Networks} 

\author{Florian List}
\email{florian.list@sydney.edu.au}
\affiliation{%
 Sydney Institute for Astronomy, School of Physics, A28, The University of Sydney, NSW 2006, Australia
}%
\author{Nicholas L. Rodd}
\affiliation{Berkeley Center for Theoretical Physics, University of California, Berkeley, CA 94720, USA}
\affiliation{Theoretical Physics Group, Lawrence Berkeley National Laboratory, Berkeley, CA 94720, USA}%
\author{Geraint F. Lewis}
\affiliation{%
 Sydney Institute for Astronomy, School of Physics, A28, The University of Sydney, NSW 2006, Australia
}%
\author{Ishaan Bhat}%
\affiliation{%
 UMC Utrecht, Image Sciences Institute, 3508 GA Utrecht, The Netherlands
}%

\date{\today}

\begin{abstract}
\PRLarxiv
{
    A fundamental question regarding the Galactic Center Excess (GCE) is whether the underlying structure is point-like or smooth, often framed in terms of a millisecond pulsar or annihilating dark matter (DM) origin for the emission. We show that Bayesian neural networks (NNs) \orange{have the potential to} resolve this debate. In simulated data, the method is able to predict the flux fractions from inner Galaxy emission components to on average $\sim$0.5\%. When applied to the \emph{Fermi} photon-count map, the NN identifies a smooth GCE in the data, suggestive of the presence of DM, with the estimates for the background templates being consistent with existing results.
}
{
    A fundamental question regarding the Galactic Center Excess (GCE) is whether the underlying structure is point-like or smooth. This debate, often framed in terms of a millisecond pulsar or annihilating dark matter (DM) origin for the emission, awaits a conclusive resolution. In this work we weigh in on the problem using Bayesian graph convolutional neural networks. In simulated data, our neural network (NN) is able to reconstruct the flux of inner Galaxy emission components to on average $\sim$0.5\%, comparable to the non-Poissonian template fit (NPTF). When applied to the actual \emph{Fermi}-LAT data, we find that the NN estimates for the flux fractions from the background templates are consistent with the NPTF; however, the GCE is almost entirely attributed to smooth emission. While suggestive, we do not claim a definitive resolution for the GCE, as the NN tends to underestimate the flux of point-sources peaked near the 1$\sigma$ detection threshold. Yet the technique displays robustness to a number of systematics, including reconstructing injected DM, diffuse mismodeling, and unmodeled north-south asymmetries. So while the NN is hinting at a smooth origin for the GCE at present, with further refinements we argue that Bayesian Deep Learning is well placed to resolve this DM mystery.
}
\end{abstract}

\maketitle

\count\footins = 500  

\mysec{Introduction}
\label{sec:intro}
Dark Matter (DM) annihilation into highly energetic standard model particles is a central prediction of well-motivated theories, including weakly interacting massive particles (WIMPs) \cite{Bertone2005}. Due to its high DM density and relative proximity, the center of the Milky Way is a promising location to search for the annihilation products, and in fact, the \emph{Fermi} $\gamma$-ray telescope observes excess emission from the inner region of the Milky Way that could be accommodated by annihilation from $\mathcal{O}(10 - 100 \ \text{GeV})$ thermal WIMPs~\cite{Goodenough2009,Hooper:2010mq,Hooper:2011ti,Abazajian:2012pn,Hooper:2013rwa,Gordon:2013vta,Abazajian2014,Daylan2016,Calore2015,Abazajian:2014hsa,Ajello2016,Linden2016,Macias:2016nev,Clark2018}. Since its discovery in 2009, a DM origin of this Galactic Center Excess (GCE) has been disputed, and arguments have been raised in favor of a population of millisecond pulsars (e.g. \cite{Hooper:2010mq, Abazajian:2012pn, Mirabal2013, Abazajian2014,Petrovic:2014xra,Yuan:2014yda, OLeary2015, Brandt:2015ula}), emission from unmodeled cosmic rays (e.g. \cite{Carlson2014, Petrovic2014, Cholis2015}), and that the emission may be more accurately correlated with stellar overdensities towards the Galactic Center (GC; e.g. ~\cite{Macias:2016nev,Ploeg2017,Bartels:2017vsx, Macias:2019omb, Abazajian2020}). The case for millisecond pulsars is based on evidence from two methods: wavelet techniques~\cite{Bartels2016,McDermott2016,Balaji2018} and the non-Poissonian template fit (NPTF)~ \cite{Malyshev:2011zi,Lee:2014mza,Lee2016,Mishra-Sharma2017}. It has been strongly argued that both methods prefer the emission to possess the small scale structure indicative of a point-source (PS) origin, rather than the smooth photon map (up to Poisson fluctuations) predicted by annihilating DM. Yet as with any inner Galaxy analysis, dependence upon systematic uncertainties is key. When masking more recently identified PSs near the GC, additional wavelet peaks associated with the GCE flux are no longer detected~\cite{Zhong:2019ycb} (cf.~\cite{Buschmann2020}). 
\orange{Further, the NPTF can be biased in favor of PSs when there is an unmodeled north-south asymmetry~\cite{Leane2020, Leane2020a}. The method also appeared unable to recover an injected DM signal~\cite{Leane2019a}, although such behavior can be expected to a degree when the flux is a mixture of smooth and PS emission~\cite{Chang2019}, and can largely be resolved through the use of either improved diffuse models or harmonic marginalization~\cite{Buschmann2020}. The more recent studies have emphasized that there is an ambiguity inherent in the physics: emission from sufficiently dim PSs is {\it exactly} Poissonian, and hence formally indistinguishable from the expected DM signal. Whether the GCE has a DM or PS origin is therefore only a well defined question for sufficiently bright PSs. In short, while we continue to learn more about the GCE and the systematic dependence of existing methods, the possibility that there is a hint of the particle nature of DM in the inner Galaxy remains.}

\par An alternative powerful approach to the problem that intrinsically utilizes inter-pixel correlations and autonomously determines the characteristic features of each template is given by convolutional neural networks (CNNs). \orange{Since a natural format for the \emph{Fermi}-LAT data is the HEALPix tessellation of the sphere \cite{Gorski2005}, we base our NN on the DeepSphere architecture \cite{Perraudin2019a, Defferrard2020}. By training on a large number of diverse photon-count maps, our NN learns to predict the flux fractions associated with the different emission components. We implement the approach of Ref.~\cite{Kendall2017} to render our NN a Bayesian Graph Convolutional Neural Network (Bayesian GCNN); thus, the NN learns a \emph{distribution} of each NN weight, rather than only a single value, which enables the estimation of uncertainties.}

\orange{Using this framework, we demonstrate that CNNs are a significant competitor for resolving the GCE debate. The method is capable of learning the central physics of template fitting: accurately estimating the flux fractions of \emph{all} the templates that compose the photon-count map (cf.~\cite{Caron2018} where CNNs were used to estimate contributions to the GCE alone). In most cases our NN recovers the GCE flux contributions to the percent level and is robust to key systematics, including a degree of diffuse mismodeling and unmodeled asymmetries. When applied to the real {\it Fermi} data, the method reports flux fractions consistent with the the NPTF for all components, differing only in the composition of the GCE: while the NPTF attributes 100\% of the GCE flux to PSs for our modeling choices, the CNN prefers an almost entirely smooth origin. While suggestive, our results also exhibit signs of the inherent PS and DM degeneracy at play, and in systematic studies we find cases where there can be bias towards the smooth template. As such, we stop short of declaring a resolution of the GCE origin with this method, but instead claim the method as having the clear potential to do so.}

\par This \emph{Letter} is structured as follows. \PRLarxiv{First, we describe the \emph{Fermi} data set and the generation of mock data used for training the NN, and we introduce our Bayesian GCNN.}{First, we briefly describe the \emph{Fermi} data set and the generation of mock data used for training the NN. Then, we summarize the DeepSphere framework that forms the backbone of our Bayesian GCNN. We introduce the concepts of aleatoric and epistemic uncertainty in the context of Bayesian Deep Learning and discuss how their estimation can be naturally embedded into the training process of the NN. Moreover, we outline the architecture of our NN.} We evaluate the performance of our NN in a proof-of-concept example with simulated maps composed of \arxiv{photon counts from} \orange{smooth and/or PS GCE} emission together with four background templates and compare our results with those from the \texttt{NPTFit}~\cite{Mishra-Sharma2017} implementation of the NPTF. Finally, we analyze the predictions of our NN for simulated mock maps of the GC that correspond to the best-fit parameters as determined by \texttt{NPTFit} and for the real \emph{Fermi} data.
Supporting results and a detailed discussion of the systematics of the method can be found in the Supplementary Material (SM).

\mysec{Data selection and generation}
\label{sec:data}
\PRLarxiv{We use the same data set that was employed in \cite{Mishra-Sharma2017, Leane2019, Buschmann2020}, containing 8 years of \emph{Fermi} data between $2 - 20 \ \text{GeV}$.}{We use the Pass 8 \emph{Fermi} photon counts within a reconstructed energy range of $2 - 20 \ \text{GeV}$ that have been detected between Aug 4, 2008 and July 7, 2016. We select events within the highest cosmic-ray rejection class, \texttt{UltraCleanVeto}, and apply the quality cuts \texttt{DATA\textunderscore QUAL==1}, \texttt{LAT\textunderscore CONFIG==1}, and zenith angle $\leq 90^\circ$. This exact data set has previously been employed in \cite{Mishra-Sharma2017, Leane2019, Buschmann2020}.} We model the inner Galaxy using the following Poissonian templates: (1) isotropic background accounting for extragalactic emission and cosmic ray contamination, (2) uniform emission from the \emph{Fermi} bubbles \cite{Su2010}, (3) Galactic diffuse background from neutral pions decaying into photons and from bremsstrahlung, (4) Galactic diffuse background from inverse Compton scattering, and (5) GCE due to annihilating DM. For the diffuse background components, by default we choose Model O, which was introduced in \cite{Buschmann2020} (building \orange{on} \cite{Macias:2016nev,Macias:2019omb}), and has been shown to remedy mismodeling concerns raised in \cite{Leane2019a}.
The template for the GCE is taken to be the line-of-sight integral through the square of a generalized Navarro--Frenk--White (NFW) profile \cite{Navarro1997}, i.e. $J_\text{DM} \propto \int \rho^2 \, ds$, with inner slope $\gamma = 1.2$ and scale radius $r_s = 20 \ \text{kpc}$ (see e.g. \cite{Daylan2016}). Additionally, we assume contributions from two \orange{PS} templates: (6) NFW-squared \orange{PSs} that serve as an alternative explanation for the GCE, and (7) disk-correlated PSs described by a doubly exponential disk with scale height $z_s = 0.3 \ \text{kpc}$ and radius $r_s = 5 \ \text{kpc}$. \orange{All maps are given resolution $N_\text{side} = 128$.}

\par CNNs are supervised learning methods, and determine a generalizable mapping between input (photon-count maps) and output (flux fraction of each template) by seeing a large number of input training samples alongside the corresponding output. For both the proof-of-concept example and the realistic scenario, we create 600,000 training maps using the publicly available tool \texttt{NPTFit-Sim} \cite{NPTFit-Sim} for the PS templates (accounting for the \emph{Fermi} point spread function) and sampling from Poisson distributions with pixel-wise means set by each exposure-corrected Poissonian template. We randomly draw all template parameters from wide prior distributions. In this way, our NN is trained to estimate the flux fractions in \emph{arbitrarily} composed maps\arxiv{ (which may contain PSs described by \emph{any} source count distribution (SCD) whose NPTF parameters lie within the prior cube)}\orange{, but narrower priors around the expected \emph{Fermi} values are considered in the SM, giving very similar results for the \emph{Fermi} map. We model the PSs with singly-broken power laws with variable slopes, spanning the entire relevant brightness range from the 3FGL detection threshold down to very faint PSs. Just like the NPTF, our method is agnostic about the physical nature of the PSs (see the SM for an application to a SCD derived from millisecond pulsars, which are the most popular candidate for PS emission).} 

\mysec{Bayesian graph convolutional neural networks}
\label{sec:bgcnns}
CNNs~\cite{Lecun1998}, \orange{and more recently Bayesian variants, have been applied to a range of problems in cosmology, see e.g.~\cite{PerreaultLevasseur2017, Hortua2019, Petroff2020}.}
They consist of multiple layers that successively map \orange{an} input $\mathbf{x}$ to an output $\mathbf{y}$. The core layers possess \orange{weights $\bm{\omega}$,  free parameters} that are updated \PRLarxiv{during the}{in the course of the} NN training by means of a stochastic gradient descent algorithm in order to minimize the expected loss. Thus, the objective of the training is to find 
\begin{equation}
\begin{aligned}
    \bm{\omega}^* &= \argmin_{\bm{\omega}} \mathbb{E}\left[\mathcal{L}(f^\mathbf{\bm{\omega}}(\mathbf{x}), \mathbf{y})\right] \\ 
    &\approx \argmin_{\bm{\omega}} \frac{1}{N} \sum_{i=1}^N \mathcal{L}(f^\mathbf{\bm{\omega}}(\mathbf{x}_i), \mathbf{y}_i),
\end{aligned}
\label{eq:training_objective}
\end{equation} 
where $\{\mathbf{x}_1, \ldots \mathbf{x}_N\} =: \mathbf{X}$ and $\{\mathbf{y}_1, \ldots \mathbf{y}_N\} =: \mathbf{Y}$ denote the samples and true labels of the training data, respectively, and $\mathcal{L}$ is a loss function that measures the fidelity of the NN output $f^{\bm{\omega}}(\mathbf{x})$ with respect to the truth $\mathbf{y}$ for an arbitrary input map $\mathbf{x}$ given the weights $\bm{\omega}$. The expectation value in the first line is taken over the joint distribution of $(\mathbf{x}, \mathbf{y})$,
\orange{which in practice is approximated by} 
minimizing the average loss across a large number of training samples $(\mathbf{x}_i, \mathbf{y}_i)$ drawn from the joint distribution of $(\mathbf{x}, \mathbf{y})$. The main building block of CNNs is the eponymous convolution operation, which \orange{in the DeepSphere approach is defined in Fourier space by resorting to the graph Laplacian operator.}
During the NN training, these convolutional kernels are gradually updated and learn to extract \orange{relevant} features in the data.
\arxiv{While the first convolutional layer tends to detect low-level features such as gradients and edges, stacking several convolutional layers (as done in ``Deep Learning'') enables the NN to recognize more involved structures in the data.} 

\begin{figure}
\centering
  \noindent
   \resizebox{1\columnwidth}{!}{
    \includegraphics{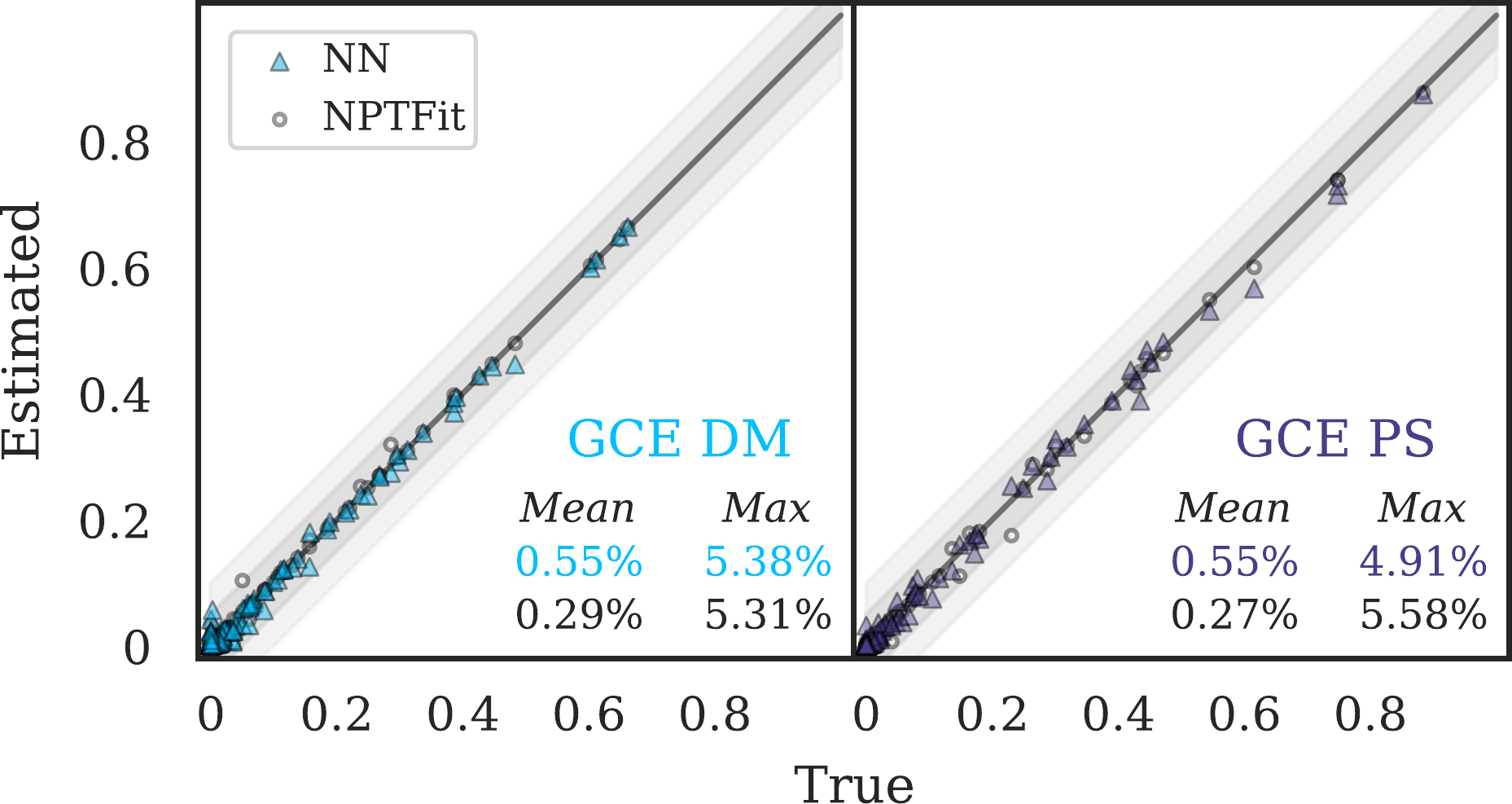}}
    \caption{True vs. estimated GCE flux contributions 
    for the NN (colored triangles) and \texttt{NPTFit} (black circles),
    \orange{evaluated on}
    172 maps from the test data set. The gray shaded regions correspond to errors of $5$ and $10 \%$, respectively. The mean and maximum absolute errors are specified in the lower right.}
    \label{fig:toy_example}
\end{figure}

\mysubsec{Estimating uncertainties}
\label{subsec:uncertainties}
While finding $\bm{\omega}$ such that $f^{\bm{\omega}}(\mathbf{x}) \approx \mathbf{y}$ is sufficient in many applications, estimating the uncertainties of the flux fractions predicted 
is vital \orange{in interpreting the results}. Following 
\orange{Refs.~}\cite{Gal2016, Gal2017, Kendall2017}, we distinguish between \orange{data-inherent} aleatoric (statistical) uncertainty, which is innate in the photon-count maps due to the stochasticity of the $\gamma$-ray emission from different origins that is detected in each pixel, 
and \orange{model-related} epistemic (systematic) uncertainty, which describes the ignorance about how well the NN approximates the correct input-to-output mapping.

\arxiv{Aleatoric uncertainty is inherent to each photon-count map and hence cannot be reduced by increasing the number of training samples. Since we expect the noise level to vary among the samples depending on the number of photons and the contributing templates (for instance, PS emission becomes degenerate with Poissonian emission in the ultrafaint limit), we model the aleatoric uncertainty to be heteroscedastic, i.e. dependent on the specific photon-count map ($\sigma = \sigma(\mathbf{x})$).} Rather than assuming a noise level \emph{a priori}, we train the NN to \emph{predict} the aleatoric noise. For this purpose, we assume that the uncertainty of each flux fraction can be reasonably described by a multivariate Gaussian with diagonal covariance matrix $\bm{\Sigma}(\mathbf{x}) = \text{diag}[\sigma_1^2(\mathbf{x}), \ldots, \sigma_T^2(\mathbf{x})]$, where $T$ stands for the number of templates.\PRLarxiv{ The extension to non-diagonal uncertainty covariance matrices is treated in the SM.}{\footnote{The extension to non-diagonal uncertainty covariance matrices is treated in the SM.}} We take the negative maximum log-likelihood estimate for the pair $(f^{\bm{\omega}}(\mathbf{x}), \bm{\Sigma}(\mathbf{x}))$ as the loss function in Eq.~\eqref{eq:training_objective}, given by
\begin{align}
    \mathcal{L}(f^{\bm{\omega}}(\mathbf{x}), \mathbf{y}) &= \sum_{t=1}^{T} \left( \frac{1}{2\sigma_t^2(\mathbf{x})} \left(f^{\bm{\omega}}(\mathbf{x})_t - \mathbf{y}_t \right)^2 + \frac{\log \sigma_t^2(\mathbf{x})}{2} \right) \nonumber \\
    &\qquad + \frac{T}{2} \log(2 \pi),
\label{eq:aleatoric_llh}
\end{align}
and use a final NN layer with output dimension $2 \times T$ for predicting $\{(f^{\bm{\omega}}(\mathbf{x})_t, s_t(\mathbf{x}))\}_{t=1}^T$, where $s_t := \log \sigma_t^2$. As the NN itself estimates $\bm{\Sigma}(\mathbf{x})$, $\bm{\Sigma}$ also depends on the weights $\bm{\omega}$. The terms with $\sigma_t$ in the numerator and denominator in \eqref{eq:aleatoric_llh} favor small and large values for $\sigma_t$, respectively, with the optimal $\sigma_t$ balancing the two.

For estimating the epistemic uncertainty, we use \emph{Concrete Dropout} \cite{Gal2017}. While the original motivation of randomly zeroing NN units is to prevent overfitting \cite{Hinton2012, srivastava2014dropout}, it has been shown that this Dropout can be viewed as a Bayesian approximation of the posterior distribution for the NN weights $p(\bm{\omega} | \mathbf{X}, \mathbf{Y})$ \cite{Gal2016}.\arxiv{ Since this distribution is in general analytically intractable, it is approximated by a simpler distribution $q(\bm{\omega})$ that minimizes the Kullback--Leibler divergence $\text{KL}\left(q(\bm{\omega}) \, || \, p(\bm{\omega} | \mathbf{X}, \mathbf{Y})\right)$, or equivalently maximizes the evidence lower bound. 
Applying Dropout at test time can then be interpreted as sampling weights from $q(\bm{\omega})$.} Whereas well-calibrated Dropout probabilities can be determined with a grid search, \emph{Concrete Dropout} lets the NN autonomously adapt the Dropout probabilities during the training.

\mysubsec{Neural network architecture and training}
\PRLarxiv
{
The key elements of our NN are graph convolutional layers, which act as feature extractors, followed by fully connected layers that infer the associated flux fractions. We use a Softmax activation for the means in order to enforce the flux fractions to lie in $[0, 1]$ and to sum to unity.
}
{
Our NN processes the input maps with 7 consecutive graph convolutional layers that act as feature extractors, each of which is followed by a batch normalization operation \cite{Ioffe2015}, a ReLU non-linearity, and a max-pooling operation, which reduces the $N_\text{side}$ parameter by a factor of 2 and therefore the number of pixels by 4, whereas the number of channels (also called feature maps, which are associated with individual sets of convolutional kernels) gradually increases from 32 after the first graph convolutional layer to a maximum of 256. Thereafter, we employ two fully connected layers with ReLU activation and a final fully connected layer with $2 \times T$ output neurons for the means and log-variances, where we use a Softmax activation for the means in order to enforce the flux fractions to lie in $[0, 1]$ and to sum to unity.}
The details are supplied in the SM.
\PRLarxiv
{
In total, our implementation of the Bayesian GCNN in \texttt{Tensorflow} \cite{Abadi2016} has $\sim 4 \times 10^6$ trainable parameters.
}
{
In total, our implementation of the Bayesian GCNN in \texttt{Tensorflow} \cite{Abadi2016} has roughly $4 \times 10^6$ trainable parameters. We use a batch size of $64$ and take an Adam optimizer \cite{Kingma2014} with learning rate $5 \times 10^{-4}$ decaying with a rate of $2.5 \times 10^{-4}$ with respect to mini-batch iterations. We train the NN by performing 30,000 mini-batch iterations (25,000 for the proof-of-concept example) on a single Nvidia Tesla Volta V100 GPU on the supercomputer Gadi, which is located in Canberra and is part of the National Computational Infrastructure (NCI), taking roughly two hours for the realistic scenario.
}
\begin{figure*}
\centering
  \noindent
  \PRLarxiv
  {
     \resizebox{0.7\textwidth}{!}{
     \includegraphics{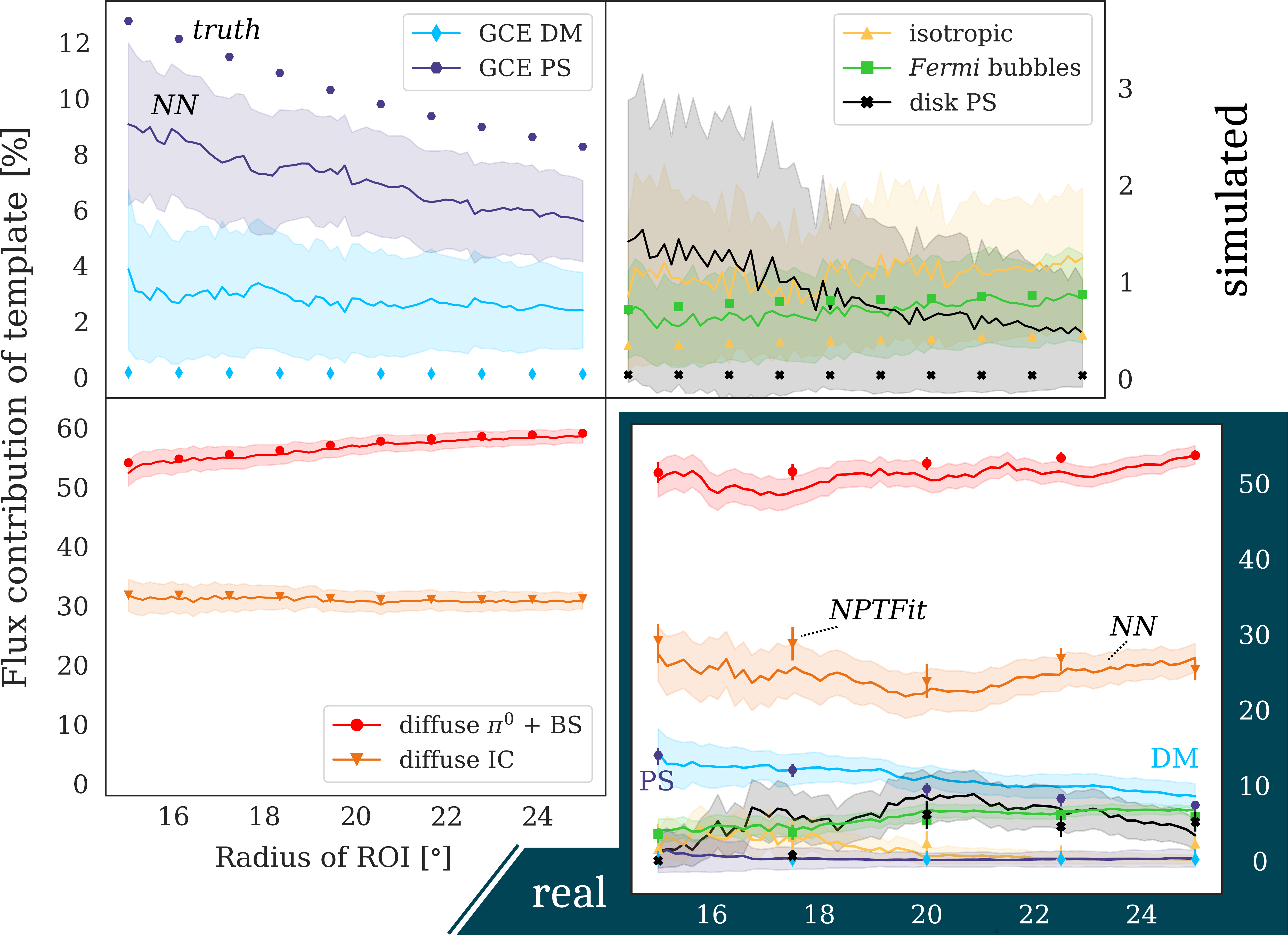}} 
  }
  {
     \resizebox{0.76\textwidth}{!}{
     \includegraphics{Fig_2_integers_new.pdf}}
  }
\caption{\emph{Upper left panels:} \orange{Flux estimates from the Bayesian GCNN on simulated data within the ROIs delimited by the outer radii given on the $x$-axis. The 250 Monte Carlo mock maps correspond to the best-fit parameters as determined by \texttt{NPTFit} in a $10^{\circ}$ ROI, within which it recovers negligible \emph{Fermi} bubbles and disk PS flux.} The flux fractions are predicted with respect to each ROI, implying that the fractions are expected to vary as a function of radius, rather than to remain constant. The shaded regions show the 1$\sigma$ scatter computed over the 250 mock samples. The markers indicate the correct flux fraction for each radius, averaged over the samples. While the NN misattributes a fraction of the GCE to DM (\orange{as expected given their inherent degeneracy}), it recognizes that the majority of GCE flux is due to PS. \emph{Lower right panel:} Prediction of the Bayesian GCNN for the real \emph{Fermi} photon-count map. The shaded regions show the predictive (aleatoric and epistemic summed in quadrature) 1$\sigma$ uncertainty. \orange{The markers with error bars ($68\%$ credible intervals) indicate the \texttt{NPTFit} estimates for comparison}. 
The \orange{NN predictions for the} GCE flux in the \emph{Fermi} map are \orange{similar in magnitude to those of \texttt{NPTFit}, but the GCE is} almost entirely attributed to DM. \orange{An interactive version can be found at \href{https://zenodo.org/record/4044689}{https://zenodo.org/record/4044689}.}
}
\label{fig:fermi_results}
\end{figure*}

\mysec{Proof-of-concept example: recovering the GCE flux fractions from simulated maps}
\label{sec:toy_example}
In this example, we consider the recovery of the flux fractions for the following scenario: all the Poissonian templates are present and we include the GCE PS as the only PS template. \orange{Importantly, this setup is representative of the challenges we will encounter in the real data as we never encounter systematic confusion between the GCE and disk PS models, as demonstrated in the SM.} We take a fixed ROI of $25 ^\circ$ around the GC, with the Galactic Plane masked at latitudes $|b| \leq 2 ^\circ$. 
We do not estimate uncertainties in this benchmark example but rather compare the NN predictions with the true values and with the estimates from \texttt{NPTFit}. 
Therefore, we train the NN with a simple $l^2$ loss. 
The nested sampling in \texttt{NPTFit} is done using \texttt{MultiNest} \cite{Feroz2009, Buchner2014} with 500 live points. To speed up \texttt{NPTFit}, we choose a uniform exposure map with \emph{Fermi} mean exposure in this example. 
\par Fig.~\ref{fig:toy_example} depicts the true vs. estimated flux fractions for the NN and \texttt{NPTFit}, for 172 out of 256 randomly composed maps from the testing data set (not seen by the NN during the training) for which \texttt{NPTFit} converged.\arxiv{\footnote{This criterion was established to account for simulated maps with a large number of counts-per-pixel, where the current implementation of the NPTF in \texttt{NPTFit} fails. Here we imposed a simple cut of rejecting cases where \texttt{NPTFit} failed to converge in a fixed run time. This is sufficient for a qualitative comparison between the method and the NN output. Nevertheless, we caution that this has undoubtedly biased the detailed results depicted, likely in favor of \texttt{NPTFit}. We discuss this point further in the SM.}} 
For \texttt{NPTFit}, we report the medians of the flux fraction posterior distributions. 
The errors for both the NN and \texttt{NPTFit} estimates are on average well below one per cent for all the templates (\orange{including} those not shown), and the maximum errors are comparable.
In the SM, we demonstrate that the NN displays robustness in distinguishing GCE DM and PS in the presence of diffuse mismodeling and an unmodeled north-south asymmetry of the GCE (a case where \texttt{NPTFit} can be biased~\cite{Leane2020, Leane2020a}).  Nonetheless, we caution that when the peak of the PS distribution tends towards the $1\sigma$ detection threshold, there can be systematic confusion between GCE DM and PS, particularly if the flux arises from a mixture of both, \orange{as seen also with \texttt{NPTFit}~\cite{Chang2019}.}

\mysec{Application to the \emph{F\lowercase{ermi}}-LAT count map}
\label{sec:fermidata}
Now, we train a Bayesian GCNN on photon-count maps consisting of all the templates: those used in the proof-of-concept example above, plus the disk PS template. We mask again the inner $|b| \leq 2 ^\circ$ around the Galactic Plane and train on photon-count maps covering the inner $15 - 25 ^\circ$ around the GC. Moreover, we mask the resolved 3FGL sources~\cite{Acero2015} at $95 \%$ containment in order for the flux within the ROI not to be dominated by known bright PSs. The training maps account for the non-uniform \emph{Fermi} exposure. 
\orange{In order to obtain realistic simulated maps, }
we fit the \emph{Fermi} data using \texttt{NPTFit} within a $10 ^\circ$ ROI (the region that drives the NPTF evidence for GCE PS) and generate 250 mock realizations that correspond to the medians of the model parameter \orange{posteriors}. For this setting, \texttt{NPTFit} finds a point-like GCE, and the identified GCE DM flux is consistent with zero. \orange{In this ROI, the isotropic, bubbles, and disk PS fluxes are negligible ($\lesssim 1\%$).} We apply our NN to each of the 250 mock maps for 64 ROIs, delimited by a radius that monotonically increases in equal steps from $15$ to $25 ^\circ$ around the GC. Figure \ref{fig:fermi_results} (\emph{upper left \orange{panels}}) shows the predictions of the NN (solid lines) and the 1$\sigma$ scatter (shaded regions) over the mock samples, as a function of the ROI radius. The fluxes of the non-GCE templates are accurately predicted, but a fraction of the GCE PS flux is incorrectly attributed to the GCE DM template. \orange{This is a manifestation of the PS-DM degeneracy expected for the simulated PS distribution, and reflective of the dim nature of the GCE PSs preferred by \texttt{NPTFit}.}
\arxiv{We assess the correlation between the source-count distribution for the GCE PSs and the degree of misattribution between the GCE templates in a dedicated experiment in the SM.} Nonetheless, the NN correctly identifies the GCE PS template to be the main constituent of the GCE in $90 \%$ of the maps. \orange{Further, in all cases the sum of GCE DM and PS contribution is consistent with the GCE flux injected, indicating the absence of confusion with additional templates such as the disk PSs that were absent in our proof-of-concept tests.} 
In the SM, we consider the same experiment for mock maps corresponding to the NPTF best-fit parameters determined using a larger ROI radius of $25 ^\circ$ \orange{with higher disk and bubble fluxes, which the NN correctly detects.}
\par In Fig.~\ref{fig:fermi_results} (\emph{lower right}), we plot the predictions of the NN for the real \emph{Fermi} data, \orange{evaluated in the same 64 ROIs, together with those of \texttt{NPTFit} for 5 ROIs}.
Now, the shaded regions indicate the $1\sigma$ predictive uncertainties of the NN (aleatoric and epistemic added in quadrature). While the NN predictions for the diffuse templates resemble those for the mock data, the GCE flux is almost entirely attributed to annihilating DM, with GCE PS being consistent with zero flux. As expected, the predicted GCE DM contribution decreases as the ROI is enlarged. The credible intervals are similar in size to the scatter of the NN predictions over the mock samples, decreasing in general as the ROI radius increases (more pixels are at the disposal of the NN), and are smaller for templates that are easy for the NN to predict thanks to their distinct shape such as the \emph{Fermi} bubbles. \arxiv{In the SM, we show that the estimated uncertainties are generally larger for maps and templates where the NN estimates are less accurate and vice versa.}
The flux contributions and in particular the total GCE flux predicted by \texttt{NPTFit} and the NN \orange{mostly} agree with each other, with the exception of the GCE flux being ascribed to DM by the NN instead of PS: the NN / \texttt{NPTFit} predictions for the GCE templates within $25^\circ$ are \orange{(in $\%$): $8.6 \pm 1.7$ / $0.2^{+1.4}_{-0.2}$ (GCE DM) and $0.3 \pm 1.2$ / $7.4^{+0.6}_{-1.2}$ (GCE PS), respectively, while the predictions for the other templates  are $53.8$/$53.7$ (diffuse $\pi^0 + \text{BS}$), $27.0$/$25.4$ (diffuse IC), $0.2$/$2.3$ (isotropic), $6.8$/$5.9$ (\emph{Fermi} bubbles), and $3.4$/$5.2$ (disk PS)}. This preference for GCE DM over GCE PS persists for three additional diffuse models and when replacing the thin disk PS template by a thick disk PS template. Yet we also find that the NN \orange{can} underestimate the flux associated with additional GCE PSs injected directly into the data, particularly if they are peaked near the 1$\sigma$ threshold, suggesting if such sources are present their flux is likely underestimated. \orange{Accordingly, we refrain from interpreting these results as a definitive statement on the origin of the GCE.} 
A detailed discussion of these points is presented in the SM.

\par With this work, we establish Bayesian Deep Learning as a powerful tool for disentangling the GCE into its individual components. It is suggestive that our NN identifies a GCE with smooth origin in all our experiments. Nonetheless, as with the NPTF, it will be crucial to further substantiate the robustness of Deep Learning methods to systematics such as diffuse mismodeling and the DM density profile. We present several encouraging results in the presence of either of these two mismodeling sources in the SM. In addition to new measurements and improved diffuse emission models, Deep Learning has the potential to critically contribute to the unraveling of the GCE mystery within the coming years.

\begin{acknowledgments}
Our work benefited from the feedback of Mariangela Lisanti, Ben Safdi, and Tracy Slatyer.
The authors acknowledge the National Computational Infrastructure (NCI), which is supported by the Australian Government, for providing services and computational resources on the supercomputer Gadi that have contributed to the research results reported\PRLarxiv{ herein}{ within this paper}.
We further acknowledge the technical assistance provided by the Sydney Informatics Hub, a Core Research Facility of the University of Sydney, and the generous allocation of resources through the Computational Grand Challenge program.
The authors thank the \emph{Fermi} collaboration for making the data publicly available.
FL is supported by the University of Sydney International Scholarship (USydIS).
NLR is supported by the Miller Institute for Basic Research in Science at the University of California, Berkeley.
This work made use of the free Python packages \texttt{matplotlib} \cite{mpl}, \texttt{seaborn} \cite{seaborn}, \texttt{numpy} \cite{npy}, \texttt{scipy} \cite{scipy}, \texttt{healpy} \cite{healpy}, \texttt{Tensorflow} \cite{Abadi2016}, \texttt{ray} \cite{ray}, \texttt{NPTFit} \cite{Mishra-Sharma2017}, \texttt{NPTFit-Sim} \cite{NPTFit-Sim}, \texttt{dill} \cite{dill}, and \texttt{cloudpickle}.\footnote{\href{https://github.com/cloudpipe/cloudpickle}{https://github.com/cloudpipe/cloudpickle}} Also, we intensively used the \texttt{arXiv} preprint repository and the free software \texttt{Inkscape}.\footnote{\href{https://inkscape.org/}{https://inkscape.org/}}
\end{acknowledgments}




%

\end{document}


\title{SUPPLEMENTARY MATERIAL \\ ``The GCE in a New Light: Disentangling the $\gamma$-ray Sky \\ with Bayesian Graph Convolutional Neural Networks''}
\author{F. List, N. L. Rodd, G. F. Lewis, and I. Bhat}
\date{\today}
\maketitle
\vspace{-0.5cm}
In this Supplementary Material, we provide further details that the readers of our \emph{Letter} ``The GCE in a New Light: Disentangling the $\gamma$-ray Sky with Bayesian Graph Convolutional Neural Networks'' may find useful. First, we gather our results from all the experiments that we perform for the \emph{Fermi} photon-count map in a table. Then, we describe our modeling of the GCE and briefly summarize the templates that we use for the various $\gamma$-ray sources expected to contribute to the \emph{Fermi} count map. Next, we tabulate the architecture of our Bayesian GCNN and introduce the individual layers; moreover, we specify the relevant loss functions with particular emphasis on their connection with the estimation of aleatoric (statistical) uncertainties, and explain the Bayesian Deep Learning technique \emph{Concrete Dropout}, which we harness for estimating the model-related epistemic uncertainties. Afterwards, we discuss how the NN estimates depend on the source count distribution (SCD) function describing the GCE PSs. Also, we examine diffuse mismodeling by applying a NN trained with the diffuse Model~O on maps containing a diffuse component given by the \emph{Fermi} model \texttt{p6v11}, we assess the performance of our NN in the case of GCE mismodeling by means of an example inspired by \cite{Leane2020} that deals with an unmodeled north-south asymmetry of the GCE, and we consider mismodeling of the \emph{Fermi} bubbles. We supply additional material for the proof-of-concept example and the application of the full Bayesian GCNN to the \emph{Fermi} count map, including the calibration of the uncertainties, the extension to a full uncertainty covariance matrix, \orange{and the application of the NN to a SCD derived from millisecond pulsars (MSPs)}. We check the consistency of the NN results for the \emph{Fermi} map when replacing the thin disk PS template by a thick disk PS template, when using different models for the Galactic foreground emission, and when considering the northern and southern hemispheres individually, demonstrating that \textbf{a GCE consistent with smooth emission is identified in all our tests}. \orange{Furthermore, we train a NN with a tight prior range around the expected parameters for the \emph{Fermi} map, and we additionally consider training on maps whose diffuse flux contributions are random linear combinations of 3 different diffuse templates, in which case the NN is trained to estimate the composition of the diffuse fluxes as an extra output.}  Finally, we study the recovery of artificially injected GCE flux from the \emph{Fermi} map. While the material herein is meant to provide additional background information and to corroborate the robustness of our findings through various experiments, we remark that \orange{more exhaustive studies} will be carried out in future work.
\orange{\par The following table provides an overview of the different scenarios that we consider in this Supplementary Material. It contains the non-Poissonian templates that we use, whether or not the NN is Bayesian, the outer radius of the ROI, masking of the 3FGL sources, and the exposure map, as well as the corresponding sections in this document. In the main body, Fig.~\ref{fig:toy_example} shows the results for the GCE templates in the proof-of-concept example, and the predictions for simulated mock data and the \emph{Fermi} map plotted in Fig.~\ref{fig:fermi_results} belong to the realistic scenario.}

\begin{table}[h]
\orange{
\begin{tabular}{@{}lcccccc@{}}
\toprule
 & Non-Poiss. templates & Bayesian & ROI & mask 3FGL & Exposure & Sections \\ \midrule
Proof-of-concept example & GCE PS & $-$ & 25$^\circ$ & $-$ & uniform & \ref{sec:SCDs} $-$ \ref{sec:add_mat_toy} \\
Realistic scenario & GCE PS, disk PS & \checkmark & $15 - 25^\circ$ & \checkmark & \emph{Fermi} & \ref{sec:add_mat_realistic}, \ref{sec:add_tests} \& \ref{sec:injection_test} \\
Narrow priors & GCE PS, disk PS & \footnote{\orange{Bayesian when using only diffuse Model O, non-Bayesian when using mixtures of Models A, F \& O.}} & 25$^\circ$ & \checkmark & \emph{Fermi} & \ref{sec:narrow_priors} \\ \bottomrule
\end{tabular}
}
\end{table}

\orange{\par Since the GCE flux is the most interesting component given its unknown origin, we place special emphasis on disentangling a smooth from a PS-like GCE in this Supplementary Material, with a dedicated experiment that assesses the extent of DM-PS confusion as a function of the PS brightness, for different compositions of the GCE. Since dim point sources are formally degenerate with Poisson emission, the question of whether emission is Poissonian (DM-like) or non-Poissonian (PS-like) in nature is inherently ambiguous in anything other than the limit where all sources produce an expected count $\gg 1$; a condition not satisfied by the potential PS population that could explain the GCE. As such, a degree of confusion between the GCE DM and PS templates is unavoidable -- this phenomenon is observed when using \texttt{NPTFit} even on simulated data (see in particular \cite[Figs. 4 \& 5]{Chang2019}), and our NN shows the same behavior, see Sec.~\ref{sec:SCDs}. In that experiment, as well as when exploring different sources of mismodeling, we consider a slightly simplified scenario (``proof-of-concept example'' in the table above), in which the NN is non-Bayesian, we fix the ROI to an outer radius of $25^\circ$ around the GC, we take a uniform \emph{Fermi} mean exposure, and we do not model disk PS emission. However, we emphasize that we did not observe systematic confusion between GCE PS and disk PS in our experiments, which can be seen e.g. in Figs.~\ref{fig:fermi_mock_histogram}, \ref{fig:fermi_results_25}, \ref{fig:error_hist_bartels}, and \ref{fig:narrow_priors_hist_best_fit_data}, for which reason our findings in the respective sections can be expected to carry over to more general modeling choices.}

{\hypersetup{linkcolor=black}
\tableofcontents
}

\section{Summary of our results for the \emph{Fermi} photon-count map}
\label{sec:summary_results}
\begin{table}[htb]
\caption{Estimated flux fractions for the \emph{Fermi} map in per cent, for all the experiments presented in this work. We report the results for an outer ROI radius of $25 ^\circ$, with latitudes $|b| \leq 2 ^\circ$ and 3FGL pixels masked.}
\begin{tabular}{@{}llccccccc@{}}
\toprule
Experiment & Method & diffuse $\pi^0$ + BS & diffuse IC & isotropic & \emph{Fermi} bubbles & \textbf{GCE DM} & \textbf{GCE PS} & disk PS \\ \midrule
\multirow{2}{*}{Default} & NN & $53.8 \pm 1.2$ & $27.0 \pm 1.9$ & $0.2 \pm 0.6$ & $6.8 \pm 0.6$ & $\mathbf{8.6} \pm 1.7$ & $0.3 \pm 1.2$ & $3.4 \pm 1.9$ \\
 & \texttt{NPTFit} & $53.7 \pm 0.7$ & $25.4 \pm 1.4$ & $2.3^{+0.9}_{-1.1}$ & $5.9 \pm 0.5$ & $0.2^{+1.4}_{-0.2}$ & $\mathbf{7.4}^{+0.6}_{-1.2}$ & $5.2 \pm 1.3$ \\ \cmidrule(l){2-9} 
\multirow{2}{*}{Thick disk} & NN & $55.1 \pm 1.2$ & $30.0 \pm 1.6$ & $0.1 \pm 0.5$ & $6.5 \pm 0.6$ & $\mathbf{7.8} \pm 1.6$ & $0.3 \pm 1.6$ & $0.1 \pm 1.0$ \\
 & \texttt{NPTFit} & $54.9 \pm 0.6$ & $29.5^{+0.8}_{-0.9}$ & $0.3^{+0.9}_{-0.3}$ & $6.4^{+0.3}_{-0.4}$ & $0.1^{+0.5}_{-0.1}$ & $\mathbf{8.2}^{+0.5}_{-0.6}$ & $0.6^{+0.5}_{-0.2}$ \\ \cmidrule(l){2-9} 
\multirow{2}{*}{\texttt{p6v11}} & NN & $87.7 \pm 1.2$\footnote{For the diffuse model \texttt{p6v11}, the two diffuse flux components are described by a single template.} & $-$ & $0.2 \pm 1.0$ & $6.9 \pm 1.1$ & $\mathbf{4.7} \pm 1.7$ & $0.5 \pm 1.4$ & $0.1 \pm 0.6$ \\
 & \texttt{NPTFit} & $88.8 \pm 0.6$ & $-$ & $0.0^{+0.2}_{-0.0}$ & $6.4 \pm 0.2$ & $0.0^{+0.1}_{-0.0}$ & $\mathbf{4.2} \pm 0.4$ & $0.9^{+0.5}_{-0.4}$ \\ \cmidrule(l){2-9} 
\multirow{2}{*}{Model~A} & NN & $47.7 \pm 1.5$ & $39.2 \pm 2.4$ & $0.5 \pm 1.2$ & $5.4 \pm 0.9$ & $\mathbf{6.7} \pm 1.4$ & $0.2 \pm 1.0$ & $0.4 \pm 1.2$ \\
 & \texttt{NPTFit} & $49.4^{+0.7}_{-0.6}$ & $35.7^{+1.2}_{-1.4}$ & $0.1^{+0.7}_{-0.1}$ & $4.9 \pm 0.3$ & $0.1^{+0.5}_{-0.1}$ & $\mathbf{6.1}^{+0.5}_{-0.6}$ & $3.7^{+1.1}_{-1.0}$ \\ \cmidrule(l){2-9} 
\multirow{2}{*}{Model~F} & NN & $57.1 \pm 1.8$ & $31.2 \pm 2.7$ & $1.2 \pm 1.7$ & $4.0 \pm 1.1$ & $\mathbf{5.6} \pm 2.0$ & $0.6 \pm 1.8$ & $0.2 \pm 0.9$ \\
 & \texttt{NPTFit} & $55.5 \pm 0.7$ & $25.7^{+1.7}_{-1.9}$ & $4.7 \pm 1.1$ & $3.7 \pm 0.4$ & $0.1^{+0.7}_{-0.1}$ & $\mathbf{4.6}^{+0.5}_{-0.7}$ & $5.8^{+1.2}_{-1.1}$ \\ \cmidrule(l){2-9} North only & NN & $63.0 \pm 2.2$ & $22.4 \pm 2.4$ & $0.1 \pm 0.5$ & $6.8 \pm 1.2$ & $\mathbf{5.2} \pm 3.1$ & $1.9 \pm 3.1$ & $0.6 \pm 1.2$ \\ \cmidrule(l){2-9}
South only & NN & $52.8 \pm 2.3$ & $28.4 \pm 2.7$ & $1.4 \pm 1.4$ & $6.7 \pm 0.9$ & $\mathbf{8.1} \pm 1.7$ & $0.2 \pm 1.3$ & $2.4 \pm 2.7$ \\ \cmidrule(l){2-9} 
\orange{Narrow priors O} & NN & $54.4 \pm 0.9$ & $26.3 \pm 1.6$ & $1.9 \pm 1.0$ & $6.1 \pm 0.6$ & $\mathbf{6.4} \pm 1.1$ & $1.1 \pm 1.1$ & $3.8 \pm 1.5$ \\ \cmidrule(l){2-9} 
\orange{Narrow priors AFO} & NN & $53.7$ & $29.9 $ & $1.9$ & $6.4$ & $\mathbf{5.6}$ & $1.7$ & $0.9$ \\
\bottomrule
 \end{tabular}
\label{table:results_summary}
\end{table}
In Table \ref{table:results_summary}, we summarize the best-fit estimates from our Bayesian GCNN and from \texttt{NPTFit} and their associated uncertainties in our default ROI (outer radius $25 ^\circ$, $|b| > 2 ^\circ$, 3FGL PSs masked). For \texttt{NPTFit}, we report the $68 \%$ credible intervals around the median whereas for the NN, we state the predicted means and $1 \sigma$ predictive uncertainties (aleatoric and epistemic uncertainties summed in quadrature) as the predictive uncertainties are close to Gaussian due to the Gaussianity of the aleatoric uncertainties, which account for the bulk of the uncertainties.
\par Our default scenario is described in the main body of this work, with Model~O as the diffuse model and a thin disk template for the disk PSs (see Fig.~\ref{fig:fermi_results} for the results as a function of the ROI radius). For the ``Thick disk'' experiment, the thin disk template (scale height $z_s = 0.3 \ \text{kpc}$) is replaced by a thicker disk ($z_s = 1 \ \text{kpc})$, see Sec.~\ref{subsec:fermi_thick_disk}. For the experiments ``\texttt{p6v11}'', ``Model~A'', and ``Model~F'', we use the respective diffuse model instead of Model~O. These experiments are presented in Sec.~\ref{subsec:fermi_p6v11}. \orange{The ``north / south only'' rows} contain the NN estimates when considering the two hemispheres individually (see Sec.~\ref{subsec:hemispheres}), which is mainly a consistency check for our NN, for which reason we did not run this test for \texttt{NPTFit}. \orange{The last two rows show the results from the NNs in Sec.~\ref{sec:narrow_priors}, for which the training maps are generated using a narrow parameter range around the expected \emph{Fermi} map values. For the variant ``O'', we take Model O as the diffuse model, whereas we allow arbitrary mixtures of Models A, F, and O for the variant ``AFO'', which the NN is trained to estimate in addition to the flux fractions. Our NN is non-Bayesian in the latter case, for which reason no uncertainties are provided in the table.} 
\par The estimated diffuse fluxes from pion decay and bremsstrahlung are consistent between the two methods, while the NN generally assigns more flux in the \emph{Fermi} map to the diffuse IC template in comparison with \texttt{NPTFit} (particularly for Models A and F). The estimates for the isotropic flux are mostly consistent with zero, with the exception of the \texttt{NPTFit} estimates for the default case ($2.3 \%$) and for Model~F ($4.7 \%$). The NN prefers slightly higher fluxes from the \emph{Fermi} bubbles than \texttt{NPTFit}, with the best-fit values ranging from $\sim~4$ to $7 \%$. In all our experiments, the NN identifies a flux fraction associated with the GCE DM template, whereas the GCE DM flux ascertained by \texttt{NPTFit} is consistent with zero in all the cases. In contrast, the NN attributes \orange{less than $2 \%$ of the flux in this ROI to GCE PS for all the experiments}, while \texttt{NPTFit} always finds a GCE with PS origin (which mostly accounts for slightly less flux than the corresponding NN estimate for GCE DM). The estimates for the disk PS flux lie between $\sim~0$ and $6 \%$ and are smallest with the thick disk template for both methods. For the estimated flux fractions within smaller ROIs, we refer the reader to the respective sections where we discuss the experiments in more detail.

\section{Modeling the GCE}
\begin{figure}[htb]
\centering
  \noindent
  \resizebox{1\textwidth}{!}{
\includegraphics{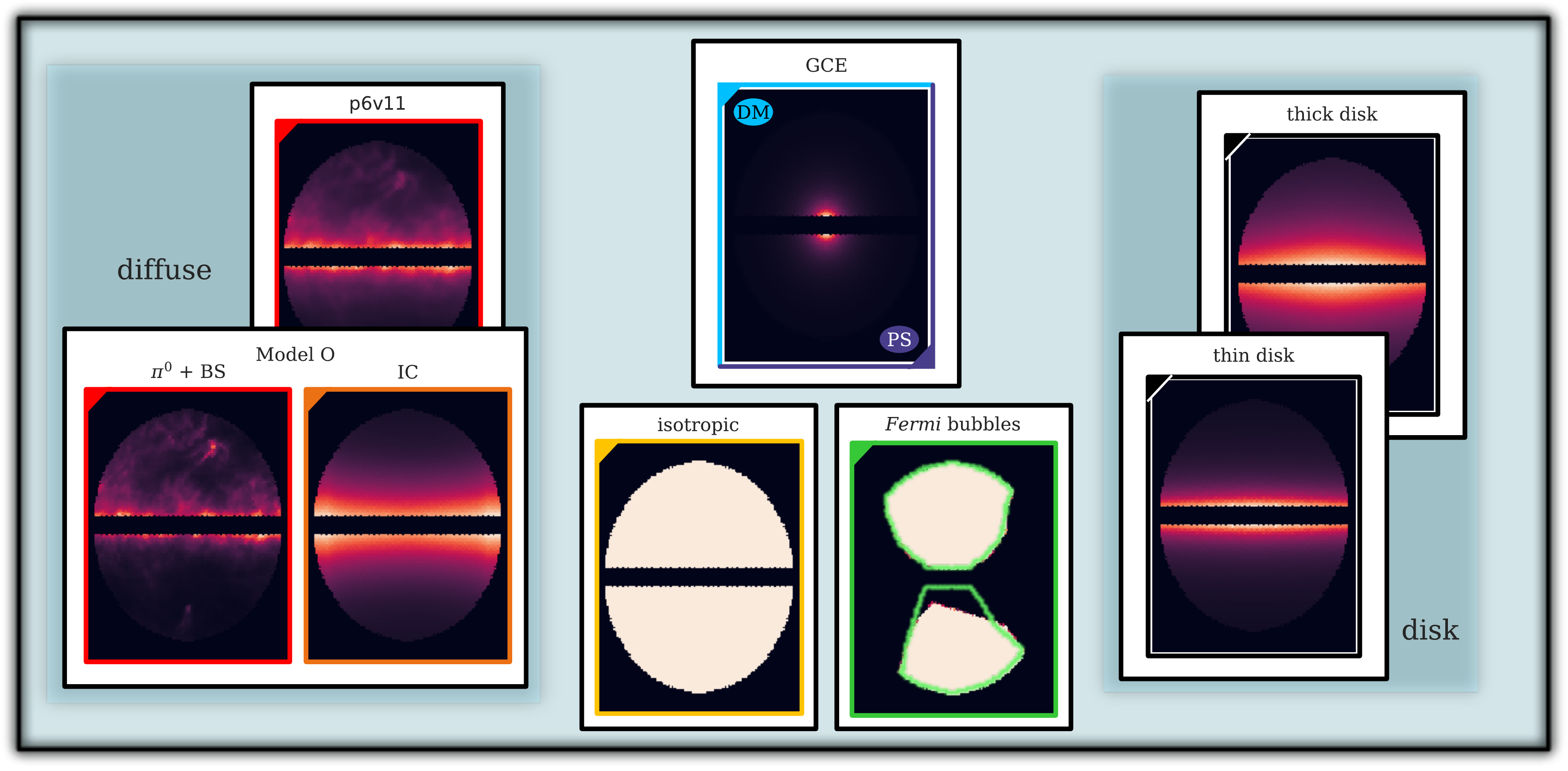} 
}
\caption{Illustration of all the templates that we use, within our largest ROI (radius $\leq 25 ^\circ$ around the GC, with $|b| \leq 2 ^\circ$ masked). The template intensity increases from black to light yellow. The exposure correction of all the templates has been removed, resulting in uniform emission from the isotropic template and from within the \emph{Fermi} bubbles. The color of each frame is associated with the respective template throughout this work. If not stated otherwise, we use the diffuse Model~O, consisting of contributions from pion decay \& bremsstrahlung ($\pi^0 + \text{BS})$ and from IC scattering. We model disk-correlated PS with the thin disk template except in Sec. \ref{subsec:fermi_thick_disk}, where we employ the thick disk template. The green contours in the sketch of the \emph{Fermi} bubbles show the extent of the alternate template that we use for a cross-check in Section \ref{subsec:bub_mismodelling}. The diffuse models A and F, which we consider in Secs. \ref{subsec:fermi_p6v11} and \ref{sec:injection_test}, are qualitatively similar to Model~O and are not shown for simplicity.}
\label{fig:templates}
\end{figure}
In this section, we describe our modeling of the GCE. We also list the parameter ranges that were used for the generation of training and test data. For comparability, we closely follow the modeling for a typical \texttt{NPTFit} analysis of the GCE and refer to \cite{Mishra-Sharma2017} for details. All our spatial templates (depicted in Fig.~\ref{fig:templates}) are normalized to have mean 1 within the inner $30 ^\circ$ around the GC when the inner $|b| \leq 2 ^\circ$ around the Galactic Plane are masked. The ROI considered throughout this work comprises an outer radius of $15 - 25 ^\circ$ around the Galactic Center, with latitudes $|b| \leq 2 ^\circ$ masked. Additionally, we mask the known bright 3FGL PSs at $95 \%$ containment for the realistic scenario (but not in the proof-of-concept example). We choose the 3FGL catalog instead of the newer 4FGL catalog \cite{Abdollahi2020} for the following two reasons: first, the 4FGL catalog contains $\sim~70 \%$ more sources that the 3FGL catalog and therefore, masking all these PSs would significantly decrease the number of pixels available to the NN. Secondly, much of the evidence for PSs from NPTF stems from analyses where a 3FGL mask was employed. For example, \cite{Buschmann2020} find negligible evidence for GCE PS when using a 4FGL mask, with a Bayes factor of only $2.6$ in favor of PSs as compared to $\sim$~2,500 in the 3FGL case, and \cite{Zhong:2019ycb} report no excess of bright regions on small angular scales in a wavelet-based analysis with a 4FGL mask, giving rise to constraints for the luminosity function of potential GCE PSs. Thus, masking only the 3FGL PSs enhances the comparability to previous literature in which a PS-dominated GCE is preferred. 
\par Throughout this work, we use the Pass 8 \emph{Fermi}-LAT photon counts within a reconstructed energy range of $2 - 20 \ \text{GeV}$ that have been detected between Aug 4, 2008 and July 7, 2016. We select events within the highest cosmic-ray rejection class, \texttt{UltraCleanVeto}, and apply the quality cuts \texttt{DATA\textunderscore QUAL==1}, \texttt{LAT\textunderscore CONFIG==1}, and zenith angle $\leq 90^\circ$. This exact data set has previously been employed in \cite{Mishra-Sharma2017, Leane2019, Buschmann2020}.
In what follows, we briefly introduce the templates and explain their physical origin. 

\subsection{Diffuse template: Model~O}
For the description of the diffuse $\gamma$-ray emission from the Milky Way, we employ the recent Model~O \cite{Buschmann2020}. While an injection test carried out in \cite{Leane2019a} showed that artificial DM flux added to the \emph{Fermi} map is not recovered by \texttt{NPTFit} below a certain flux threshold with the widely employed diffuse model \texttt{p6v11} (\texttt{gll\textunderscore iem\textunderscore v02\textunderscore P6\textunderscore V11\textunderscore DIFFUSE}\footnote{\url{https://fermi.gsfc.nasa.gov/ssc/data/access/lat/ring_for_FSSC_final4.pdf}}), it was demonstrated in \cite[e.g. Fig. 12]{Buschmann2020} that Model~O resolves this issue.\footnote{The \texttt{NPTFit} results can be further improved by applying a spherical-harmonic marginalization procedure.} With diffuse emission accounting for the bulk of the $\gamma$-ray emission measured by \emph{Fermi}-LAT, the diffuse template is the most critical and is expected to be the major source of mismodeling (see the references within this section and the references therein). Model~O consists of two individual components that model distinct physical processes: 1) gas-correlated processes: cosmic-ray protons that hit the interstellar gas produce neutral pions which then decay to photons via the process $pp \to X + \pi^0 \to X + \gamma\gamma$. Furthermore, the interaction of cosmic-ray electrons with the gas produces bremsstrahlung, although this process is subdominant to the pion decay. Since these two processes trace the distribution of the interstellar gas (mostly $H_\text{I}$ and $H_\text{II}$), they are described by a joint template. 2) Photons from the CMB and interstellar radiation fields are upscattered by cosmic-ray electrons via the inverse Compton (IC) effect and can reach $\gamma$-ray energies. The spatial distribution of the arising $\gamma$-ray emission is correlated with that of electrons and radiation fields and is therefore modeled with a separate template. In addition, we use the \texttt{GALPROP}-based Model~A and Model~F \cite{Ackermann2012, Calore2015}, and the model \texttt{p6v11}, which is the last official diffuse model from \emph{Fermi} that does not incorporate large-scale structures such as the \emph{Fermi} bubbles. The model \texttt{p6v11} accounts for the previously discussed processes with a single template, for which reason only one normalization parameter $A_\text{dif}$ is needed, compared with two in the case of Models~O, A, and F. It has been pointed out that the \texttt{p6v11} template may lead to over-subtraction in the data \cite{Calore2015, Buschmann2020}, possibly in part due to its hard IC component. Yet, considering various diffuse models and discussing the effects of diffuse mismodeling is of paramount importance in order to substantiate the preference for GCE DM that our NN finds \orange{(see Sections \ref{subsec:diffuse_mismodelling}, \ref{subsec:fermi_p6v11}, and \ref{sec:narrow_priors})}. This is because none of the diffuse models provide a perfect description of the actual data, for which reason robustness with respect to mismodeling errors is crucial for any fitting method.

\subsection{Isotropic emission}
The spatial distribution of $\gamma$-ray photons from sources outside the Milky Way can be assumed to be approximately uniform on the sky, for which reason we take a uniform Poissonian template for modeling the extragalactic background, whose intensity is set by the normalization parameter $A_\text{iso}$. Additionally, this template can absorb cosmic-ray contamination. 

\subsection{\emph{Fermi} bubbles}
The \emph{Fermi} bubbles \cite{Su2010} are a large structure with sharp edges, extending $8 - 10 \ \text{kpc}$ north and south of the GC. They emit $\gamma$-rays with a harder spectrum than the emission arising from the processes described by the diffuse template. While the exact origin of these structures is not well understood to date (see e.g. \cite{Bordoloi2017, Zhang2020}), mechanisms related to a recent explosive outburst from Milky Way's supermassive black hole, or the high star formation activity around it, might provide an explanation. Since the $\gamma$-ray emission from the \emph{Fermi} bubbles does not show large spatial variations, we model them with a spatially uniform Poissonian template (identical to the one used in \cite{Buschmann2020}) with normalization parameter $A_\text{bub}$. We do not invoke hypothetical non-Poissonian emission from the \emph{Fermi} bubbles that was used in \cite{Leane2019a} in a proof-of-concept example. In Sec. \ref{subsec:bub_mismodelling}, we analyze the effects of mismodeled \emph{Fermi} bubbles on the NN estimates by means of an alternate \emph{Fermi} bubble template with a different spatial morphology (see Fig.~\ref{fig:templates}).

\subsection{GCE emission}
The shape of the GCE has been found to be well described by a generalized NFW profile \cite{Navarro1997} parametrized as
\begin{equation}
    \rho(r) \propto \frac{\left(r / r_{s}\right)^{-\gamma}}{\left(1+r / r_{s}\right)^{3-\gamma}},
\end{equation}
regardless of its origin. Here, $\rho(r)$ is the DM density at distance $r$ from the GC, $r_s = 20 \ \text{kpc}$ is the scale radius, and $\gamma$ is a shape parameter, equal to $1$ for the canonical NFW profile. In line with previous literature (e.g. \cite{Daylan2016, Linden2016, Buschmann2020}), we adopt a slightly larger value of $\gamma = 1.2$. 
For DM pair annihilation, the resulting $\gamma$-ray flux $J(\phi)$ detected on Earth at an angle $\phi$ from the GC is proportional to the line-of-sight integral of the square of this distribution, i.e.
\begin{equation}
    J(\phi) \propto \int_{\text{l.o.s}} \rho^2(r(s)) \, ds.
\end{equation}
The strength of the Poissonian emission from the (generalized) NFW-shaped DM halo is modulated by the free parameter $A_\text{gce}$. For investigating the hypothesis that the GCE results from unresolved PS, we define a non-Poissonian template with the same spatial distribution, described by the normalization parameter $A_\text{gce}^\text{PS}$, the source-count break $S_1^\text{gce}$, and power law indices $n_1^\text{gce}$ and $n_2^\text{gce}$ above and below the count break $S_1^\text{gce}$, respectively (see Sec. \ref{subsec:params_and_priors}). We leave the investigation of different values for $\gamma$ as well as PS emission correlated to the ``Boxy Bulge'' \cite{Macias:2016nev} to future work.

\subsection{Disk-correlated PS}
For the application of our Bayesian GCNN to the \emph{Fermi} map, we additionally include a non-Poissonian template for PSs correlated with the thin disk of the Milky Way, whose spatial distribution $n(z, R)$ in cylindrical coordinates $z$ and $R$ is given by
\begin{equation}
    n(z, R) \propto \exp \left(\frac{-R}{r_s}\right) \exp \left(\frac{-|z|}{z_s}\right),
\label{eq:disk_PS}
\end{equation}
where we take the scale radius to be $r_s = 5 \ \text{kpc}$ and the scale height $z_s = 0.3 \ \text{kpc}$ as in \cite{Lee2016}, matching the best-fit parameter $z_s$ for the description of the \orange{MSPs} correlated with the Galactic Disk obtained in \cite{Lorimer2006}. The case of a thicker disk with $z_s = 1 \ \text{kpc}$ that was employed in e.g. \cite{Leane2019a, Buschmann2020} is considered in Sec. \ref{subsec:fermi_thick_disk}. In order to obtain the final spatial template for the photon count distribution with free parameters $A_\text{disk}^\text{PS}$, $S_1^\text{disk}$, $n_1^\text{disk}$, and $n_2^\text{disk}$, we integrate $n(z, R)$ along each line-of-sight direction.

\subsection{Model parameters and prior ranges}
\label{subsec:params_and_priors}
\begin{table}[ht]
\caption{Prior ranges for the generation of training and test data, for the Poissonian and non-Poissonian templates. The prior ranges for $A_\text{dif}$ and for the disk-related parameters are used for all the diffuse templates and disk templates that we consider, respectively.}
\centering
\begin{tabular}{@{}lc@{}}
\toprule
Model parameter                     & Prior range  \\ \midrule
$\log_{10} A_\text{dif}$            & $[0, 2]$     \\
$\log_{10} A_\text{bub}$            & $[-3, 2]$    \\
$\log_{10} A_\text{iso}$            & $[-3, 2]$    \\
$\log_{10} A_\text{gce}$            & $[-3, 2]$    \\
$\log_{10} A_\text{gce}^\text{PS}$  & $[-6, 1]$    \\
$\log_{10} A_\text{disk}^\text{PS}$ & $[-6, 2]$    \\
$S_1^\text{gce}$                    & $[0.05, 60]$ \\
$S_1^\text{disk}$                   & $[0.05, 60]$ \\
$n_1^\text{gce}$                    & $[5, 60]$    \\
$n_2^\text{gce}$                    & $[-3, 0.95]$ \\
$n_1^\text{disk}$                   & $[2.05, 60]$ \\
$n_2^\text{disk}$                   & $[-3, 0.95]$ \\ \bottomrule
\end{tabular}
\label{table:priors}
\end{table}

\par Table \ref{table:priors} lists the prior ranges of the uniform distributions from which we draw the parameters when generating the mock maps. For Poissonian templates, there is only a single parameter, namely the normalization parameter $A$ that determines the strength of each template: the mean number of counts for template $t$ in pixel $p$ is given as $\mu_{p, t} = A_t T_{p, t}^{(S)}$, where $T_{p, t}^{(S)}$ is the value of the spatial template $t$ in pixel $p$ in terms of counts (i.e. with exposure correction applied). For the non-Poissonian templates that model unresolved PSs whose exact locations within the template are not known, we take a broken power law for the SCD function with negative slopes $n_1$ and $n_2$ above and below the count break $S_1$. To be specific, the differential number of sources $dN_{p, t}$ from template $t$ per pixel per differential flux interval $dF$ is given as
\begin{equation}
\frac{d N_{p, t}}{d F}(F) = A_t T_{p, t}^{(\mathrm{PS})}\left\{\begin{array}{ll}
\left(\frac{F}{F_{1}}\right)^{-n_{1}}, & F \geq F_{1}, \\
\left(\frac{F}{F_{1}}\right)^{-n_{2}}, & F < F_{1},
\end{array}\right.
\end{equation}
where $T_{p, t}^{(PS)}$ is the spatial template for the PS distribution, and the flux $F$ and flux break $F_1$ can be recast into counts $S$ and count break $S_1$, respectively, via the \emph{Fermi} exposure map $E_p$ (which has units $\text{cm}^2 \ \text{s}$):
\begin{equation}
    \frac{d N_{p, t}}{d S}(S) = \frac{1}{E_{p}} \frac{d N_{p, t}}{d F} \left(F = S / E_{p}\right).
\end{equation}
When generating realizations of the non-Poissonian templates using \texttt{NPTFit-Sim} \cite{NPTFit-Sim}, the \emph{Fermi} PSF is taken into account.

\subsection{Data generation}
\label{subsec:data_gen}
In order to generate a sufficient amount of training data, we create a catalog of 602,500 count maps from each of the Poissonian templates by randomly sampling from Poissonian distributions with pixel-wise means set by the respective template (accounting for the \emph{Fermi} exposure map). For the generation of non-Poissonian data, we use the publicly available tool \texttt{NPTFit-Sim} \cite{NPTFit-Sim} and take the \emph{Fermi}-LAT point spread function into account. Since each non-Poissonian map generation requires running a rejection sampling algorithm and is therefore computationally more expensive, we create 37,500 maps for each non-Poissonian template. We perform this procedure twice; once for the \emph{Fermi} exposure map and once for uniform \emph{Fermi} mean exposure. While the NN performance did not deteriorate in the presence of a non-uniform exposure map in our experiments (and the \emph{Fermi} exposure map is very homogeneous within our ROI in any case), a uniform exposure map allows us to choose a single exposure region in \texttt{NPTFit} and to set the parameter $n_\text{exp} = 1$, speeding up the template fitting. These data sets provide the basis for all our experiments, and maps from templates that are not present in a certain experiment are simply discarded before executing the steps that follow. We randomly lay aside 2,500 maps from each template and sum them up to obtain a testing data set. For the generation of training data, we use each of the remaining non-Poissonian template maps multiple times and randomly combine them with the 600,000 unused maps from each Poissonian template, yielding in total 600,000 training samples. Note that the photon-count maps do not obey any symmetries in the realistic scenario due to the non-uniform sky coverage by the \emph{Fermi}-LAT instrument, for which reason data augmentation by techniques such as mirroring is not an option. Although mirroring could be used for the experiments with uniform exposure map, we refrain from doing so \orange{as} we are confident that 600,000 training maps are sufficient, \orange{which is suggested by the evolution of the NN accuracy with increasing number of training maps (discussed in Sec.~\ref{subsec:no_training_maps}), and by the very small epistemic uncertainties that we observe. As an additional test, we train our NN on a narrower prior range around the expected parameters for the \emph{Fermi} map, in which case we generate the Poissonian template maps \emph{on-the-fly} during the NN training, resulting in \emph{each training map to be unique}, and we combine them with 100,000 realizations of each non-Poissonian template (see Sec.~\ref{sec:narrow_priors} for the results).}  
\par As discussed in the main body, in order to compare the performance of the NN and NPTF measurements of the flux fraction in the proof-of-concept example, we took 256 maps from the testing set, and ran the two techniques on each map. The NN produces \emph{some} output for every input map vector with the correct shape (even for inputs that do not even correspond to a photon-count map at all, in which case the NN predictions are meaningless of course), for which reason convergence considerations that are vital for sampling-based methods such as NPTF do not apply for the NN. However, the NPTF as implemented in \texttt{NPTFit} ran into difficulty in a number of cases. The source of the problem is that by drawing random values from the priors in Table \ref{table:priors}, maps can be generated with a significantly larger number of counts than in the actual \emph{Fermi} data considered in this work. In the case where the maximum photon-per-pixel count is a factor of a few higher than in the current data, \texttt{NPTFit} is unable to evaluate the non-Poissonian likelihood for certain choices of the model parameters. This issue is not a fundamental limitation of the NPTF method, but rather with the present implementation in \texttt{NPTFit}. As the issue is a function of the maximum counts-per-pixel and also the model parameters, we cannot simply resolve it by removing simulated datasets above a certain count threshold. As the goal is to obtain a rough comparison between the two methods, instead we adopted the simple criterion of discarding maps where \texttt{NPTFit} was unable to converge on a run of $n_\text{live} = 50$ live points within an hour. We caution that this cut undoubtedly introduces a bias away from cases where the NPTF likelihood may be evaluated correctly, but the fit is having difficulty converging, and thus is likely to overstate the performance of \texttt{NPTFit}. For this reason, a detailed comparison of the performance of the two methods is left to future work. We stress also that this issue has no bearing on the performance on the NN, or on the conclusion the two methods reach in the GC. It only impacts the presentation of the comparison shown in the main body.

\section{Neural network architecture}
\begin{figure}[htb]
\centering
  \noindent
  \resizebox{1\textwidth}{!}{
\includegraphics[trim=8.5cm 3.5cm 8.5cm 6cm,clip]{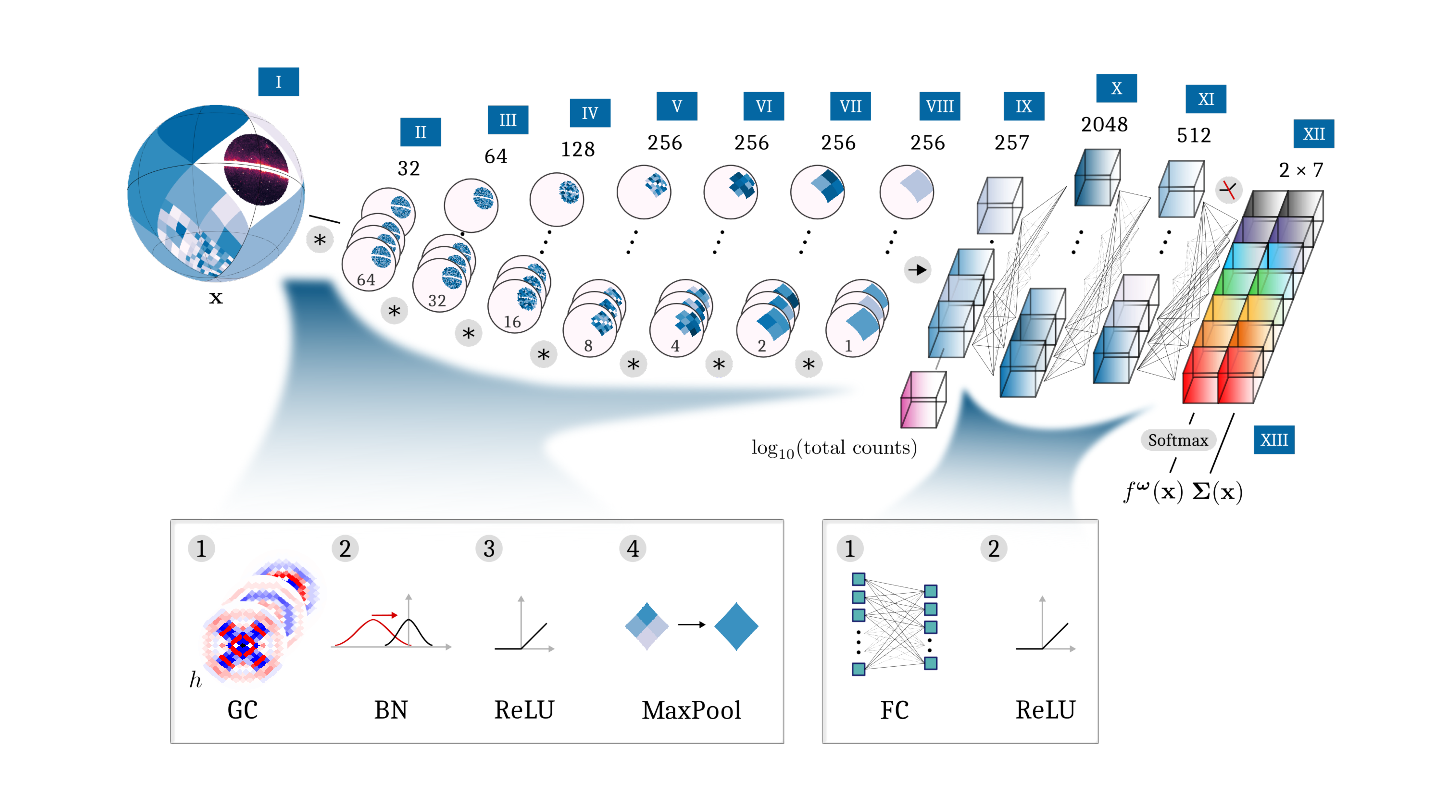}
}
\caption{
Sketch of our GCNN structure. The layer numbers are indicated by white roman numerals (more details for each layer can be found in Table \ref{table:NN}). The sphere in Layer I shows our ROI embedded within the central coarse pixel at $N_\text{side} = 1$ (in white). The coarse pixel to the lower left of the ROI is subdivided into pixels at resolutions $N_\text{side} = 2, 4, 8,$ and $16$ for illustration purposes. The spheres in Layers II -- VIII stand for the channels (or feature maps) of the convolutional layers and exhibit how the spatial resolution is gradually reduced due to the maximum pooling operations (MaxPool). Each convolutional block \raisebox{.5pt}{\textcircled{\raisebox{-.4pt} {$\ast$}}} consists of the four operations depicted in the box below (GC: graph convolutional layer, for which three exemplary kernels $h$ are shown, BN: batch normalization, ReLU: Rectified Linear Unit activation function, and maximum pooling). The numbers inside the spheres give the $N_\text{side}$ parameter that belongs to the respective spatial resolution, while the Arabic numerals above the spheres indicate the number of channels for each layer. In Layer VIII, only a single pixel value remains for each channel ($256$ in total). The logarithm of the total number of photons in the map is then appended, and two fully connected (FC) layers with ReLU activation follow. Layer XII is a fully connected layer whose output dimension is $2 \, T$ (for $T$ templates, e.g. 7 in the realistic scenario), after which no ReLU is applied. Instead, the output is reshaped to $2 \times T$ (means and log-variances), and a Softmax activation function is used in order to obtain the flux fraction means. The remaining $T$ output elements form the vector $(\log \sigma^2_t(\mathbf{x}))_{t=1}^T$, with $\bm{\Sigma}(\mathbf{x}) = \text{diag}[\sigma_1^2(\mathbf{x}), \ldots, \sigma_T^2(\mathbf{x})]$.}
\label{fig:NN_sketch}
\end{figure}
\begin{table}[ht]
\caption{Tabulated architecture of the Bayesian GCNN used for the \emph{Fermi} map (also see the sketch in Fig.~\ref{fig:NN_sketch}). 
The trainable parameters are split up into the matrix weights, biases, and trainable Dropout probabilities.}
\centering
\begin{tabular}{@{}lllcc@{}}
\toprule
Layer & Operations & Tensor output shape & \multicolumn{1}{l}{Output $N_\text{side}$} & Trainable parameters \\ \midrule
I & Input & 16,384 $\times$ 1 & 128 & $-$ \\
II & MaxPool $\circ$ ReLU $\circ$ BN $\circ$ GC & 4,096 $\times$ 32 & 64 & 320 + 32 + 1 \\
III & MaxPool $\circ$ ReLU $\circ$ BN $\circ$ GC & 1,024 $\times$ 64 & 32 & 20,480 + 64 + 1 \\
IV & MaxPool $\circ$ ReLU $\circ$ BN $\circ$ GC & 256 $\times$ 128 & 16 & 81,920 + 128 + 1 \\
V & MaxPool $\circ$ ReLU $\circ$ BN $\circ$ GC & 64 $\times$ 256 & 8 & 327,680 + 256 + 1 \\
VI & MaxPool $\circ$ ReLU $\circ$ BN $\circ$ GC & 16 $\times$ 256 & 4 & 655,360 + 256 + 1 \\
VII & MaxPool $\circ$ ReLU $\circ$ BN $\circ$ GC & 4 $\times$ 256 & 2 & 655,360 + 256 + 1 \\
VIII & MaxPool $\circ$ ReLU $\circ$ BN $\circ$ GC & 1 $\times$ 256 & 1 & 655,360 + 256 + 1 \\
IX & Append $\log_{10}$(total photon counts) & 1 $\times$ 257 &  & $-$ \\
X & ReLU $\circ$ FC & 1 $\times$ 2,048 &  & 526,336 + 2,048 + 1 \\
XI & ReLU $\circ$ FC & 1 $\times$ 512 &  & 1,048,576 + 512 + 1 \\
XII & Reshape $\circ$ FC & 2 $\times$ 7 &  & 7,168 + 0 + 1 \\
XIII & Softmax (means only) & 2 $\times$ 7 &  & $-$ \\
 & &  &  & 3,982,378 \\ \bottomrule
\end{tabular}
\label{table:NN}
\end{table}

A NN is assembled as a sequence of building blocks (so-called layers), each of which transforms its input tensor to an output tensor (that possibly has different dimensions). Thus, each input map $\mathbf{x}$ is gradually processed by the individual NN layers, and the final NN layer produces the output, which is in our case given by the estimated flux fractions $f^{\bm{\omega}}(\mathbf{x})$ and their associated aleatoric uncertainties $\bm{\Sigma}(\mathbf{x})$. An illustrative sketch of our NN is provided in Fig.~\ref{fig:NN_sketch}. In Table \ref{table:NN}, we list the individual layers of the Bayesian GCNN used for the \emph{Fermi} map. The input $\mathbf{x}$ for the NN is given by photon-count maps, from which we remove the exposure correction (that is, we divide the counts in each pixel $p$ by $E_p / \bar{E}$, where $\bar{E}$ stands for the mean exposure) and subsequently scale the pixel values such that they sum to unity since NNs prefer normalized inputs. Hence, the NN input consists of a single channel (in contrast to e.g. color images with 3 channels for RGB values). An interesting extension could be gathering the \emph{Fermi} counts in multiple energy bins each of which would be treated as an individual channel. We leave this for future work. The 16,384 pixels at the initial resolution of $N_\text{side} = 128$ are those that lie within the HEALpix pixel at the coarsest resolution ($N_\text{side} = 1$) which contains the GC (the white coarse pixel in Layer I in Fig.~\ref{fig:NN_sketch}). This pixel is delimited by the four corners at $(l, b) = (0 ^\circ, 41.8 ^\circ), (315 ^\circ, 0 ^\circ), (0 ^\circ, -41.8 ^\circ), (45 ^\circ, 0)$. Thus, an input size of 16,384 $= N_\text{side}^2 = 1/12 \ N_\text{pix}$ pixels is sufficient for all ROIs at resolution $N_\text{side} = 128$ that fit within this coarse $N_\text{side} = 1$ pixel. For larger ROIs, our NN architecture would need to be adapted to include the neighboring coarse pixels (for considering the entire sky, the number of pixels at $N_\text{side} = 128$ would be $12 \ N_\text{side}^2 = N_\text{pix} =$ 196,608). 
\par 
The input maps are then processed by 7 consecutive blocks of graph convolutions, each of which is followed by a batch normalization operation \cite{Ioffe2015}, a ReLU non-linearity, and a maximum pooling operation, which reduces the $N_\text{side}$ parameter by a factor of 2 (and therefore the number of pixels by 4). For the graph convolutions, we start with $32$ feature maps and double the number with every subsequent graph convolution up to a maximum of $256$ feature maps, which we use for the last 4 graph convolutions. The receptive field is set by the degree of the Chebyshev polynomials, which we take to be $9$ ($4$ for the proof-of-concept experiment). The graph convolutional layers act as feature extractors that detect characteristics such as edges of the templates (for instance, the sharp borders of the \emph{Fermi} bubbles) and gradients. After applying the maximum pooling 7 times, the spatial dimension has reduced to $1$ since the ROI of $\leq 25 ^\circ$ around the GC considered here fits within one of the twelve pixels of the coarsest HEALPix tessellation ($N_\text{side} = 1$). Convolving the input maps so often that only one pixel remains whose receptive field extends over the \emph{entire} ROI turned out to be crucial in our experiments. This is intuitive since although the NN bases its inference on the \emph{local} pixel information in the input maps, it outputs a single number per template for the corresponding (\emph{global}) flux fraction within the entire ROI. We append an additional channel with the logarithm of the total number of counts in the respective map (with exposure correction removed) as a supplementary criterion for the NN to render its judgment. Finally, we employ fully connected layers (whose task is to interpret the 256(+1)-dimensional output from the convolutional layers) with 2,048 and 512 neurons followed by ReLU activations, and a final fully connected layer with $2 \times T$ output neurons (means and log-variances), where we use a Softmax activation for the means in order to enforce the flux fractions to be in $[0, 1]$ and to sum up to one. In what follows, we briefly introduce the building blocks of our Bayesian GCNN, summarize the loss functions that we use in this work, introduce \emph{Concrete Dropout}, and provide the details of the NN training.

\subsection{Architectural building blocks}
In order for the NN to learn from the training data, it requires trainable parameters that are gradually adjusted during the training process using a stochastic gradient descent method (in our case an Adam optimizer \cite{Kingma2014}) in order to minimize a user-defined loss function (see Eq.~\eqref{eq:training_objective}). Our Bayesian NN features two different types of layers with trainable parameters: graph convolutional layers, and fully connected layers. 

\subsubsection{Graph convolutional layers}
The graph convolutional layers form the core of our GCNN and act as feature extractors, detecting salient features in the photon-count maps. The sphere is described by a graph in DeepSphere with one vertex located at each pixel center, connected to the vertices that represent the neighboring pixels. Hence, the convolution operation differs from the classical notion of sliding a convolutional kernel over the image and instead leverages the fact that the convolution operation is tantamount to a multiplication in the spectral domain. Thus, a signal $\mathbf{f} \in \mathbb{R}^{n_{\text{pix}}}$ can be convolved with a kernel (also known as filter) $h$ in the spectral domain as $\mathbf{f} \mapsto \mathcal{F}^{-1}_\mathcal{G} \left\{h \, \mathcal{F}_\mathcal{G}\{\mathbf{f}\} \right\}$, where $\mathcal{F}_\mathcal{G}$ and $\mathcal{F}^{-1}_\mathcal{G}$ denote the graph Fourier transform and its inverse on the weighted undirected graph $\mathcal{G}$, and $n_\text{pix} \leq N_\text{pix}$ is the spatial dimension of the signal. This Fourier transform can be made explicit by observing that the Fourier basis on $\mathcal{G}$ is given by the eigenvectors $\{\mathbf{u}_i\}_{i=1}^{n_\text{pix}}$ of the symmetric normalized graph Laplacian $\mathbf{L}$ defined as $\mathbf{L} = \mathbf{I} - \mathbf{D}^{-1/2} \mathbf{W} \mathbf{D}^{-1/2}$, where $\mathbf{W} = (w_{ij}) \in \mathbb{R}^{n_\text{pix} \times n_\text{pix}}$ is the weighted adjacency matrix, $\mathbf{D} = (d_{ij})$ is the diagonal matrix with $d_{ii} = \sum_j w_{ij}$, and $\mathbf{I} \in \mathbb{R}^{n_\text{pix} \times n_\text{pix}}$ is the identity matrix. The eigenvectors of the Laplacian on the DeepSphere graph constructed on the HEALPix sphere have a tight connection with the spherical harmonics (see \cite[Fig. A.12]{Perraudin2019a}). We use the choice of weights proposed in \cite{Perraudin2019a} as we did not notice an improvement with the updated weighting scheme presented in \cite{Defferrard2020}, that is
\begin{equation}
w_{ij}= \left\{\begin{array}{ll}
\exp \left(-\frac{\left\|\boldsymbol{r}_{i}-\boldsymbol{r}_{j}\right\|_{2}^{2}}{\rho^{2}}\right) & \text { if pixels } i \text { and } j \text { are neighbors, } \\
0 & \text { otherwise, }
\end{array}\right.
\label{eq:weight_matrix}
\end{equation}
where $\mathbf{r}_i \in \mathbb{R}^3$ is the position of pixel $i$ and $\rho$ is the average Euclidean distance between two neighboring pixels.
Diagonalizing the graph Laplacian into $\mathbf{L} = \mathbf{U} \bm{\Lambda} \mathbf{U}^\top$ with the orthonormal matrix of eigenvectors $\mathbf{U} = [\mathbf{u}_1, \ldots \mathbf{u}_{n_\text{pix}}]$ and the diagonal matrix of eigenvalues $\bm{\Lambda}$, the Fourier transform simply becomes $\hat{\mathbf{f}} = \mathcal{F}_\mathcal{G} \{\mathbf{f}\} = \mathbf{U}^\top \mathbf{f}$, and $\mathcal{F}^{-1}_\mathcal{G} \{\hat{\mathbf{f}}\} = \mathbf{U} \hat{\mathbf{f}} = \mathbf{U} \mathbf{U}^\top \mathbf{f} = \mathbf{f}$. Assuming that the convolutional kernel $h$ is given by a polynomial (or more generally can be expressed as a power series), the convolution operation takes the form
\begin{equation}
    \mathbf{f} \mapsto \mathcal{F}^{-1}_\mathcal{G} \left\{h(\bm{\Lambda}) \, \mathcal{F}_\mathcal{G}\{\mathbf{f}\} \right\} = \mathbf{U} h(\bm{\Lambda}) \mathbf{U}^\top \mathbf{f} = h(\mathbf{L}) \mathbf{f},
\label{eq:filtering}
\end{equation}
where $h(\bm{\Lambda})$ is the diagonal matrix with $h(\bm{\Lambda})_{ii} = h(\bm{\Lambda}_{ii})$. Defining the kernels $h$ as Chebyshev polynomials with coefficients $\bm{\theta} = \{\theta_{k}\}_{k=0}^{K-1}$, Eq.~\eqref{eq:filtering} yields
\begin{equation}
\mathbf{f} \mapsto h_{\bm{\theta}}(\tilde{\mathbf{L}}) \mathbf{f}=\mathbf{U}\left(\sum_{k=0}^{K-1} \theta_{k} T_{k}(\tilde{\bm{\Lambda}})\right) \mathbf{U}^{\top} \mathbf{f}=\sum_{k=0}^{K-1} \theta_{k} T_{k}(\tilde{\mathbf{L}}) \mathbf{f},
\label{eq:filtering_cheby}    
\end{equation}
with the rescaled Laplacian 
\begin{equation}
    \tilde{\mathbf{L}} = -\frac{2}{\lambda_{\max}} \mathbf{D}^{-1 / 2} \mathbf{W} \boldsymbol{D}^{-1 / 2}
\end{equation}
that has eigenvalues in $[-1, 1]$, gathered in the diagonal matrix $\tilde{\bm{\Lambda}}$. Here, $\lambda_\text{max}$ is the largest eigenvalue of $\bm{\Lambda}$. The Chebyshev polynomial $T_k$ of degree $k$ is defined recursively as $T_0(\tilde{\mathbf{L}}) = \mathbf{I}$, $T_1(\tilde{\mathbf{L}}) = \tilde{\mathbf{L}}$, and $T_k(\tilde{\mathbf{L}}) = 2 \tilde{\mathbf{L}} T_{k-1}(\tilde{\mathbf{L}}) - T_{k-1}(\tilde{\mathbf{L}})$. Hence, the filtering in Eq.~\eqref{eq:filtering_cheby} only requires $\mathcal{O}(K)$ matrix multiplications, each of which can be done in $\mathcal{O}(n_\text{pix})$ since each vertex is connected to at most 8 neighbors and the weights are chosen such that $w_{ij} = 0$ whenever vertex $i$ is not a neighbor of vertex $j$ (see Eq.~\eqref{eq:weight_matrix}). This implies that the graph convolution operator on the full sphere scales as $\mathcal{O}(N_\text{pix})$ and therefore more efficiently than a spherical harmonic transform on the HEALPix sphere, which has a computational cost of $\mathcal{O}(N_\text{pix}^{3/2})$. Note that although the spectral domain is exploited for the definition of the graph convolution operation, the explicit dependence on the spectrum has been eliminated in Eq.~\eqref{eq:filtering_cheby}, and each term $T_k(\tilde{\mathbf{L}})$ captures $k$-neighborhoods in the pixel domain. This stems from the fact that each entry $(\mathbf{W}^k)_{ij}$ is the sum of all path weights for paths of length $k$ between vertices $i$ and $j$, where the path weight is defined as the product of all the edge weights along the path. The kernel $h$ can be localized at a specific vertex $i$ by means of the localization operator $\mathcal{T}_i$ defined as $\mathcal{T}_i h_{\bm{\theta}} = h_{\bm{\theta}}(\tilde{\mathbf{L}}) \bm{\delta}^i = (h_{\bm{\theta}}(\tilde{\mathbf{L}}))_i$, where the Kronecker delta $\bm{\delta}^i \in \mathbb{R}^{n_\text{pix}}$ extracts the $i$-th column of $h_{\bm{\theta}}(\tilde{\mathbf{L}})$.
\par The number of trainable matrix parameters for graph convolutional layers is $n_\text{in} \times n_\text{out} \times K$, where $n_\text{in}$ and $n_\text{out}$ are the number of input and output channels, respectively; for instance 20,480 $ = 32 \times 64 \times 10$ in Table \ref{table:NN}. An output channel $\mathbf{y}_i \in \mathbb{R}^{n_\text{pix}}$ for $i \in \{1, \ldots, n_\text{out}\}$ is computed from the input channels $\{\mathbf{x}_j\}_{j=1}^{n_\text{in}}$, where $\mathbf{x}_j \in \mathbb{R}^{n_\text{pix}}$, as
\begin{equation}
    \mathbf{y}_{i} = \sum_{j=1}^{n_{\text{in}}} h_{\bm{\theta}_{ij}}(\tilde{\mathbf{L}}) \mathbf{x}_{j} + b_{i},
\end{equation}
where $\mathbf{b} = (b_i) \in \mathbb{R}^{n_\text{out}}$ is a (trainable) bias vector, and each graph convolution $h_{\bm{\theta}_{ij}}(\tilde{\mathbf{L}}) \mathbf{x}_{j}$ is computed using Eq.~\eqref{eq:filtering_cheby}.
While DeepSphere by construction requires the filters to be radially symmetric (each coefficient $\theta_k$ corresponds to a $k$-neighborhood, not to a specific neighboring vertex; see the examples for $h$ depicted in Fig.~\ref{fig:NN_sketch}), this lack of non-symmetric filters did not seem to impair the NN performance when we compared our DeepSphere-based NN architecture to a canonical 2D CNN applied to projected photon-count maps, for which reason we decided to proceed with the DeepSphere-based NN, which readily takes the spherical photon-count maps as an input without the need for any projections. We refer the interested reader to \cite{Perraudin2019a} for further details on the graph convolution operation.

\subsubsection{Fully connected layers}
For rotationally symmetric problems on the HEALPix sphere, it is common to successively apply several graph convolutions and to average the outcome over the spatial dimension in order to obtain the NN output, giving rise to a so-called fully convolutional network. Since the $\gamma$-ray emission from the GC is not symmetric, however, and the exact location of the detected photons within a map is crucial, we use fully connected layers to process the output of the graph convolutional layers. Their task is to interpret the results from the convolution process and to infer the associated flux fractions. Fully connected layers map an input vector $\mathbf{x} \in \mathbb{R}^{n_\text{in}}$ to an output vector $\mathbf{y} \in \mathbb{R}^{n_\text{out}}$ via the affine mapping
\begin{equation}
    \mathbf{y} = \mathbf{W} \mathbf{x} + \mathbf{b},
\end{equation}
where the matrix $\mathbf{W} \in \mathbb{R}^{n_\text{out} \times n_\text{in}}$ and the bias vector $\mathbf{b} \in \mathbb{R}^{n_\text{out}}$ are trainable parameters (see the last column in Table \ref{table:NN}). For fully connected layers, the number of trainable matrix weights is simply $n_\text{in} \times n_\text{out}$; for example 526,336 $= 257 \times$ 2,048. The length of the bias equals the number of output channels, and we use no bias after the last fully connected layer before reshaping the tensor from $1 \times 14$ to $2 \times 7$. For the estimation of the epistemic uncertainties via \emph{Concrete Dropout} \cite{Gal2017}, the probabilities for the Dropouts after each fully connected and graph convolutional layer are trainable parameters as well and are adjusted during the training stage (see Sec. \ref{subsec:concrete_dropout}). 

\subsubsection{Additional building blocks}
\par Batch normalization fixes the means and variances within each mini-batch during training, while the population statistics are used for the scaling when applying batch normalization at the inference stage \cite{Ioffe2015}. This has been shown to improve the performance of NNs. In our case, the maximum pooling operation computes the maximum of the four values in each patch of adjacent pixels that together form the larger pixel of the next coarser HEALPix hierarchy level, thereby coarsening its input by a factor of four (\cite[Fig. 5]{Perraudin2019a}, see also Fig.~\ref{fig:NN_sketch}).
\par Activation functions are a key element in NNs since they introduce non-linearities that enable the NN to learn complex mappings. We use the following two activation functions: the ReLU activation function is defined element-wise as $\text{ReLU}(x) = \max\{x, 0\}$, and the Softmax function is written as
\begin{equation}
    \text{Softmax}(\mathbf{x})_n = \frac{e^{x_n}}{\sum_{m=1}^M e^{x_m}},
\end{equation}
for a vector $\mathbf{x} = (x_1, \ldots, x_M) \in \mathbb{R}^M$. The Softmax activation is often applied in the context of classification  tasks in connection with a cross-entropy loss. Since our task at hand is a global regression problem for the flux fractions with the additional constraint that all the flux fractions sum up to 1, we harness the Softmax function for automatically enforcing this normalization condition for the means in the final layer (as $\sum_{m=1}^M \text{Softmax}(\mathbf{x})_m = 1$), while resorting to a loss function suited for regression given by the negative log-likelihood (see Eq.~\eqref{eq:aleatoric_llh} and Sec. \ref{sec:losses} below). 
\par An architectural particularity in our NN consists in appending the (logarithm of the) total number of counts in the map (after removing the exposure correction) as an additional channel in Layer IX. Our motivation for this is the following: since we normalize the maps before showing them to the NN, each pixel contains the \emph{relative} flux within the map; therefore, the information about how many counts / how much flux was originally in the map is lost. However, feeding this information to the NN might provide additional information. Consider the following example for illustration: a uniformly exposed photon-count map consisting of 5 pixels with 1 detected count in each pixel looks identical to a map with 10 detected counts in each pixel after the normalization. However, the probability of the first map arising from 5 independent Poissonian sources (one per pixel) with expectation value $\mu_p = 1$ is $0.67 \%$, whereas the probability of the second map arising from one Poissonian source in each pixel with expectation value $\mu_p = 10$ is only $0.0031 \%$ (since the variance of a Poissonian distribution equals its mean). Thus, the likelihoods of the flux fractions may depend on the total number of photons in the map, and the NN might adapt its estimates of the means and uncertainties accordingly. We will explore in future work whether this additional channel has an appreciable effect on the NN accuracy.

\subsection{Aleatoric uncertainty estimation and the choice of the loss function}
\label{sec:losses}
The choice of the loss function is crucial since the loss function epitomizes what the NN is meant to learn, or more specifically, which metric between the true and estimated flux fractions is supposed to be minimized. For our full Bayesian GCNN applied to the \emph{Fermi} map, we assume a Gaussian likelihood for the NN estimates, that is $p(\mathbf{y} | f^{\bm{\omega}}(\mathbf{x})) \sim \mathcal{N}\left(f^{\bm{\omega}}(\mathbf{x}), \bm{\Sigma}(\mathbf{x}) \right)$, and treat the flux estimates for the individual templates as being independent from each other, resulting in the loss function given by Eq.~\eqref{eq:aleatoric_llh}.
\par In our benchmark tests (proof-of-concept example in the main body and in Sections~\ref{sec:SCDs}, \ref{sec:mismodelling}, \ref{sec:add_mat_toy}), estimating uncertainties is less important since the true flux fractions are known in these experiments, for which reason we use a non-Bayesian GCNN with a simple $l^2$ loss function for the training, that is
\begin{equation}
    \mathcal{L}(f^{\bm{\omega}}(\mathbf{x}), \mathbf{y}) = \sum_{t=1}^T \left( f^{\bm{\omega}}(\mathbf{x})_t - \mathbf{y}_t \right)^2.
    \label{eq:l2loss}
\end{equation}
Note the similarity of this loss function to Eq.~\eqref{eq:aleatoric_llh} in the Bayesian case, where the $l^2$ error for each template is scaled with the inverse (aleatoric) uncertainty standard deviation (or precision) $\sigma_t^{-1}$, and the magnitude of $\sigma_t$ is controlled by a penalty term that arises naturally from assuming Gaussian likelihoods for the flux fraction of each template. The scaling of the errors for each template with the respective $\sigma_t^{-1}$ for the Bayesian NN can be interpreted as a loss attenuation \cite{Kendall2017}. The output shape of the final non-Bayesian NN layer is therefore $1 \times T$ (cf. Table \ref{table:NN} for the Bayesian case, where the output layer is two-headed to enable the prediction of $\{\sigma_t\}_{t=1}^T$). We do not employ any Dropout when using the non-Bayesian GCNN since we did not observe any signs of overfitting in our experiments.
\par In Sec. \ref{subsec:full_covar}, we present the extension of our Bayesian GCNN to the case of full covariance uncertainty matrices, which permits to analyze how the uncertainties between the flux fractions are correlated. To this end, Eq.~\eqref{eq:aleatoric_llh} is generalized to the non-diagonal case, yielding
\begin{equation}
    \mathcal{L}(f^{\bm{\omega}}(\mathbf{x}), \mathbf{y}) = \frac{1}{2} \left(f^{\bm{\omega}}(\mathbf{x}) -\mathbf{y}\right)^\top \bm{\Sigma}(\mathbf{x})^{-1} \left(f^{\bm{\omega}}(\mathbf{x}) -\mathbf{y}\right) + \frac{1}{2} \log \left(\det \bm{\Sigma}(\mathbf{x}) \right) + \frac{T}{2} \log(2 \pi). 
    \label{eq:loss_full_covar}
\end{equation}
Hence, the required output dimension of the NN for $T$ templates becomes $T (T + 3) / 2$, namely $T$ means, $T$ variances, and $\sum_{t=1}^{T-1} t = T (T-1) / 2$ correlations. We present the results for the \emph{Fermi} map in Fig.~\ref{fig:corner}.  
\par Another interesting approach is modeling the likelihoods with a Laplace distribution in lieu of a Gaussian distribution. Taking the associated negative log-likelihood as the loss function for the NN results in an $l^1$-type loss function, where the error for each template scales inversely with the scale parameter $b$, which yields (when treating the flux fractions of the templates independently from each other):
\begin{equation}
\begin{aligned}
    \mathcal{L}(f^{\bm{\omega}}(\mathbf{x}), \mathbf{y}) &= \sum_{t=1}^{T} \left( \frac{1}{b(\mathbf{x})_t} \big{|} f^{\bm{\omega}}(\mathbf{x})_t - \mathbf{y}_t \big{|} + \log (2 b(\mathbf{x})_t) \right),
\end{aligned}
\label{eq:aleatoric_llh_l1}
\end{equation}
In our experiments, we achieved slightly smaller mean absolute errors (likely because this is the metric that the NN is trained to minimize) but observed a few more outliers (which is not surprising in view of the heavier tails of the Laplace distribution), for which reason we present the results for the Gaussian likelihoods herein. Recently, \cite{Hortua2020a} showed how Bayesian NNs can be combined with Normalizing Flows in order to capture the non-Gaussianity of the 21 cm signal from the Epoch of Reionization. Harnessing Normalizing Flows for this present task of estimating $\gamma$-ray flux fractions and thereby allowing for more general flux posterior distributions merits further investigation, which we leave for future work.

\subsection{Epistemic uncertainty estimation using \emph{Concrete Dropout}}
\label{subsec:concrete_dropout}
Now, we outline the modifications to our NN that enable the estimation of the epistemic uncertainty by modeling the NN weights within a probabilistic framework. We are interested in determining the posterior distribution of the weights $\bm{\omega}$ given the training data $\mathbf{X}$ and labels $\mathbf{Y}$, that is $p(\bm{\omega} | \mathbf{X}, \mathbf{Y})$. Placing a prior $p(\bm{\omega})$ on the weights and using Bayes' Theorem, this distribution can be found as
\begin{equation}
    p(\bm{\omega} | \mathbf{X}, \mathbf{Y}) = \frac{p(\mathbf{Y} | \mathbf{X}, \bm{\omega}) \, p(\bm{\omega})}{p(\mathbf{Y} | \mathbf{X})}.
\end{equation}
While simple to formulate, directly using this equation to compute the posterior distribution is not possible since the marginal distribution $p(\mathbf{Y} | \mathbf{X})$ is analytically intractable. Therefore, one approximates the posterior distribution $p(\bm{\omega} | \mathbf{X}, \mathbf{Y})$ by a simple variational distribution $q_{\bm{\theta}}(\bm{\omega})$ that minimizes the Kullback--Leibler (KL) divergence within a family of distributions that is characterized by a vector of variational parameters ${\bm{\theta}}$. This KL divergence can be expressed as
\begin{equation}
\begin{aligned}
    \text{KL}\left(q_{\bm{\theta}}(\bm{\omega}) \, || \, p(\bm{\omega} | \mathbf{X}, \mathbf{Y})\right) &= \int q_{\bm{\theta}}(\bm{\omega}) \log\left(\frac{q_{\bm{\theta}}(\bm{\omega})}{p(\bm{\omega} | \mathbf{X}, \mathbf{Y})} \right) \ d\bm{\omega} \\
    &= -\left(\underbrace{\mathbb{E}_{q_{\bm{\theta}}}\left[\log p(\bm{\omega}, \mathbf{Y} | \mathbf{X})\right] - \mathbb{E}_{q_{\bm{\theta}}}\left[\log q_{\bm{\theta}}(\bm{\omega})\right]}_{\text{ELBO}} \right) + \log p(\mathbf{Y} | \mathbf{X}) \\
    &= -\mathbb{E}_{q_{\bm{\theta}}} \left[\log p(\mathbf{Y} | \mathbf{X}, \bm{\omega})\right] + \text{KL}\left(q_{\bm{\theta}}(\bm{\omega}) \, || \, p(\bm{\omega})\right)  + \log p(\mathbf{Y} | \mathbf{X}).
\label{eq:KL_div}
\end{aligned}
\end{equation}
The identity in the second line shows that minimizing the KL divergence is equivalent to maximizing the evidence lower bound (ELBO), while the identity in the last line will turn out beneficial in a moment. Note that the term $\log p(\mathbf{Y} | \mathbf{X})$ is independent of the NN weights $\bm{\omega}$ and can hence be ignored when computing optimal weights using stochastic gradient descent optimization. Regardless of the particular choice for $q_{\bm{\theta}}$, the first term in the last row of Eq.~\eqref{eq:KL_div} can be formulated as a sum over the training maps and labels $\{(\mathbf{x}_n, \mathbf{y}_n)\}_{n=1}^N$:
\begin{equation}
    -\mathbb{E}_{q_{\bm{\theta}}} \left[\log p(\mathbf{Y} | \mathbf{X}, \bm{\omega})\right] = - \sum_{n=1}^N \int q_{\bm{\theta}}(\bm{\omega}) \log p(\mathbf{y}_n | \mathbf{x}_n, \bm{\omega}) \ d\bm{\omega}.
    \label{eq:KL_term_1}
\end{equation}
Using Monte Carlo (MC) sampling for the weights and drawing a single sample $\bm{\omega}_n \sim q_{\bm{\theta}}(\bm{\omega})$ for every training map, each term in the sum provides an unbiased estimate $- \log p(\mathbf{y}_n | \mathbf{x}_n, \bm{\omega}_n) = - \log p(\mathbf{y}_n | f^{\bm{\omega}_n}(\mathbf{x}_n))$, which is exactly the negative log-likelihood that we use as our loss function (Eq.~\eqref{eq:aleatoric_llh}), evaluated for the randomly sampled weights $\bm{\omega}_n$. 
\par In order to process the KL divergence of $q_{\bm{\theta}}(\bm{\omega})$ towards $p(\bm{\omega})$ in Eq.~\eqref{eq:KL_div} further, the choice for $q_{\bm{\theta}}(\bm{\omega})$ needs to be made explicit. Let $L$ be the number of NN layers with random weight matrices $\bm{\omega} = \{\mathbf{W}_l\}_{l=1}^L$, and define the family of distributions $q_{\mathbf{M}_l}(\mathbf{W}_l)$ for layer $l$ as
\begin{equation}
    q_{\mathbf{M}_l}(\mathbf{W}_l) = \mathbf{M}_l \ \text{diag}\left[\text{Bernoulli}(1 - p_l)^{n_{\text{out}, l}} \right],
\end{equation}
where $\mathbf{M}_l$ denotes the mean weight matrix for layer $l$, $p_l$ is the so-called Dropout probability, and $n_{\text{out}, l}$ is the number of output channels (thus, $\mathbf{W}_l, \mathbf{M}_l \in \mathbb{R}^{n_{\text{out}, l} \times n_{\text{in}, l}}$). Consequently, the variational parameter vector for each layer is given as $\bm{\theta}_l = \{p_l, \mathbf{M}_l\}$. Multiplying the mean weight matrices $\mathbf{M}_l$ with a diagonal matrix whose elements are drawn from a Bernoulli distribution means that weights are randomly zeroed with a probability of $p_l$. This ``Dropout'' has been previously proposed \cite{Hinton2012,srivastava2014dropout} for preventing NNs from overfitting. The KL divergence can then be written as a sum over the layers, and each term can be approximated as \cite{Gal2017}
\begin{align}
    \text{KL}\left(q_{\bm{\theta}}(\bm{\omega}) \, || \, p(\bm{\omega})\right) &= \sum_{l=1}^{L} \text{KL}\left(q_{\mathbf{M}_l}(\mathbf{W}_l) \, || \, p(\mathbf{W}_l) \right), 
    \label{eq:KL_term_2} \\
    \text{KL}\left(q_{\mathbf{M}_l}(\mathbf{W}_l) \, || \, p(\mathbf{W}_l) \right) &\propto \frac{\lambda^2 (1 - p_l)}{2} \|\mathbf{M}_l\|^2 - n_{\text{out}, l} \mathcal{H}(p_l),
\end{align}
where $\lambda$ is a prior length scale that we choose to be $0.01$ herein, and $\mathcal{H}$ denotes the entropy of a Bernoulli random variable, that is
\begin{equation}
    \mathcal{H}(p_l) = -p_l \log p_l - (1 - p) \log\left(1 - p_l\right).
\end{equation}
Hence, the KL term that pushes the distribution $q_{\bm{\theta}}(\bm{\omega})$ towards the prior $p(\bm{\omega})$ can approximately be expressed by two additional regularization loss terms, the first of which penalizes large weight magnitudes, and a second term that is independent of the weights and solely depends on the Dropout probabilities $\{p_l\}_{l=1}^L$. Note that minimizing the latter is equivalent to maximizing the entropy $\mathcal{H}(p_l)$, driving $p_l$ towards $0.5$. Dividing the terms in equations \eqref{eq:KL_term_1} and \eqref{eq:KL_term_2} by the total number of training samples $N$, it becomes apparent that the relative importance of the regularization terms decreases as the number of available training samples increases. In order to allow for the loss function to be differentiated with respect to the Dropout probabilities (which is necessary for optimizing the Dropout probabilities during the NN training), the Bernoulli distribution is replaced by the eponymous Concrete distribution, which can be regarded as a continuous relaxation of the Bernoulli distribution. 
\par At evaluation time, the predictive mean for each map $\mathbf{x}$ is found using so-called \emph{MC Dropout}, i.e. by drawing $R$ samples and averaging the predictions:
\begin{equation}
    \overline{f^{\bm{\omega}}(\mathbf{x})} \approx \frac{1}{R} \sum_{r=1}^R f^{\bm{\omega}_r}(\mathbf{x}),
\end{equation}
where $\bm{\omega}_r \sim q_{\bm{\theta}}(\bm{\omega})$ are the randomly sampled weight matrices for the $L$ NN layers in the $r$-th drawing. For the case of a full uncertainty covariance matrix (Eq.~\eqref{eq:loss_full_covar}), one obtains the predictive covariance uncertainty matrix $\bm{\Sigma}^\text{pred}(\mathbf{x})$ as
\begin{equation}
    \bm{\Sigma}^\text{pred}(\mathbf{x}) \approx \underbrace{\frac{1}{R} \sum_{r=1}^{R} \bm{\Sigma}_{r}(\mathbf{x})}_{\text {aleatoric }} + \underbrace{\frac{1}{R} \sum_{r=1}^{R}\left(f^{\bm{\omega}_r}(\mathbf{x}) - \overline{f^{\bm{\omega}}(\mathbf{x})}\right)\left(f^{\bm{\omega}_r}(\mathbf{x})-\overline{f^{\bm{\omega}}(\mathbf{x})}\right)^{\top}}_{\text {epistemic }},
\label{eq:pred_covar}
\end{equation}
which can be simplified for the case of a diagonal uncertainty covariance matrix considered in the main part of our \emph{Letter}.
More in-depth explanations of the concepts discussed within this section can be found in \cite{Gal2016, Gal2017,Kendall2017}.

\subsection{Neural network training}
We use a batch size of $64$ and take an Adam optimizer \cite{Kingma2014} with learning rate $5 \times 10^{-4}$ decaying with a rate of $2.5 \times 10^{-4}$ with respect to mini-batch iterations. We train the NN by performing 30,000 mini-batch iterations (25,000 for the proof-of-concept example) on a single Nvidia Tesla Volta V100 GPU on the supercomputer Gadi, which is located in Canberra and is part of the National Computational Infrastructure (NCI), taking roughly two hours for the realistic scenario.

\section{Dependence of the neural network estimates on the GCE PS source count distribution}
\label{sec:SCDs}
\begin{figure}[htb]
\centering
  \noindent
  \resizebox{0.6\textwidth}{!}{
\includegraphics{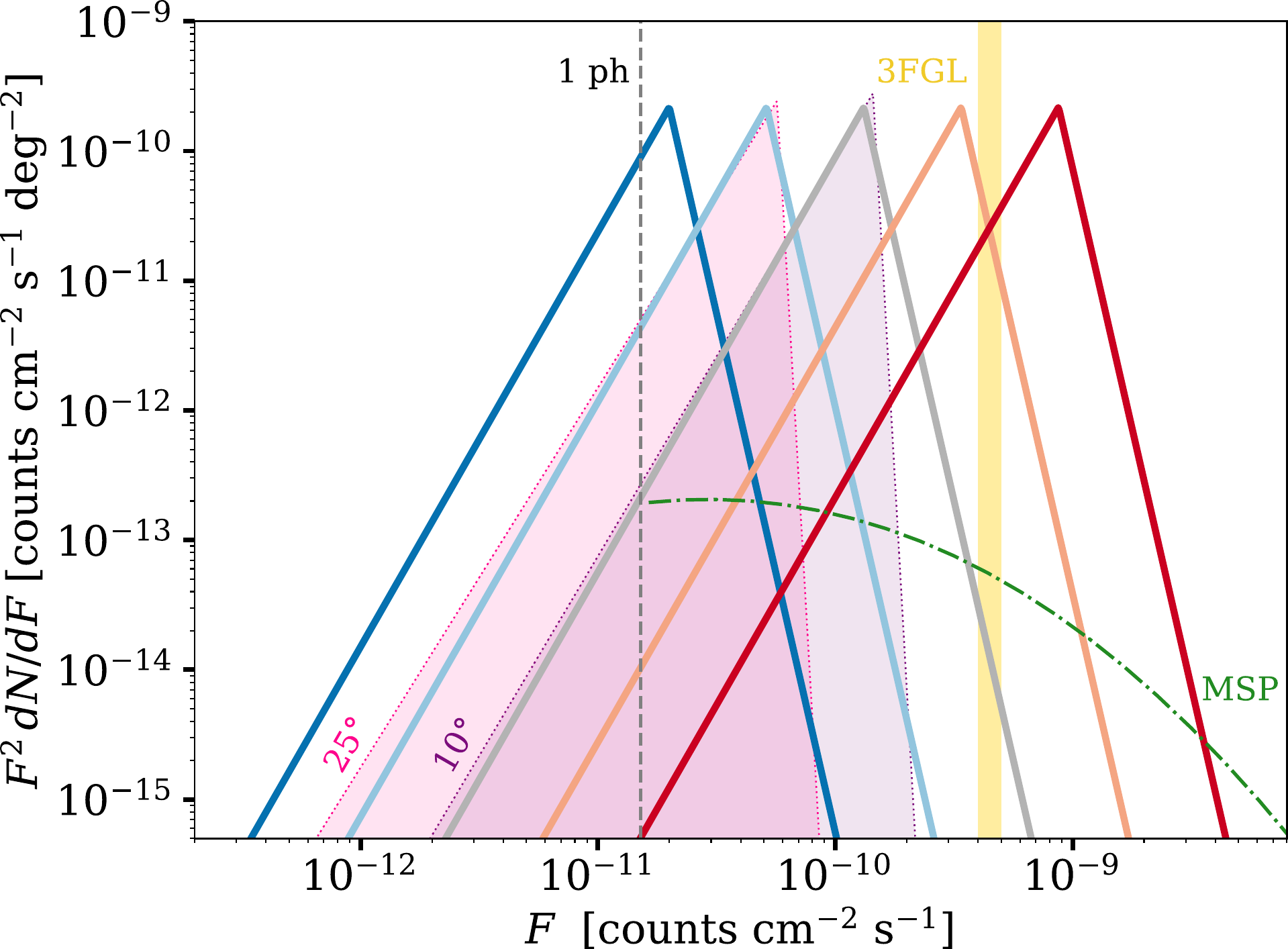}
}
\caption{The 5 SCDs used for the experiment in Sec. \ref{sec:SCDs}, from very dim \orange{(left, blue)} to very bright \orange{(right, red)}. The gray dashed line marks the flux corresponding to a single photon, which can roughly be viewed as the limit below which PS and Poissonian emission become degenerate. The yellow rectangle indicates the \emph{Fermi} 3FGL threshold of $\sim~4 - 5 \times 10^{-10} \ \text{counts} \ \text{cm}^{-2} \ \text{s}^{-1}$, above which PSs can be expected to be resolved. \orange{The two regions shaded in purple and pink indicate the best-fit SCDs for GCE PS described by a singly-broken power law as determined by \texttt{NPTFit} within $10^\circ$ and $25^\circ$, respectively, which correspond to the NPTF parameters in Table \ref{table:NPTF_mock_params}. The green dash-dotted line shows the smooth MSP SCD from \cite{Bartels2018}, cut off below the 1 photon line, which we use for an additional test in Sec.~\ref{subsec:bartels_SCD}.}}
\label{fig:SCDs}
\end{figure}
\begin{table}[ht]
\caption{Model parameters for the experiment in Sec. \ref{sec:SCDs}. The 5 values for $A_\text{gce}^\text{PS}$ and $S_1^\text{gce}$ (alongside $n_1^\text{gce}$ and $n_2^\text{gce}$) define the 5 different SCDs depicted in Fig.~\ref{fig:SCDs} (from very bright to very dim). The values labeled ``with / without DM / PS'' correspond to our choices when the respective template is modeled to be present / absent.}
\centering
\begin{tabular}{@{}lcr@{}}
\toprule
Model parameter & Value &  \\ \midrule
$\log_{10} A_\text{dif}^{\pi^0 + \text{BS}}$ & $0.91$ &  \\
$\log_{10} A_\text{dif}^{\text{IC}}$ & $0.58$ &  \\
$\log_{10} A_\text{iso}$ & $-0.30$ &  \\
$\log_{10} A_\text{bub}$ & $-0.06$ &  \\
\multirow{2}{*}{$\log_{10} A_\text{gce}$} & $0.032$ & with DM \\
 & $-1.49$ & without DM \\
\multirow{2}{*}{$\log_{10} A_\text{gce}^\text{PS}$} & $\{-3.13, -2.31, -1.49, -0.67, 0.15\}$ & with PS \\
 & $-5.5$ & without PS \\
\multirow{2}{*}{$S_1^\text{gce}$} & $\{57.57, 22.40, 8.71, 3.39, 1.31\}$ & with PS \\
 & $8.71$ & without PS \\
$n_1^\text{gce}$ & $10.0$ &  \\
$n_2^\text{gce}$ & $-1.2$ &  \\ \bottomrule
\end{tabular}
\label{table:SCD_params}
\end{table}
This section is dedicated to an analysis of the NN predictions for different SCDs of the GCE PS template. As briefly mentioned in the main part of this \emph{Letter}, ultra-faint PSs are exactly degenerate with purely Poissonian emission. For this reason, it is interesting to examine in which regimes the NN is able to discern GCE PS and GCE DM. We carry out an experiment inspired by the benchmark test for \texttt{NPTFit} in \cite[Sec. IV]{Chang2019} and consider the slightly simplified setting from the proof-of-concept example (non-Bayesian GCNN, no 3FGL PSs, no disk PSs, uniform mean \emph{Fermi} exposure, fixed outer ROI radius of $25 ^\circ$) with the following templates: diffuse emission described by Model~O, isotropic emission, \emph{Fermi} bubbles, GCE DM, and GCE PS. We model the GCE PS using each of the five SCDs shown in Fig.~\ref{fig:SCDs} (cf. \cite[Fig. 1]{Chang2019}). These SCDs all yield the same total flux and span the relevant flux range for the NN: PSs brighter than the 3FGL threshold can be assumed to be resolved by \emph{Fermi}, for which reason methods such as \texttt{NPTFit} or our NN are not needed beyond this flux of $\sim~4 - 5 \times 10^{-10} \ \text{counts} \ \text{cm}^{-2} \ \text{s}^{-1}$. On the other hand, very dim PSs that emit on average $\sim~1$ photon cannot be expected to be distinguished from purely Poissonian emission, which can be interpreted as the result of multiple single-photon PSs taken in aggregate. Hence, separating GCE PS from GCE DM within the flux range between these two boundaries is where our NN comes into play. We first consider the case of the GCE consisting entirely of PS. Then, we turn towards the more challenging case of DM and PSs each contributing $50 \%$ of the GCE emission. Lastly, we address the case of $100 \%$ DM for the sake of completeness. 

\subsection{GCE: 100\% PS}
\label{subsec:SCDs_100PS}
\begin{figure}[htb]
\centering
  \noindent
  \resizebox{1\textwidth}{!}{
\includegraphics{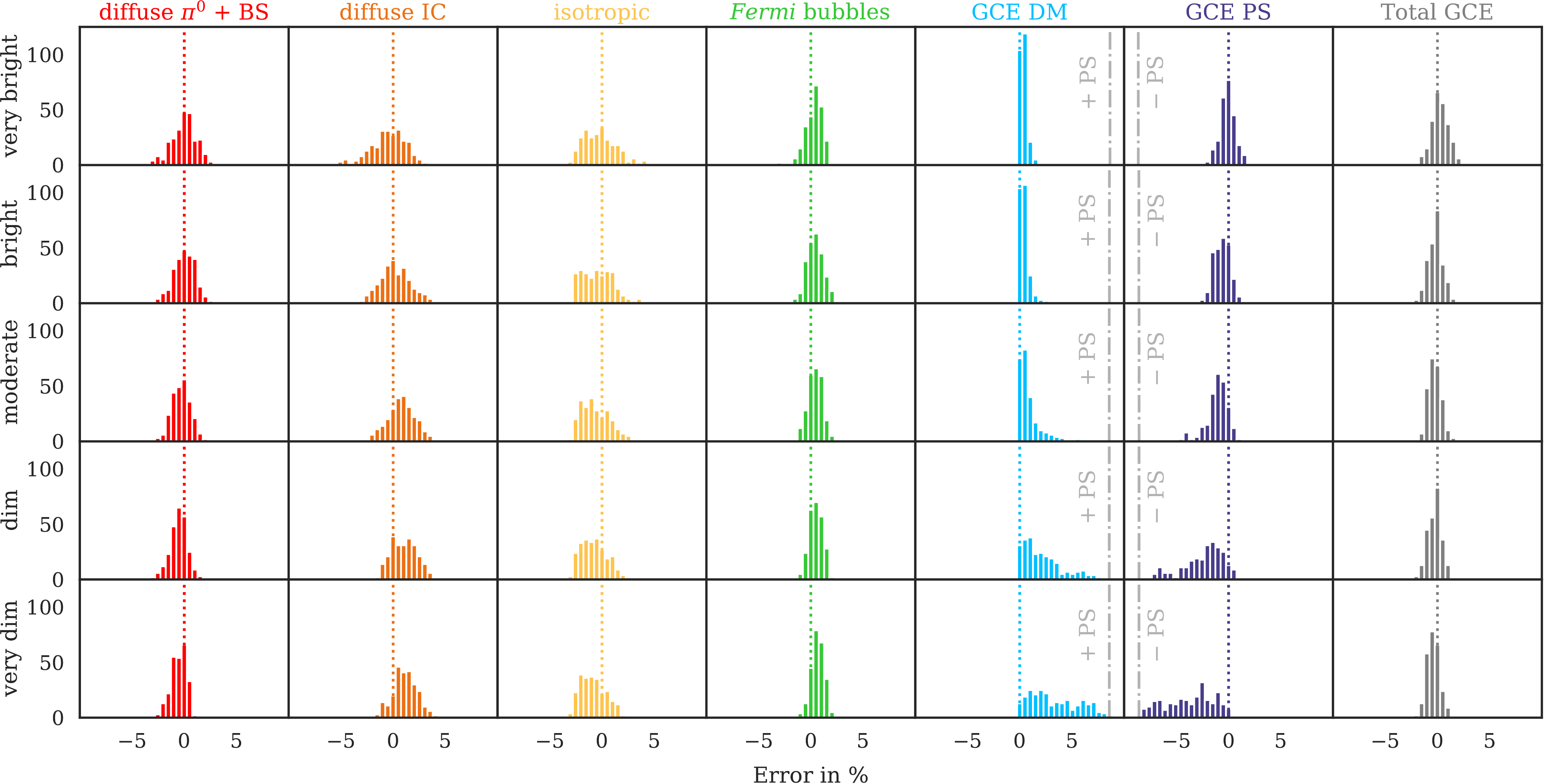} 
}
\caption{Histogram of the differences between the NN estimates and the true flux fractions in per cent, for the GCE consisting of $100 \%$ PS. The testing data set for this experiment contains 250 photon-count maps. The rows correspond to the SCDs for the GCE PS shown in Fig.~\ref{fig:SCDs}. The gray dash-dotted lines in the GCE DM and PS columns indicate a discrepancy consistent with misattribution of the entire GCE PS flux to DM.}
\label{fig:SCD_test_without_DM}
\end{figure}
\par For each of the SCDs, we generate 250 photon-count maps that have a similar composition as is expected from the \emph{Fermi} map: the total number of photons within the ROI is $\sim$ 126,000 (the \emph{Fermi} map has $\sim$ 124,000 counts within this ROI when applying the 3FGL mask), and the mean flux fractions for diffuse $\pi^0 + \text{BS}$, diffuse IC, isotropic, \emph{Fermi} bubbles, GCE DM, and GCE PS are $56.1 \%$, $26.1 \%$, $3.1 \%$, $5.8 \%$, $0.3 \%$, and $8.6 \%$ (see Table \ref{table:SCD_params} for the exact model parameters). We take the (non-Bayesian) NN from the proof-of-concept example that was trained on maps within a constant ROI of outer radius $25 ^\circ$ and evaluate it on these 250 maps for each of the SCDs (evaluating the NN on a mini-batch consisting of $64$ photon-count maps takes roughly a second on a laptop CPU). \orange{The fact that the disk PS template is not included here does not lead to a loss of generality with respect to the results presented in this section as we have not observed any substantial confusion between the GCE PS and the disk PS template in our experiments -- see for instance Fig.~\ref{fig:fermi_mock_histogram}, which shows that the errors for the disk PS template are very small, and Fig.~\ref{fig:fermi_results_25}, where it can be seen that the NN reliably recovers a non-negligible disk PS flux in \emph{Fermi} mock maps if present.}
\par A histogram of the errors for all the templates and SCDs is plotted in Fig.~\ref{fig:SCD_test_without_DM}. The non-GCE templates are largely insensitive to the respective SCD of the GCE PS, and the errors for these templates are tightly scattered around zero (the estimates for the diffuse IC and isotropic templates appear to slightly move as the SCD is varied). The total GCE flux (DM + PS) is also predicted accurately for all maps and SCDs. The columns for GCE DM and GCE PS reveal how the NN splits the GCE flux between to the two GCE templates. The gray dash-dotted lines in the GCE DM / PS plots indicate the errors corresponding to the total GCE flux being faithfully recovered but the whole GCE flux being misattributed to DM. For (very) bright GCE PS, the GCE DM flux is correctly estimated to be close to zero. For the moderate SCD (count break at $S_1^\text{gce} = 8.71$), a slight misattribution from GCE PS to GCE DM is noticeable, but the mean error for GCE DM / GCE PS is still only $0.87 \%$ / $1.21 \%$, respectively. In the dim regime ($S_1^\text{gce} = 3.39$), there are some maps for which NN misattributes the majority of GCE flux to DM; however, the mean error of $2.19 \%$ / $2.49 \%$ is only slightly larger than $25 \%$ of the total GCE flux contribution. For the very dim case ($S_1^\text{gce} = 1.31$), the degeneracy between the faint PSs and DM is reflected in the NN estimates for the two GCE templates: for some maps, the entire GCE flux is absorbed by the GCE DM template, with the mean errors of $3.37 \%$ / $3.77 \%$ being close to $50 \%$ of the total GCE flux ($4.3 \%$), which would be expected in case GCE DM and PS were completely indistinguishable and the NN divided the entire GCE flux randomly between them. \orange{The question explored in this section, namely to what extent the DM-PS degeneracy should manifest itself for a given $dN/dF$, has not been resolved analytically in the literature. We do not claim that our NN separates the two GCE components to the highest degree mathematically possible -- for example, Fig.~\ref{fig:narrow_priors_hist_best_fit_data} shows that we achieve a slight improvement in terms of PS-DM misattribution for the best-fit \emph{Fermi} mock maps with a NN training procedure optimized for maximum accuracy in the \emph{Fermi}-like parameter region, where we consider a fixed $25^\circ$ ROI and take more training maps with narrow priors around the \emph{Fermi} values. Still, Fig.~\ref{fig:SCD_test_without_DM} clearly displays the expected qualitative behavior, which resembles that of the NPTF (see \cite{Chang2019}), and provides a baseline for future Deep Learning-based fitting methods. Also, the results of this experiment are consistent with the partial misattribution from GCE PS to DM that can be seen in Fig.~\ref{fig:fermi_results} in the main body: the best-fit $dN/dF$ for GCE PS as determined by \texttt{NPTFit} has a count break at $S_b = 9.52$, similar to the moderate SCD in this experiment ($S_b = 8.71$), but it drops off more rapidly ($n_1 = 34.34$ vs. $n_1 = 10.0$) above the count break (see Fig.~\ref{fig:SCDs} and Table \ref{table:NPTF_mock_params}), for which reason the ``difficulty'' for the NN can be expected to lie between the moderate and the dim SCDs in this experiment, explaining a certain extent of confusion (see also Fig.~\ref{fig:fermi_mock_histogram} for the associated error histogram).}
From this first part of the experiment, we conclude the following: \textbf{if the templates provide a perfect description of the data and the GCE consists entirely of unresolved PSs that are not degenerate with Poissonian emission, the NN reliably identifies the PS origin of the GCE.} While we remind the reader that the results presented in this section assume no mismodeling and adopt the slightly simplified settings from the proof-of-concept example (with GCE PS being the only non-Poissonian template), we emphasize that an evaluation of the NN estimates in the realistic scenario will be presented in Sec. \ref{sec:add_mat_realistic} below.

\subsection{GCE: 50\% PS / 50\% DM}
\label{subsec:SCDs_5050}
\begin{figure}[htb]
\centering
  \noindent
  \resizebox{1\textwidth}{!}{
\includegraphics{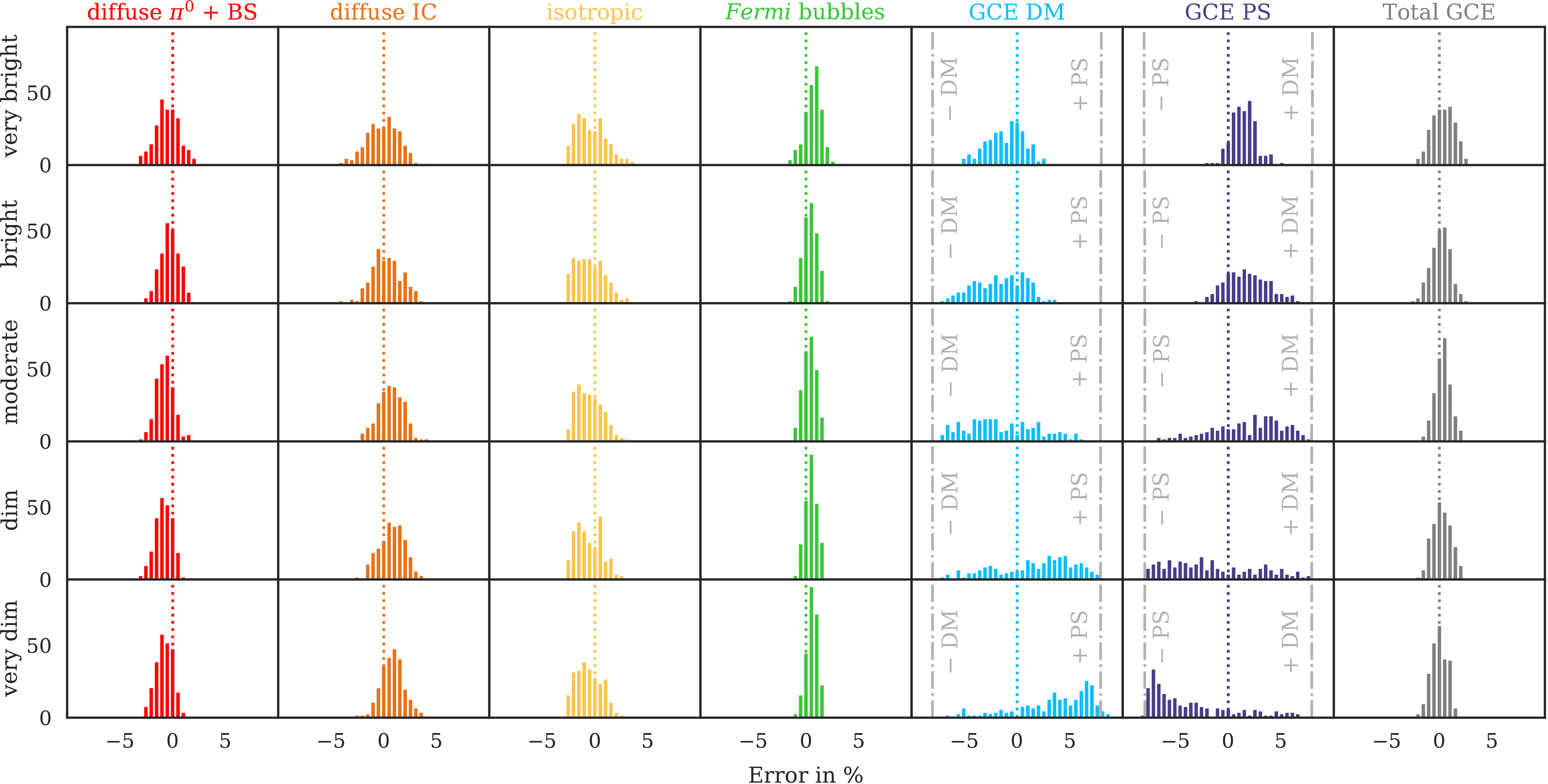} 
}
\caption{Same as Fig.~\ref{fig:SCD_test_without_DM}, but with an additional DM contribution of equal flux such that the GCE consists of $50 \%$ PS / $50 \%$ DM. The gray dashed-dotted lines indicate errors corresponding to misattribution of the entire GCE PS flux to DM and vice versa.}
\label{fig:SCD_test_with_DM}
\end{figure}
Now, we repeat the experiment with an additional GCE contribution from DM, which equals that from PSs (DM-related parameters ``with DM'' in Table \ref{table:SCD_params}). Thus, the total GCE flux contribution increases to $16 \%$, half of which is due to GCE DM and GCE PS, respectively. The resulting errors for this more challenging case are depicted in Fig.~\ref{fig:SCD_test_with_DM}. As before, we find that the NN accurately determines the flux fractions of the non-GCE templates, as well as the total GCE flux. As expected, the scatter in the error for the two GCE templates has become much larger. Since misattribution of the GCE components now occurs in both directions, we plot two gray dash-dotted lines that evidence errors corresponding to the entirety of one GCE template being attributed to the other. Whereas the mean GCE DM / PS errors for the bright SCD are still modest with $2.35 \%$ / $2.21 \%$, the error distributions flatten for moderate and dim SCDs. Interestingly, the NN preferentially errs in favor of GCE PS for (very) bright PSs, suggesting that the NN has no systematic preference for DM in ambiguous situations and (slightly) weighing against the hypothesis of there in fact being GCE PS in the \emph{Fermi} data that are missed by the NN due to systematics. On the other hand, the GCE PS flux arising from the very dim SCD tends to be absorbed by the GCE DM template. We reach the following conclusion from this second part of the experiment: \textbf{if the GCE is split up between DM and very dim to moderately bright PS, the NN may misattribute the entirety or a fraction of the DM flux to PSs and vice versa.} Our findings are similar to the conclusions in \cite[Sec. IV B]{Chang2019} for \texttt{NPTFit}. A difference is that whereas \texttt{NPTFit} seems to prefer assigning the entire GCE to one of the two templates in this situation (see the bimodal histograms for GCE DM and PS in Figs. 4 and 5 in said work), the NN assigns a non-negligible flux fraction to \emph{both} templates for many maps, resulting in the error histograms to be flattened rather than bimodal.

\subsection{GCE: 100\% DM}
\label{subsec:SCDs_100DM}
\begin{figure}[htb]
\centering
  \noindent
  \resizebox{1\textwidth}{!}{
\includegraphics{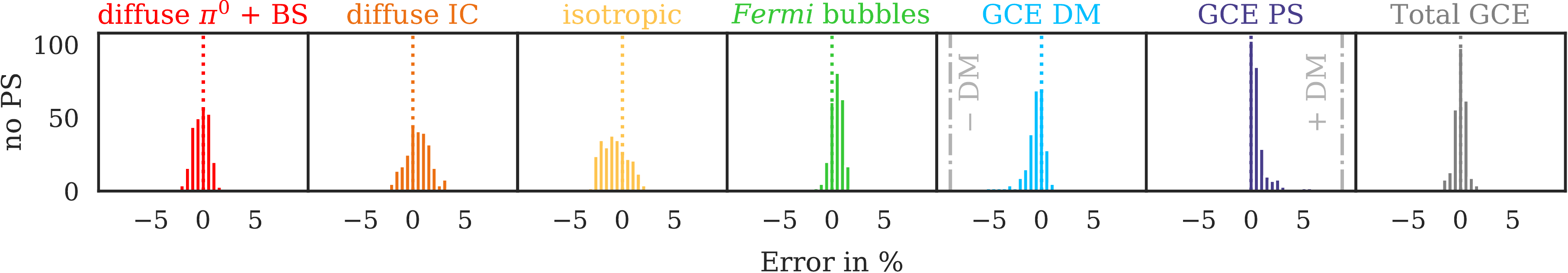}
}
\caption{Same as Fig.~\ref{fig:SCD_test_without_DM}, but for the GCE consisting of $100 \%$ DM flux. The gray dashed-dotted lines indicate errors corresponding to misattribution of the entire GCE DM flux to PS.}
\label{fig:SCD_test_only_DM}
\end{figure}
Finally, we consider the case of the the GCE being made up entirely of DM flux. To this end, we generate 250 maps with the model parameters labeled ``without PS'' / ``with DM'' in Table \ref{table:SCD_params}. This gives a mean GCE PS flux fraction $\lesssim 0.001 \%$ and a GCE DM flux of $\sim~8.7 \%$, yielding a very similar GCE flux to the $100 \%$ PS case in Sec. \ref{subsec:SCDs_100PS}. The resulting distribution of the NN errors is plotted in Fig.~\ref{fig:SCD_test_only_DM}. Although there is a slight misattribution from GCE to GCE PS for some maps, the NN accurately recovers the GCE DM flux for the vast majority of the maps: the mean absolute errors for GCE DM / GCE PS are $0.81 \%$ / $0.64 \%$, and the NN estimate for the GCE PS flux is less than $3 \%$ for $242$ out of the $250$ maps. Thus, this experiment demonstrates the following: \textbf{if the GCE is made up of DM flux only, the NN generally ascertains the DM origin of the GCE}. We remind the reader that the disk PS template is the only non-Poissonian template considered in this experiment and that we assume that the templates provide a perfect description of the data. 

\section{Neural network estimates in the case of mismodeling}
\label{sec:mismodelling}
In this section, we demonstrate the robustness of our NN against two different sources of mismodeling. Note that whereas simulating mismodeling for template fitting methods is tantamount to providing spatial templates to the fitting tool that differ from those that constitute the data, examining mismodeling in the case of the NN consists in evaluating the NN on maps that are composed of different templates from those that the NN has been trained on. 
First, we consider the accuracy of the NN when being evaluated on maps containing a diffuse emission template different from the one used during the training. Then, we analyze the robustness of the NN against an unmodeled GCE asymmetry, motivated by the recent findings of \cite{Leane2020}. Finally, we present the results for mismodeled \emph{Fermi} bubbles. For these proof-of-concept experiments, we utilize the non-Bayesian NN from the proof-of-concept example that was trained to minimize an $l^2$ loss function (Eq.~\eqref{eq:l2loss}) on maps consisting of 6 templates, namely Galactic foregrounds described by Model~O, isotropic extragalactic emission, the \emph{Fermi} bubbles, GCE DM, and GCE PS. The ROI in these examples has a fixed outer radius of $25 ^\circ$ around the GC, with $|b| \leq 2 ^\circ$ masked.

\subsection{Diffuse mismodeling}
\label{subsec:diffuse_mismodelling}
\begin{figure*}[htb]
\centering
  \noindent
  \resizebox{\textwidth}{!}{
\includegraphics{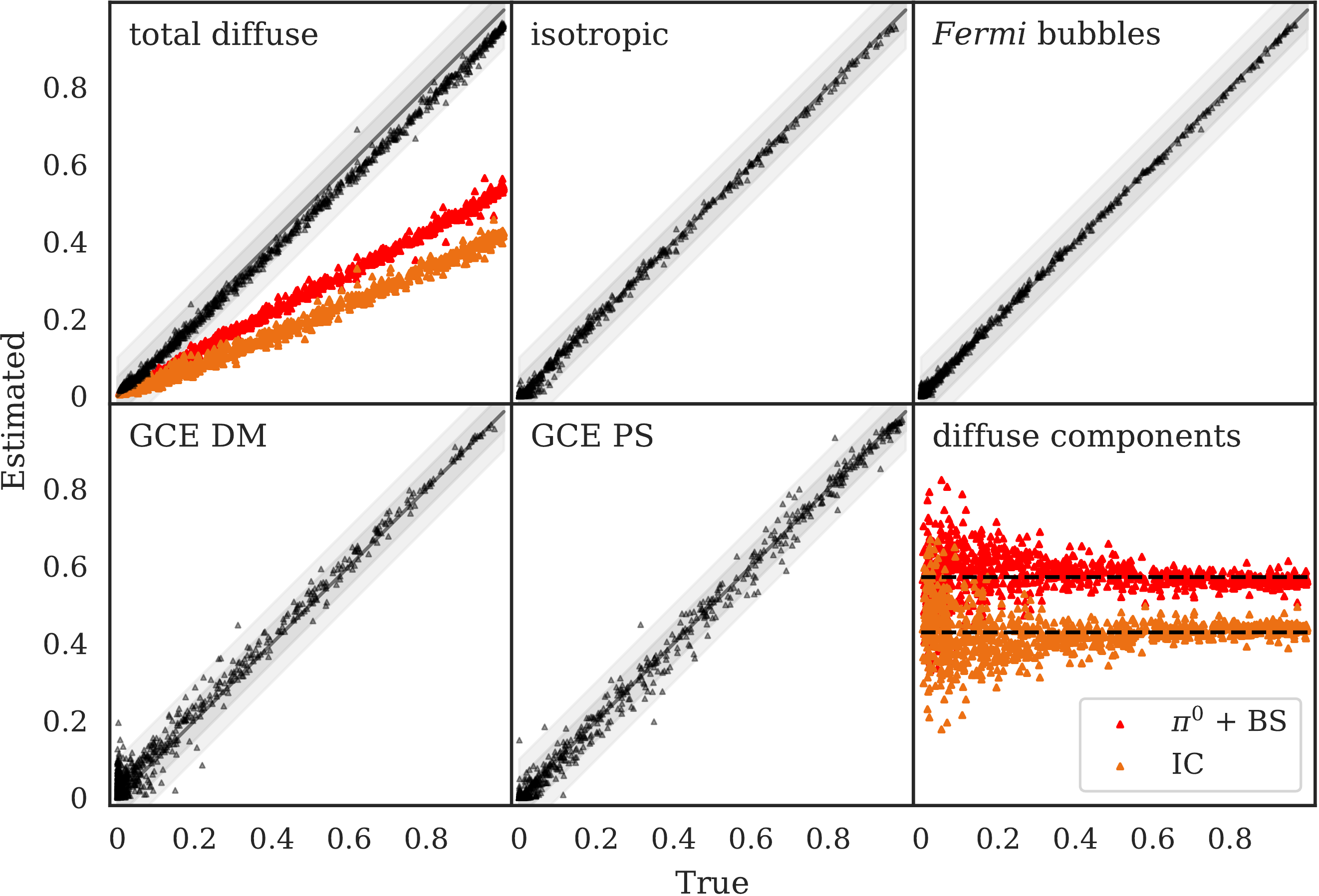} 
}
\caption{\emph{Diffuse mismodeling}: true and estimated flux fractions for 1,000 mock maps whose diffuse component is modeled by the \texttt{p6v11} template, whereas the NN was trained on maps containing diffuse flux from the pions \& bremsstrahlung ($\pi^0 + \text{BS}$) and IC constituents of Model~O. The upper left panel shows the estimates separately for the two diffuse components of Model~O and their sum (in black). In the lower right panel, we plot the fractions of these two components within the total estimated diffuse flux as a function of the true \texttt{p6v11} diffuse flux fraction in the respective map. The dashed black lines indicate the means computed over all the maps.}
\label{fig:diffuse_mismodelling}
\end{figure*}
It has been shown that incorrect modeling of the diffuse flux can result in \texttt{NPTFit} yielding statistical preference for PSs even if the GCE consists entirely of DM \cite{Leane2019a} (see also \cite[Fig. 7]{Chang2019}) and in an artificially injected GCE DM flux in the \emph{Fermi} map not being properly recovered \cite[Fig. 2]{Leane2019a} and \cite[Fig. 5]{Buschmann2020}. In order to assess the reliability of the NN predictions in the realistic situation where the true diffuse Galactic foreground differs by a certain amount from the diffuse model that the NN was trained on, we carry out the following experiment: we generate mock maps in which the diffuse contribution from Model~O (both the templates for pion decay \& bremsstrahlung and for IC scattering) is substituted by a random contribution from the \emph{Fermi} model \texttt{p6v11}. 
\par We plot the estimates of the NN for 1,000 test maps with \texttt{p6v11} diffuse flux in Fig.~\ref{fig:diffuse_mismodelling}. Recall that \texttt{p6v11} has only one free normalization parameter, and the ratio between the contributions from the gas-related process and from IC scattering is fixed. The black triangles in the upper left panel show the estimated sum of both diffuse flux constituents vs. the true \texttt{p6v11} flux fraction, whereas the red and orange triangles indicate the individual contributions from pions \& bremsstrahlung and from IC scattering, respectively. In the lower right panel, we plot the estimated fractions of the two diffuse components within the entire estimated diffuse flux against the true \texttt{p6v11} flux fraction. The black dashed lines show the estimated mean fractions for each of the two diffuse contributions. As is expected, the scatter in these fractions is larger for maps with a small diffuse flux contribution and decreases substantially for maps with a realistic diffuse flux contribution of $\gtrsim 0.7$. Interestingly, the NN finds the flux ratio between pions \& bremsstrahlung and IC for these maps to be $\sim~1.4$ while the preferred ratio for the \emph{Fermi} map is roughly $2$ (as determined by both the NN and \texttt{NPTFit}, see the main part of our \emph{Letter}) when allowing the two diffuse components of Model~O to float independently from each other, indicating that the contribution of photons from IC relative to those from pion decay is overestimated within \texttt{p6v11} (see also \cite[Fig. 6]{Calore2015}). The diffuse mismodeling causes the NN estimates to be somewhat more noisy (the mean errors for GCE DM and GCE PS in the 1,000 maps plotted in Fig.~\ref{fig:diffuse_mismodelling} are $2.36 \%$ and $1.21 \%$, respectively, as compared to mean errors $< 1 \%$ when the diffuse emission is modeled correctly), and the total estimate for the Model~O contribution accounts for slightly less than the total true \texttt{p6v11} flux fraction, with GCE DM absorbing a small fraction of the diffuse flux. Nonetheless, the NN is still able to discern GCE DM and GCE PS fairly accurately, with $95 \%$ of the maps having an error in the GCE PS flux fraction of $< 5 \%$. Note that the increased number of outliers in Fig.~\ref{fig:diffuse_mismodelling} as compared to Fig.~\ref{fig:toy_example} is partly a consequence of the larger number of plotted samples. The fact that the scatter in the estimated ratio between the constituents of Model~O is small when the diffuse (\texttt{p6v11}) contribution dominates the map (as is the case for the \emph{Fermi} map) demonstrates the robustness of the NN against small perturbations in the exact definition of the spatial templates. We summarize: \textbf{when the diffuse emission is mismodeled, the scatter of the NN estimates increases and the GCE DM template may absorb a fraction of the diffuse flux, however, there is no apparent systematic confusion between DM and PS.} \orange{A possible way to further mitigate the sensitivity of the NN to the diffuse model is to ``marginalize'' over a range of diffuse models by training the NN on maps made up of various diffuse models and/or combinations of them. We present a first example in Sec.~\ref{sec:narrow_priors} for linear combinations of diffuse Models A, F, and O, and we will continue exploring this avenue in future work.}

\subsection{GCE mismodeling}
\label{subsec:GCE_mismodelling}
\begin{figure*}[htb]
\centering
  \noindent
  \resizebox{\textwidth}{!}{
\includegraphics{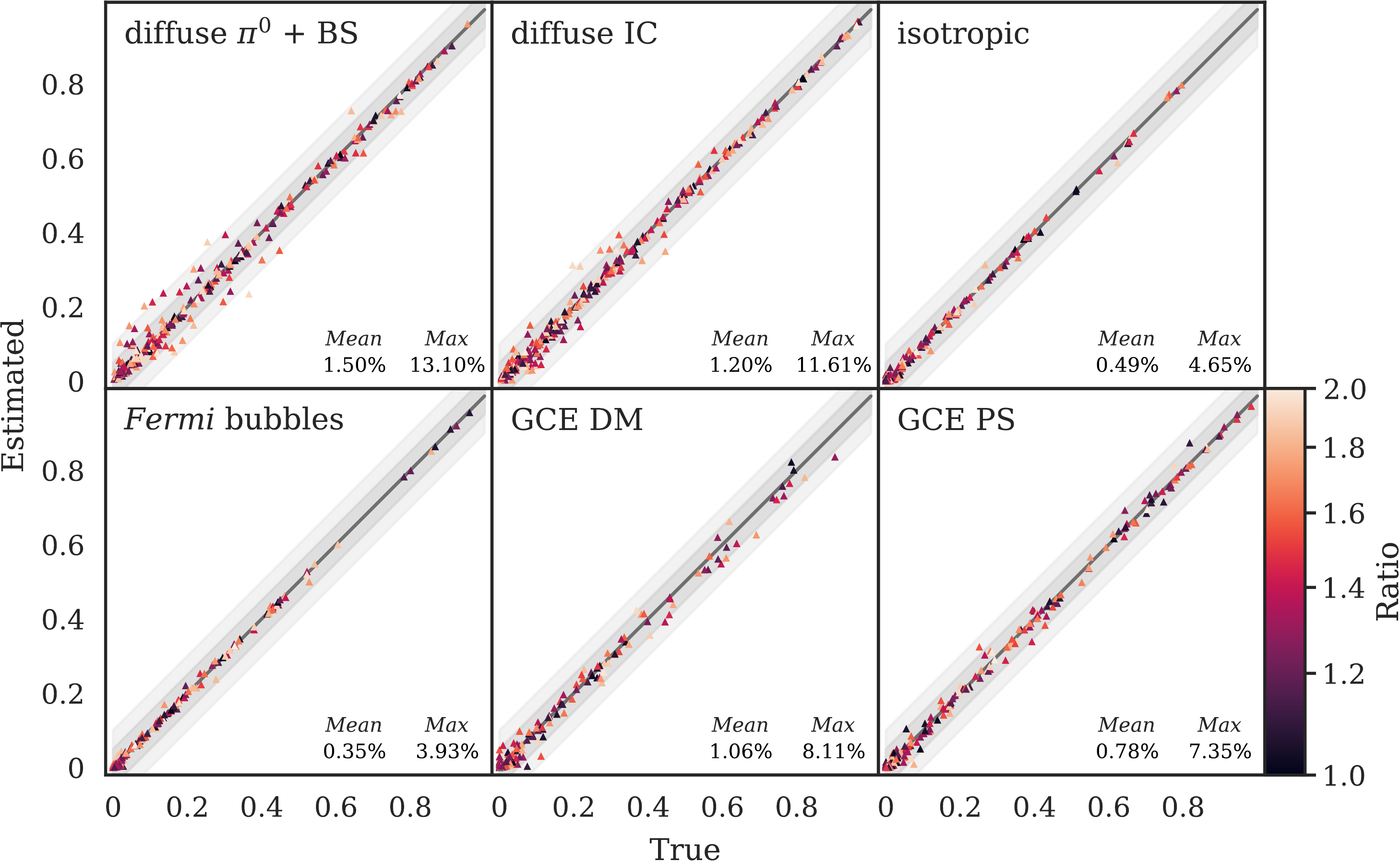} 
}
\caption{\emph{GCE mismodeling}: true and estimated flux fractions for the 305 maps from our test data set for which the north-south GCE DM flux discrepancy lies within a factor of two. The coloring of the markers corresponds to the extent of the GCE DM asymmetry in the data. 
The true flux on the $x$-axis in the GCE DM panel is the sum of the north and south contributions.}
\label{fig:gce_mismodelling_max_2}
\end{figure*}
\begin{figure*}[htb]
\centering
  \noindent
  \resizebox{\textwidth}{!}{
\includegraphics{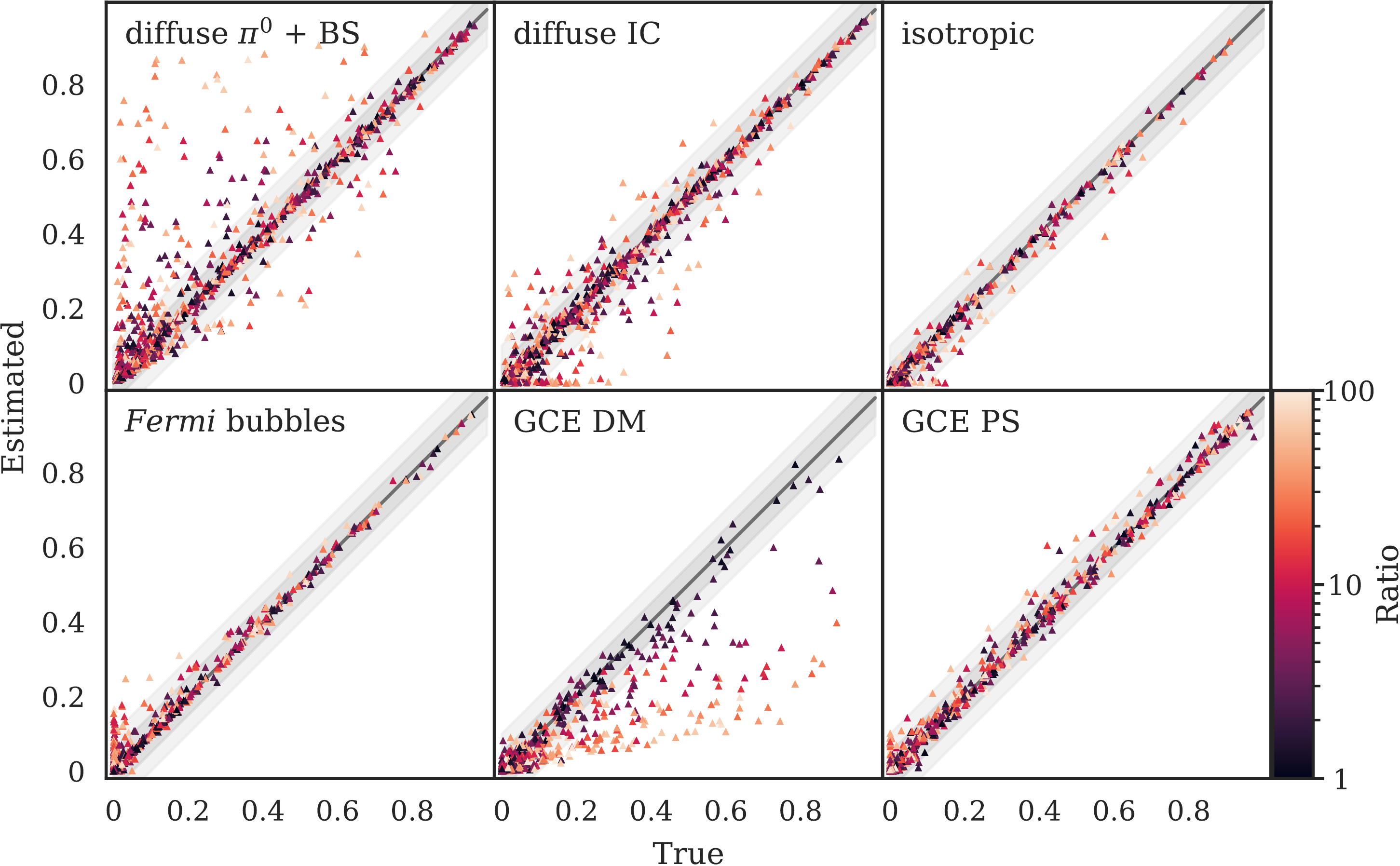} 
}
\caption{\emph{GCE mismodeling}: same as Fig.~\ref{fig:gce_mismodelling_max_2}, but now for north-south GCE DM flux ratios of up to 100.}
\label{fig:gce_mismodelling_max_100}
\end{figure*}
Recently, it has been shown \cite{Leane2020} that if individual template normalizations for the GCE in the northern and southern hemisphere are permitted, within a 10$^{\circ}$ ROI \texttt{NPTFit} prefers a DM origin of the GCE with the GCE PS flux fractions being consistent with zero, and a larger flux from GCE DM from the northern hemisphere is preferred (see \cite[Fig. 1]{Leane2020}). While it is not clear if this north-south asymmetry has a physical origin, it is interesting to examine the sensitivity of the NN estimates with respect to such an unmodeled asymmetry. For this aim, we evaluate the NN from the proof-of-concept example on photon-count maps whose GCE DM contribution has two separate normalizations $A_\text{gce}^\text{north}$ and $A_\text{gce}^\text{south}$ for the northern and southern hemispheres, respectively. We only allow for an asymmetry in the GCE DM template and use the symmetric template for the GCE PS contribution in the test case presented herein. Note that no photon-count maps with asymmetric GCE DM contributions were shown to the NN during the training, and the NN does not explicitly express any potentially detected asymmetry; hence the asymmetry is ``unmodeled''. 
\par Fig.~\ref{fig:gce_mismodelling_max_2} shows the true vs. estimated flux fractions for the 305 maps within our test data set with a GCE DM flux ratio $\leq 2$ between the two hemispheres. The color of the markers indicates the degree of asymmetry in the GCE DM emission, for values of the variable
\begin{equation}
    r^\text{N-S}_{\text{DM}} = \max \left\{  \frac{\mathbf{y}^\text{north}_\text{GCE DM}}{\mathbf{y}^\text{south}_\text{GCE DM}},
    \frac{\mathbf{y}^\text{south}_\text{GCE DM}}{\mathbf{y}^\text{north}_\text{GCE DM}} \right\}
\end{equation}
from 1 (black) to 2 (light yellow) in a logarithmic scaling.
Whereas the estimates for the diffuse templates become more noisy due to the unmodeled GCE DM asymmetry, the NN still recovers the GCE flux components well, with mean absolute errors of about $1 \%$ for the DM and PS contributions. The authors of \cite{Leane2020} found that when different north / south normalizations $A$ are allowed for both the GCE templates (DM and PS), the ratio between the maximum a posteriori (MAP) estimates for the GCE DM flux fractions in the northern and southern hemisphere is roughly $2$. Rather than demonstrating that the NN would still perform well if such a GCE asymmetry were physical (which would indeed be very surprising and challenging to accommodate by DM annihilation), the main purpose of this experiment is to underpin the robustness of the NN against modest shortcomings in the GCE modeling and the absence of systematic spurious misattribution of flux contributions. In Sec. \ref{subsec:hemispheres}, we fit the flux fractions of all the templates separately in the two hemispheres with our Bayesian GCNN while masking the other, and find that while our fits prefer a DM explanation for the GCE in both hemispheres, the DM flux fraction in the southern hemisphere is estimated to be somewhat larger than in the northern hemisphere.
\par In Fig.~\ref{fig:gce_mismodelling_max_100}, we plot the true and estimated flux fractions for 1,000 maps from the test data set with $r^\text{N-S}_{\text{DM}} \in (1, 100)$, which is also taken to be the range of the color map. Now, it becomes apparent that the NN underestimates the GCE DM flux fraction to a gradually increasing extent as the asymmetry grows. This suggests that the NN has learned during the training that the GCE DM template is approximately symmetric and now starts to assign the GCE DM flux from the dominant hemisphere to the other asymmetric templates, mainly to the (most asymmetric) diffuse pions \& bremsstrahlung component of Model~O and less so to the diffuse IC component and the \emph{Fermi} bubbles. Breaking down Fig.~\ref{fig:gce_mismodelling_max_100} into the cases where the flux from the northern / southern hemisphere dominates, respectively, reveals that for a larger GCE DM flux in the northern hemisphere, the emission from pion decay \& bremsstrahlung is preferentially overestimated while the IC flux is generally underestimated. This behavior is reversed for maps with stronger GCE DM emission in the southern hemisphere. Since the $\pi^0$ + BS template is brighter in the northern hemisphere, the NN tries to compensate for the GCE-induced asymmetry by balancing the two diffuse components. Strikingly, the GCE PS estimates are much less affected even when the north-south GCE DM flux asymmetry is as large as $100$. This suggests that the NN associates GCE PS flux with large granularity (or more mathematically pixel-to-pixel variance), preventing it from systematically confusing GCE PS flux with flux from Poissonian templates, albeit mismodeled. Our conclusion from this experiment is: \textbf{an unmodeled GCE DM asymmetry as found in the \emph{Fermi} data by \cite{Leane2020} using \texttt{NPTFit} barely affects the NN estimates for the GCE DM and GCE PS flux fractions, and the NN on average still achieves sub-percent accuracy for the GCE PS flux fraction on our test data set}.

\subsection{Mismodeling of the \emph{Fermi} bubbles}
\label{subsec:bub_mismodelling}
\begin{figure*}[h!]
\centering
  \noindent
  \resizebox{\textwidth}{!}{
\includegraphics{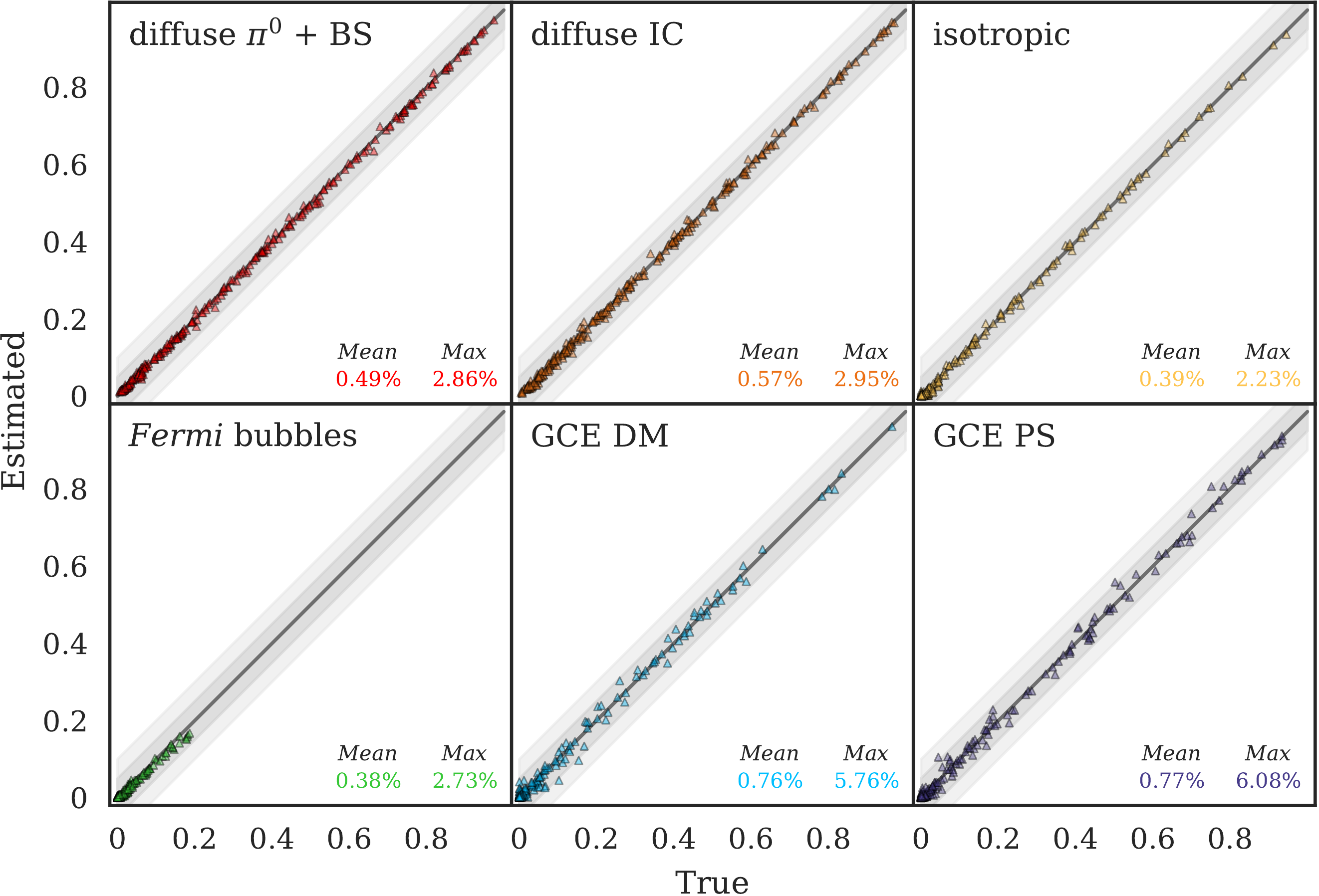} 
}
\caption{\emph{Mismodeling of the Fermi bubbles}: True and estimated flux fractions for the 256 maps from our test data set for which the flux fraction of the \emph{Fermi} bubbles (modeled with the alternate template) is $\leq 20 \%$.}
\label{fig:bub_mismodelling_max_20}
\end{figure*}
\begin{figure*}[h!]
\centering
  \noindent
  \resizebox{\textwidth}{!}{
\includegraphics{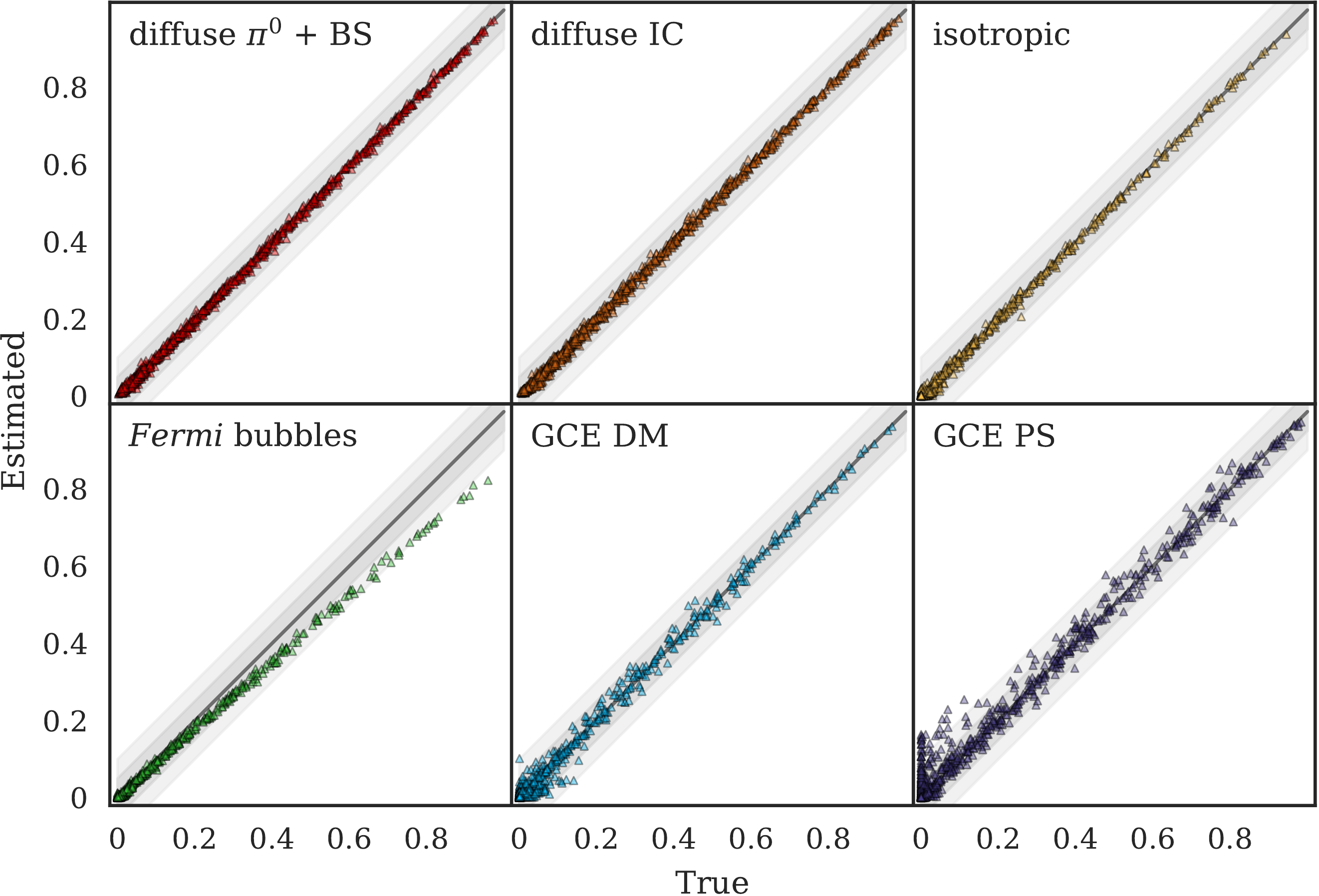} 
}
\caption{\emph{Mismodeling of the Fermi bubbles}: same as Fig.~\ref{fig:bub_mismodelling_max_20}, but now for 1,000 arbitrarily composed maps from our test data set.}
\label{fig:bub_mismodelling_any}
\end{figure*}
Given the general uncertainty around the attribution of flux in the inner Galaxy, it is perhaps unsurprising that the exact spatial distribution of the \emph{Fermi} bubbles close to the Galactic Plane is uncertain. Here we test how dependent the NN results are to variations in the assumed bubble morphology. We apply the trained NN from the proof-of-concept example in the main body of this work to maps whose \emph{Fermi} bubble contribution follows a different spatial template (indicated by the green lines in the \emph{Fermi} bubble sketch in Fig.~\ref{fig:templates}). This alternate \emph{Fermi} bubble template, which has been used by e.g. \cite{Linden2016}, models spatial uniform emission within the bubbles as well, but differs from our default \emph{Fermi} bubble template in that it has a larger support in the southern hemisphere that touches our mask for the Galactic Plane at $b = -2 ^\circ$. 
\par Fig.~\ref{fig:bub_mismodelling_max_20} shows the predictions of our NN for 256 maps whose \emph{Fermi} bubble flux contribution is described by the alternate bubble template and does not exceed $20 \%$ (as is the case for the \emph{Fermi} map). Comparing the NN estimates to the case without mismodeling of the \emph{Fermi} bubbles in Fig.~\ref{fig:toy_example} reveals that the NN accuracy is almost unaffected by the alternate \emph{Fermi} bubble template. The mean absolute errors for all the templates lie still below $1 \%$, and the maximum error for the flux fraction of the \emph{Fermi} bubbles is less than $3 \%$ among these maps, with an average error of only $0.38 \%$. 
\par It is instructive to consider the NN estimates for more extreme (and physically irrelevant) cases in order to expose systematic behavior caused by the mismodeling. In Fig.~\ref{fig:bub_mismodelling_any}, we plot the NN estimates for 1,000 random maps with mismodeled \emph{Fermi} bubble emission of arbitrary strength. Now, it becomes apparent that for \emph{Fermi} bubble flux fractions $\gtrsim 20 \%$, the larger support of the alternate \emph{Fermi} bubble template in the maps causes the NN to systematically underestimate the \emph{Fermi} bubble flux, attributing it to the GCE templates instead -- in particular to GCE PS. As the difference between the \emph{Fermi} bubble templates used for training and evaluation is asymmetric and almost exclusively concerns the southern hemisphere, the preference of the NN for attributing the unrecognized flux from the \emph{Fermi} bubbles to GCE PS over GCE DM is reminiscent of the fact that an unmodeled asymmetry can be accounted for more easily by a non-Poissonian template, as recently shown in \cite{Leane2020a}. However, we recall that our NN appears to be robust to modest GCE DM north-south asymmetries as demonstrated in Fig.~\ref{fig:gce_mismodelling_max_2}. The mismodeling of the \emph{Fermi} bubbles has no apparent effect on the NN estimates for the other non-GCE templates. We conclude from this experiment: \textbf{for maps with realistic \emph{Fermi} bubble flux fractions, the NN estimates for the GCE are insensitive to varying the spatial morphology of the \emph{Fermi} bubbles for the templates considered herein.}

\section{Additional material for the proof-of-concept example}
\label{sec:add_mat_toy}
In this section, we provide additional material for the proof-of-concept example in the main body of this work. First, we present the NN estimates shown in Fig.~\ref{fig:toy_example} for all the templates that we use for the modeling of the inner Galaxy in this example. Then, we consider the predictions for the first 9 photon-count maps in our testing data set in detail.

\subsection{NN estimates for all the templates}
\begin{figure*}
\centering
  \noindent
    \resizebox{1\textwidth}{!}{
     \includegraphics{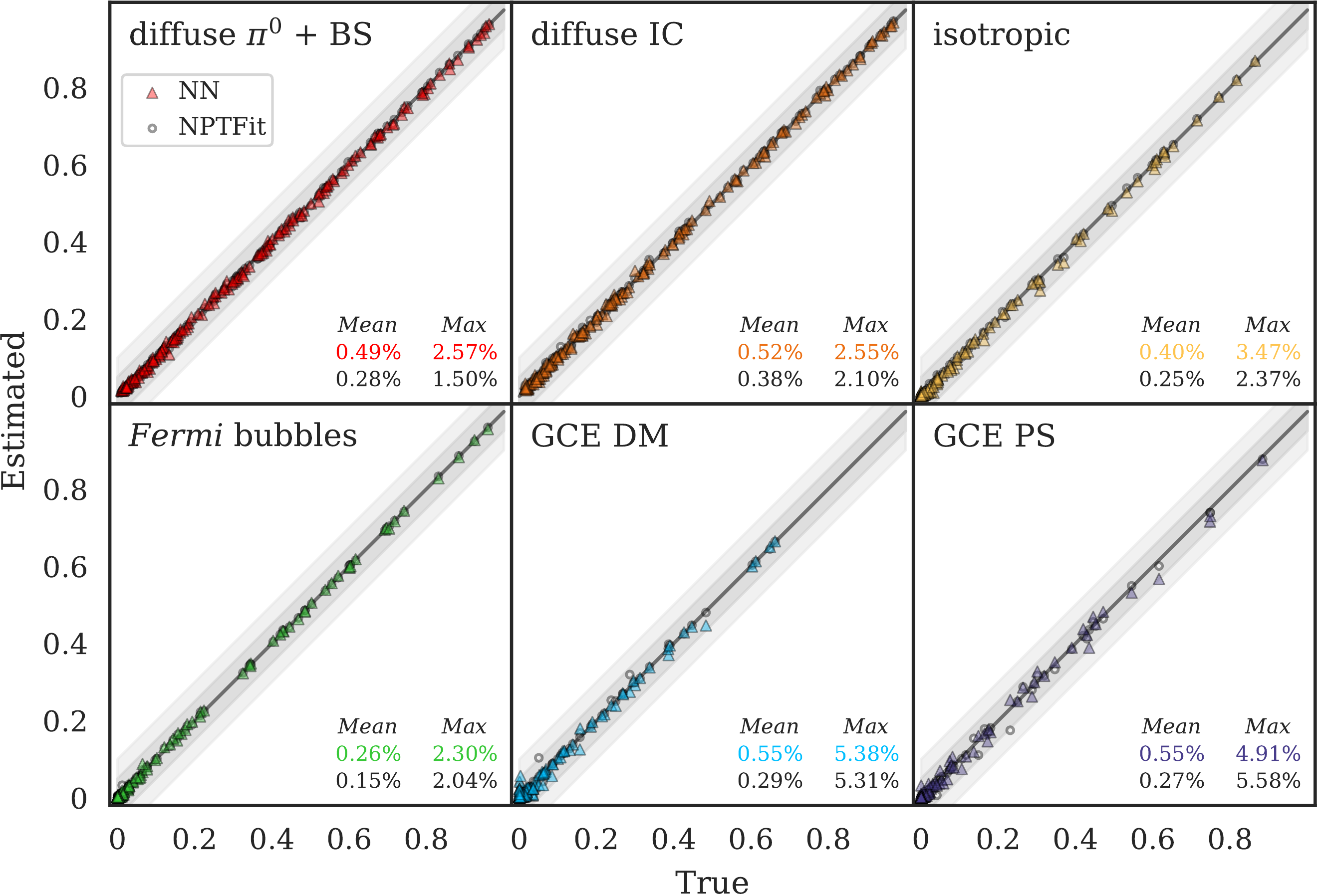}}
     \caption{True vs. estimated flux contributions within the ROI for the NN (colored triangles) and \texttt{NPTFit} (black circles) for each of the templates used in the proof-of-concept example. Plotted are the estimates for 172 maps from the test data set. The gray shaded regions correspond to errors of $5$ and $10 \%$, respectively. The mean and maximum absolute errors are specified in the lower right corner for each template.}
     \label{fig:toy_example_all}
\end{figure*}

Fig.~\ref{fig:toy_example_all} depicts the true vs. estimated flux fractions for the 172 randomly generated maps from the test data set plotted in Fig.~\ref{fig:toy_example}, but now for \emph{all} the templates present in the proof-of-concept example. The colored triangles and black circles indicate the NN and \texttt{NPTFit} predictions, respectively. The \texttt{NPTFit} estimates are generally somewhat more accurate than their NN counterparts, but the mean errors for both methods lie well below one per cent in this example, while the maximum errors are $\lesssim 3 \%$ for the non-GCE templates and $\sim 5 \%$ for each of the two GCE templates. We expect that the NN accuracy can be further improved in future by exploiting more elaborate Deep Learning techniques.

\begin{figure*}[htb]
\centering
  \noindent
  \resizebox{0.7\textwidth}{!}{
\includegraphics{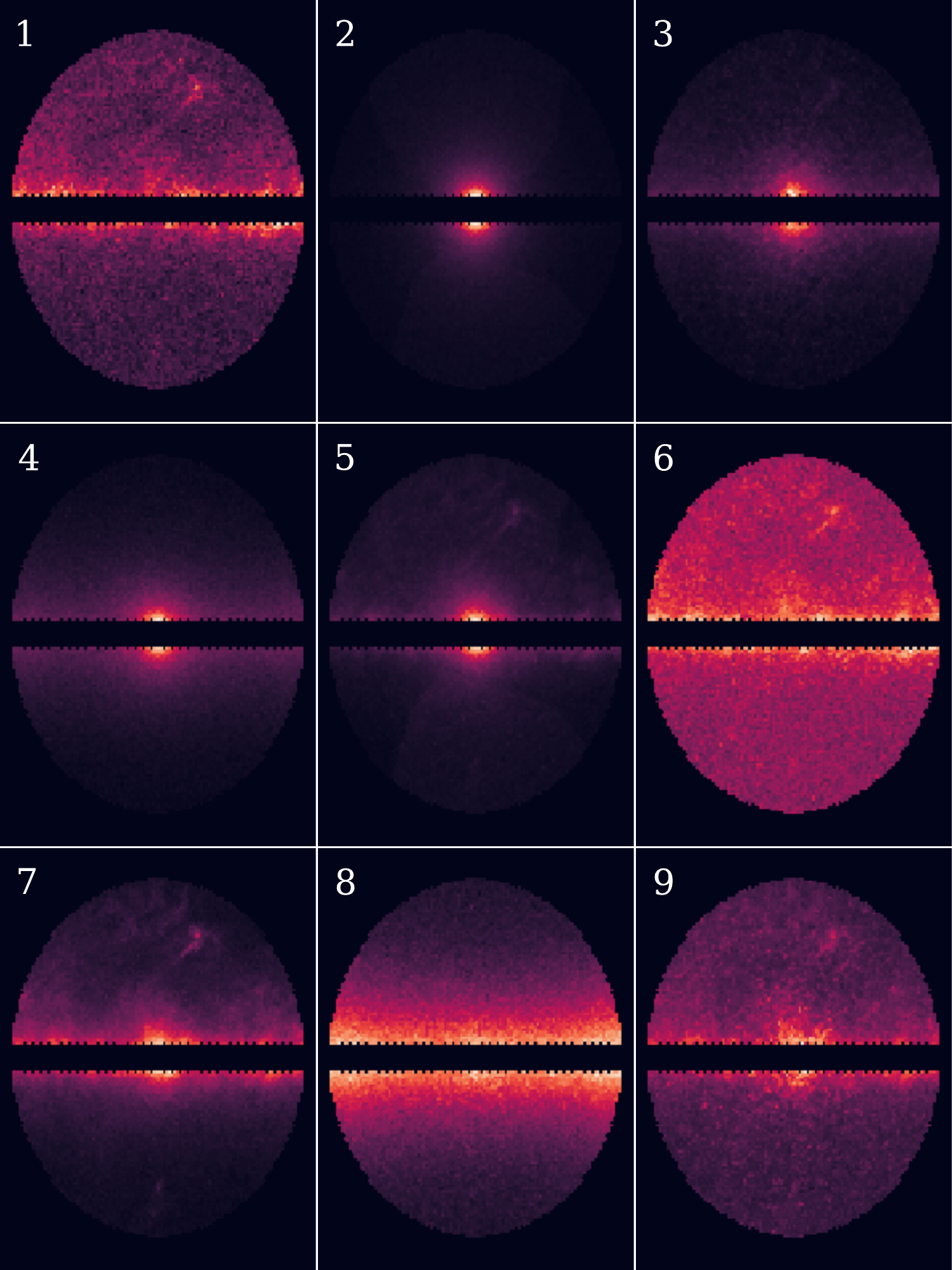} 
}
\caption{Projection of the first 9 photon-count maps from the test data set for the scenario in the proof-of-concept example. The true and estimated flux fractions for each of the maps, as well as the maximum number of photon counts in a single pixel $s_\text{max}$ (which defines the upper (bright) limit of the color map for each map), are listed in Table \ref{table:flux_fractions_toy}.}
\label{fig:gce_and_background_examples}
\end{figure*}
\begin{table}[htb]
\caption{True and predicted flux fractions for the 9 photon-count maps depicted in Fig.~\ref{fig:gce_and_background_examples} \orange{in per cent}. 
The last column contains the maximum number of photon counts in a single pixel $s_\text{max}$. The maximum error is $2.9 \%$, which occurs for the GCE DM template in the last map, where the GCE has contributions from both GCE DM and GCE PS.}
\orange{
\begin{tabular}{@{}ccccccccc@{}}
\toprule
\multicolumn{1}{l}{Map} &  & $\pi^0$ + BS & IC & isotropic & bubbles & GCE DM & GCE PS & \multicolumn{1}{l}{$s_\text{max}$} \\ \midrule
\multirow{2}{*}{1} & true & 44.3 & 6.1 & 49.4 & 0.1 & 0.1 & 0.1 & \multirow{2}{*}{89} \\
 & NN & 45.5 & 6.0 & 48.0 & 0.3 & 0.1 & 0.1 &  \\ \cmidrule(lr){2-8}
\multirow{2}{*}{2} & true & 2.2 & 1.3 & 3.2 & 13.0 & 80.3 & 0.0 & \multirow{2}{*}{1,737} \\
 & NN & 1.9 & 1.7 & 2.3 & 14.0 & 79.8 & 0.1 &  \\ \cmidrule(lr){2-8}
\multirow{2}{*}{3} & true & 22.4 & 27.4 & 6.4 & 0.0 & 0.2 & 43.6 & \multirow{2}{*}{1,410} \\
 & NN & 23.1 & 27.0 & 5.7 & 0.4 & 1.1 & 42.7 &  \\ \cmidrule(lr){2-8}
\multirow{2}{*}{4} & true & 4.1 & 62.1 & 0.0 & 0.1 & 33.7 & 0.0 & \multirow{2}{*}{696} \\
 & NN & 3.7 & 62.2 & 0.1 & 0.1 & 33.8 & 0.1 &  \\ \cmidrule(lr){2-8}
\multirow{2}{*}{5} & true & 40.2 & 5.6 & 0.0 & 12.5 & 40.8 & 0.9 & \multirow{2}{*}{1,373} \\
 & NN & 40.3 & 5.0 & 0.2 & 12.7 & 40.2 & 1.5 &  \\ \cmidrule(lr){2-8}
\multirow{2}{*}{6} & true & 22.6 & 2.1 & 71.4 & 2.7 & 1.2 & 0.0 & \multirow{2}{*}{154} \\
 & NN & 23.9 & 1.9 & 71.4 & 2.3 & 0.4 & 0.2 &  \\ \cmidrule(lr){2-8}
\multirow{2}{*}{7} & true & 54.3 & 26.6 & 0.5 & 0.1 & 18.5 & 0.0 & \multirow{2}{*}{672} \\
 & NN & 54.3 & 26.7 & 0.3 & 0.2 & 18.4 & 0.1 &  \\ \cmidrule(lr){2-8}
\multirow{2}{*}{8} & true & 1.3 & 86.4 & 9.3 & 0.0 & 0.0 & 3.1 & \multirow{2}{*}{351} \\
 & NN & 2.0 & 85.9 & 8.4 & 0.3 & 0.4 & 3.1 &  \\ \cmidrule(lr){2-8}
\multirow{2}{*}{9} & true & 36.2 & 2.7 & 48.3 & 1.1 & 3.5 & 8.1 & \multirow{2}{*}{381} \\
 & NN & 36.5 & 2.9 & 48.6 & 0.8 & 0.6 & 10.6 &  \\ \bottomrule
\end{tabular}
}
\label{table:flux_fractions_toy}
\end{table}

\subsection{Evaluation of the NN on test data}
In order to make the flux fraction estimation more illustrative, we consider the NN estimates for the first 9 (randomly composed) maps from the testing data set in detail. These maps are depicted in Fig.~\ref{fig:gce_and_background_examples} within our ROI of $\leq 25 ^\circ$ around the GC, with $|b| \leq 2 ^\circ$ masked. The color scale ranges from 0 (black) to the maximum number of counts within the ROI of each map (light yellow). Some of the maps do not contain a GCE (e.g. map 1), whereas the GCE in other maps consists of DM flux (e.g. map 2), PS flux (e.g. map 3), or both (map 9). The template that contributes the bulk of the flux differs among the maps and is given, for instance, by diffuse $\pi^0 + \text{BS}$ in map 7, diffuse IC in map 8, GCE PS in map 3, GCE DM and diffuse $\pi^0 + \text{BS}$ in equal parts in map 5, and isotropic emission in map 6. Note that this variety among the maps is also present in the training data set since we believe that training our NN on diverse maps causes it to learn the entire parameter space and to understand the distinct characteristics of each individual template, leading to robust estimates. 
\par The true flux fractions as well as the NN estimates for the plotted maps are listed in Table \ref{table:flux_fractions_toy}. The diffuse flux fractions in the maps are estimated accurately and all the errors lie below $1.5 \%$. The same holds for the isotropic template and the \emph{Fermi} bubbles. For the GCE, we confirm the results from the experiment for different SCDs in Sec. \ref{sec:SCDs}: if the GCE is made up entirely of DM or PS, the NN predictions are very accurate ($< 1 \%$ for these maps). For map 9 with a GCE consisting of $8.1 \%$ PS and $3.5 \%$, the NN attributes the GCE flux solely to PS. For the NN estimates on a more representative set of count maps for this scenario, we refer the reader to Fig.~\ref{fig:toy_example}.

\section{Additional material for the realistic scenario}
\label{sec:add_mat_realistic}
In this section, we provide further insights into the Bayesian GCNN that we use for estimating the flux fractions in the \emph{Fermi} map in the main part of our \emph{Letter}. First, we tabulate the NPTF parameters determined by \texttt{NPTFit} for the \emph{Fermi} map, for outer ROI radii of $10 ^\circ$ and $25 ^\circ$. Moreover, we show the error histograms for the simulated maps in the realistic scenario since the non-Gaussianity of the NN errors is not apparent in Fig.~\ref{fig:fermi_results}. We repeat the experiment from the main body with simulated maps corresponding to the $25 ^\circ$ NPTF parameters. Then, we present the predictions of the NN for each of the first 9 maps from our testing data set and discuss some of the scenarios that pose difficulties to the NN. Next, we show an error plot with the NN estimates for the flux fractions and their uncertainties for 200 maps from the test data sets. Furthermore, we address the calibration of the NN uncertainty estimates and present the predictive uncertainty covariances for the \emph{Fermi} map when modeling the full uncertainty covariance matrix using the loss function in Eq.~\eqref{eq:loss_full_covar}. \orange{Finally, we consider the dependence of the NN accuracy on the number of training maps, and we evaluate our NN on maps with a more realistic MSP SCD.}

\subsection{NPTF parameters for the \emph{Fermi} map}
\begin{table}[ht]
\caption{Model parameters for the simulated \emph{Fermi} mock data used in Fig.~\ref{fig:fermi_results} ($10 ^\circ$ ROI for the NPTF) and in Fig.~\ref{fig:fermi_results_25} ($25 ^\circ$ ROI for the NPTF). They correspond to the medians of the marginalized posterior distributions determined by \texttt{NPTFit} within the respective ROI.}
\centering
\begin{tabular}{@{}lcc@{}}
\toprule
\multirow{2}{*}{Model parameter} & \multicolumn{2}{c}{Value} \\ \cmidrule(l){2-3} 
 & \multicolumn{1}{c}{$10 ^\circ$ ROI} & $25 ^\circ$ ROI \\ \midrule
$\log_{10} A_\text{dif}^{\pi^0 + \text{BS}}$ & $0.94$ & $0.90$ \\
$\log_{10} A_\text{dif}^{\text{IC}}$ & $0.66$ & $0.58$ \\
$\log_{10} A_\text{iso}$ & $-1.15$ & $-0.43$ \\
$\log_{10} A_\text{bub}$ & $-0.90$ & $-0.06$ \\
$\log_{10} A_\text{gce}$ & $-1.84$ & $-1.57$ \\
$\log_{10} A_\text{gce}^\text{PS}$ & $-1.46$ & $-0.71$ \\
$\log_{10} A_\text{disk}^\text{PS}$ & $-4.50$ & $-1.47$ \\
$S_1^\text{gce}$ & $9.52$ & $3.75$ \\
$S_1^\text{disk}$ & $21.71$ & $4.71$ \\
$n_1^\text{gce}$ & $34.34$ & $33.97$ \\
$n_2^\text{gce}$ & $-1.08$ & $-0.93$ \\
$n_1^\text{disk}$ & $31.58$ & $3.32$ \\
$n_2^\text{disk}$ & $-1.13$ & $-0.97$ \\ \bottomrule
\end{tabular}
\label{table:NPTF_mock_params}
\end{table}
For completeness, we supply the model parameters that we use for the simulated mock \emph{Fermi} data in Fig.~\ref{fig:fermi_results} in Table \ref{table:NPTF_mock_params}, which are given by the medians of the posterior distributions determined by \texttt{NPTFit} for a ROI radius of $10 ^\circ$, with $|b| \leq 2 ^\circ$ as well as 3FGL sources masked. Furthermore, we list the parameters obtained from \texttt{NPTFit} when selecting a larger ROI with an outer radius of $25 ^\circ$. This case is considered below.
The $10 ^\circ$ ROI is too small for the NPTF to detect the flux from the \emph{Fermi} bubbles (see Fig.~\ref{fig:fermi_results}), and the ascertained disk PS flux is negligible. The source count break $S_1^\text{gce}$ in this case is similar to the ``moderate'' case in the SCD experiment in Sec. \ref{sec:SCDs}. However, the large value of $n_1^\text{gce} = 34.34$ implies a steeper slope of the SCD function for fluxes above the break, causing the GCE PS to be dimmer on average and making the prediction for the NN more challenging, resulting in a fraction of the GCE flux being mistakenly attributed to DM. 

\subsection{Error histogram for the simulated \emph{Fermi} best-fit mock maps}
\begin{figure*}
\centering
  \noindent
  \resizebox{1\textwidth}{!}{
\includegraphics{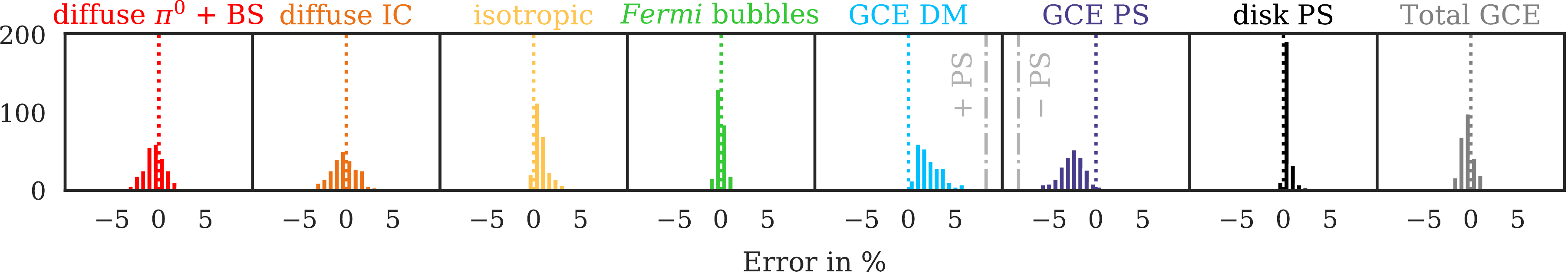} 
}
\caption{Histogram of the NN errors for our largest ROI (outer radius $25 ^\circ$, see Fig.~\ref{fig:fermi_results} for the estimates as a function of the ROI radius) for the 250 simulated maps corresponding to the best-fit parameters for the \emph{Fermi} map as determined by \texttt{NPTFit} within a ROI of outer radius $10 ^\circ$. While the errors for the two diffuse components are scattered in an approximately symmetric way around zero, the distributions for the templates with small flux fractions are skewed away from zero (most notably GCE DM, and isotropic, disk PS, and the \emph{Fermi} bubbles to a lesser extent) since the NN outputs are forced to be positive due to the use of a Softmax activation function for the mean estimates. The error distribution for the GCE PS flux peaks at approximately $-2 \%$, evidencing that the NN underestimates the GCE PS flux fraction by a few per cent for most of the maps. Still, for $85 \%$ of the maps, the GCE PS flux fraction estimates lie not more than $4 \%$ below the true value of $8.3 \%$.}
\label{fig:fermi_mock_histogram}
\end{figure*}
Although we model the uncertainties of the flux fractions to be Gaussian in this work, the actual distribution of the NN predictions is non-Gaussian to a certain extent, in particular in the low-flux region where the distribution is bounded from below by zero and therefore often positively skewed. Hence, the $1 \sigma$-region of the scatter in the NN estimates for the simulated maps in Fig.~\ref{fig:fermi_results} does not contain the full information about the error distribution. In Figure \ref{fig:fermi_mock_histogram}, we plot the NN errors towards the true values for all the templates, evaluated in a ROI with outer radius $25 ^\circ$. As already apparent from Fig.~\ref{fig:fermi_results}, a fraction of the GCE PS is misattributed to GCE DM, and the error distribution for DM exhibits a positive skewness, with a mean absolute error of $2.3 \%$. The distribution of the errors for the GCE PS flux appears to be roughly Gaussian (albeit slightly negatively skewed), however centered at a negative value (mean: $-2.7 \%$, median: $-2.5 \%$, mean absolute error: $2.7 \%$). Nonetheless, the NN assigns a larger flux to GCE PS than to GCE DM for $88.8 \%$ of the maps (222 out of 250). The mean absolute error for the total GCE flux fraction is $0.64 \%$, and for the non-GCE templates $1.0 \%$ (diffuse $\pi^0$ + BS), $1.2 \%$ (diffuse IC), $0.82 \%$ (isotropic), $0.37 \%$ (\emph{Fermi} bubbles), and $0.44 \%$ (disk PS).

\subsection{Evaluation of the NN on simulated \emph{Fermi} best-fit maps as determined within 25$^{\boldsymbol{\circ}}$}
\begin{figure*}
\centering
  \noindent
  \resizebox{1\textwidth}{!}{
\includegraphics{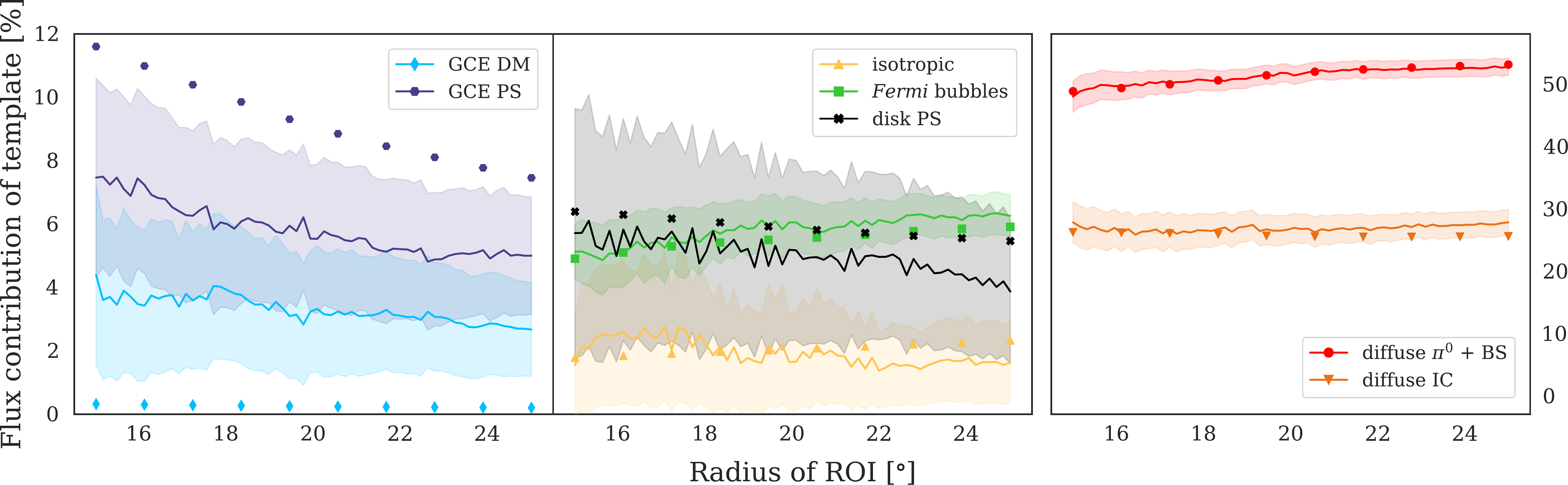} 
}
\caption{Flux estimates from the Bayesian GCNN for 250 mock maps corresponding to the best-fit parameters as determined by \texttt{NPTFit} for an outer ROI radius of $25 ^\circ$ (cf. Fig.~\ref{fig:fermi_results} for the best-fit parameters within a radius of $10 ^\circ$), within the ROIs delimited by the outer radii given on the $x$-axis. The flux fractions are predicted with respect to each ROI, implying that the fractions are expected to vary as a function of radius, rather than to remain constant. The shaded regions show the 1$\sigma$ scatter computed over the 250 mock samples. The markers indicate the correct flux fraction for each radius, averaged over the samples.}
\label{fig:fermi_results_25}
\end{figure*}
We revisit the experiment for our Bayesian GCNN on simulated best-fit \emph{Fermi} maps depicted in Fig.~\ref{fig:fermi_results} (\emph{upper left}), but now we determine the NPTF parameters for the simulated maps by running \texttt{NPTFit} on the \emph{Fermi} map for a larger ROI radius of $25 ^\circ$ around the GC, as usual with $|b| \leq 2 ^\circ$ and 3FGL sources masked. The resulting medians of the marginalized posterior distributions are tabulated in Table \ref{table:NPTF_mock_params}. For this ROI, \texttt{NPTFit} finds the source count break to be $S_1^\text{gce} = 3.75$, which is similar to the value for the ``dim'' case in the SCD experiment in Sec. \ref{sec:SCDs}. Again, $n_1^\text{gce}$ is large, for which reason only few GCE PS that are brighter than the flux associated with the count break are expected to be present in the 250 simulated \emph{Fermi} mock maps that we generate. 
\par The NN estimates and the correct flux fractions (averaged over the maps) are shown in Fig.~\ref{fig:fermi_results_25}. Now, a flux contribution from the \emph{Fermi} bubbles is identified by \texttt{NPTFit}, which is accurately recovered by the NN (see also Fig.~\ref{fig:fermi_results} for the flux contribution from the \emph{Fermi} bubbles that the NN finds in the real \emph{Fermi} map). Also, the normalization of the disk PS template $A_\text{disk}^\text{PS}$ is three orders of magnitude larger than for the $10 ^\circ$ case, resulting in a disk PS flux fraction of roughly $5 \%$. Due to the fainter GCE PS, the fraction of GCE flux that is misattributed to DM is larger in this case, and the $1\sigma$ scatter regions overlap for all ROI radii. Still, the NN correctly infers on average that the GCE in the simulated maps are dominated by GCE PS.   

\subsection{Evaluation of the NN on test data}
\begin{figure}[htb]
\centering
  \noindent
  \resizebox{0.3\textwidth}{!}{
\includegraphics{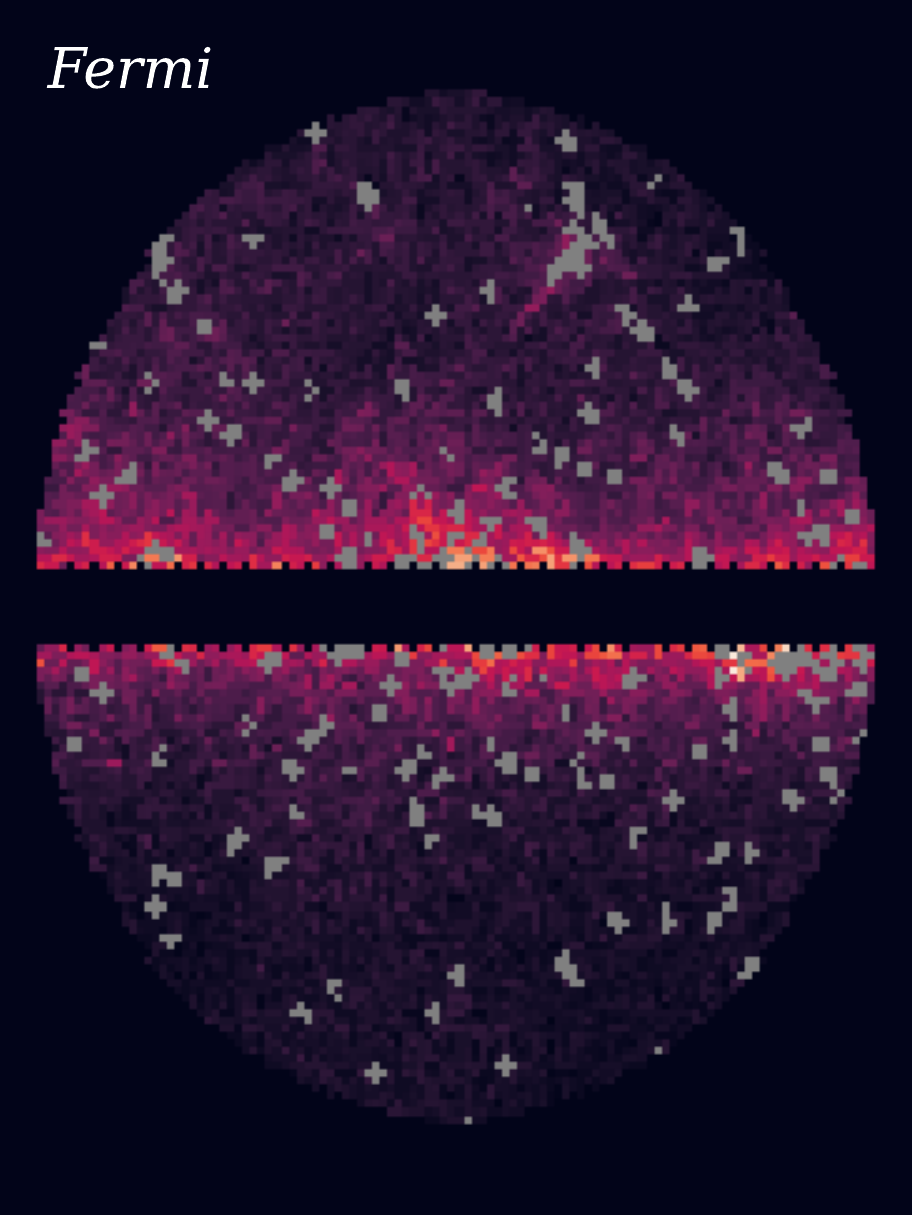} 
}
\caption{Projection of the \emph{Fermi} photon-count map used in this work, within our maximum ROI of radius $25 ^\circ$. The largest number of counts in a single pixel is 164, but we choose the range for the color map to be $[0, 100]$ (from black to light yellow) to enhance the contrast. Masked 3FGL sources are colored gray.}
\label{fig:fermi_count_map}
\end{figure}
\begin{figure*}[htb]
\centering
  \noindent
  \resizebox{0.7\textwidth}{!}{
\includegraphics{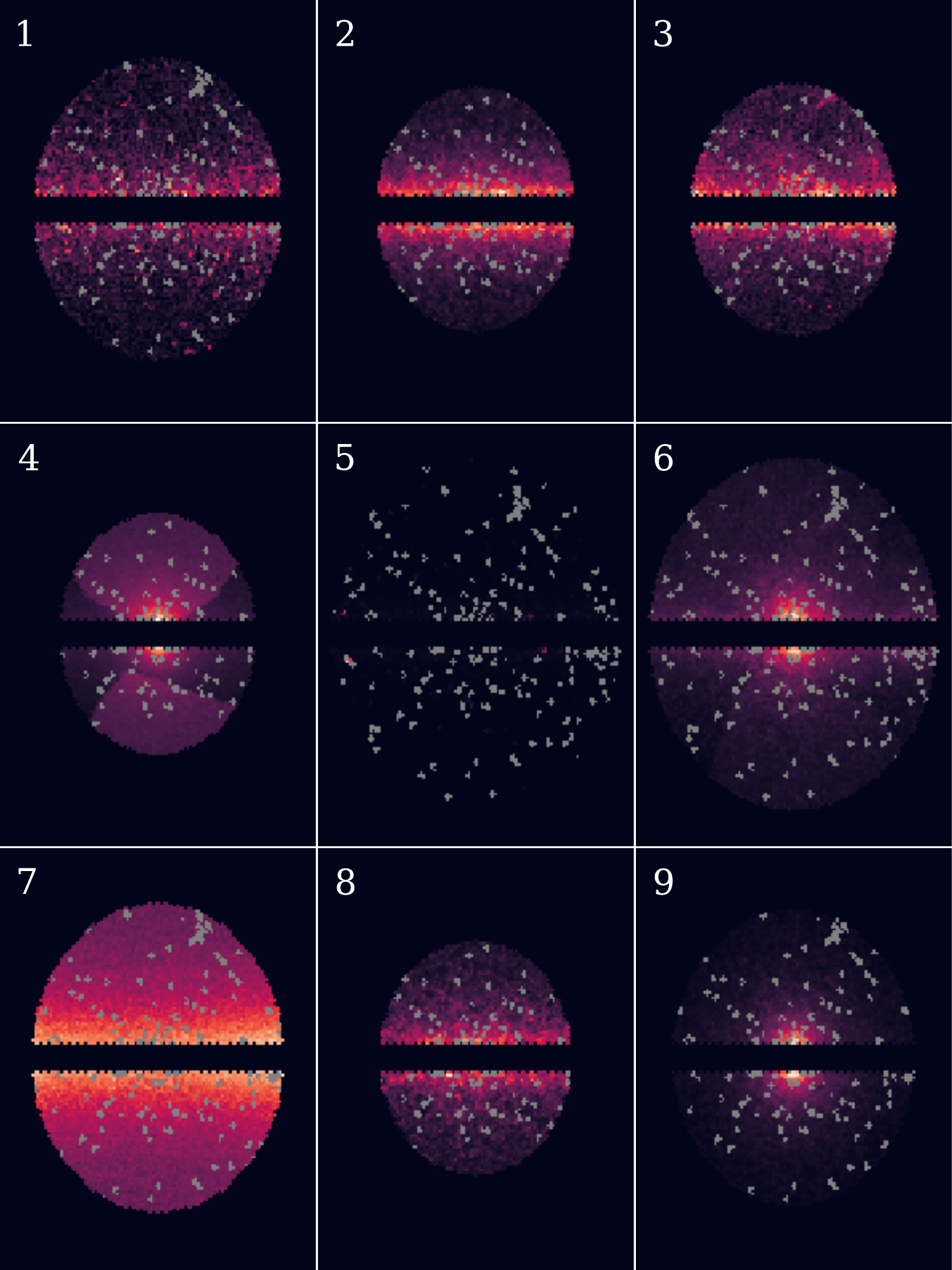} 
}
\caption{Projection of the first 9 photon-count maps from the test data set for the realistic scenario. The true and estimated flux fractions for each of the maps, as well as the maximum number of photon counts in a single pixel $s_\text{max}$ (which defines the upper (bright) limit of the color map for each map), and the respective ROI radii $\phi_\text{ROI}$ are listed in Table \ref{table:flux_fractions_fermi}. Masked 3FGL sources are colored gray.}
\label{fig:gce_for_letter_examples}
\end{figure*}
\begin{table}[hbt]
\caption{True and predicted flux fractions for the 9 photon-count maps depicted in Fig.~\ref{fig:gce_for_letter_examples} \orange{in per cent}. 
The second last column contains the maximum number of photon counts in a single pixel $s_\text{max}$, and the last column lists the ROI radii.}
\orange{
\begin{tabular}{@{}ccccccccccc@{}}
\toprule
\multicolumn{1}{l}{Map} &  & $\pi^0$ + BS & IC & isotropic & bubbles & GCE DM & GCE PS & disk PS & \multicolumn{1}{l}{$s_\text{max}$} & \multicolumn{1}{l}{$\phi_\text{ROI}$ [$ ^\circ$]} \\ \midrule
\multirow{2}{*}{1} & true & 27.8 & 56.8 & 6.3 & 0.1 & 0.1 & 0.0 & 8.8 & \multirow{2}{*}{30} & \multirow{2}{*}{21.15} \\
 & NN & 29.3 & 57.1 & 6.6 & 0.5 & 0.3 & 0.1 & 6.0 &  &  \\ \cmidrule(lr){2-9}
\multirow{2}{*}{2} & true & 14.8 & 2.8 & 0.7 & 0.0 & 4.1 & 0.7 & 76.9 & \multirow{2}{*}{1,132} & \multirow{2}{*}{16.99} \\
 & NN & 13.6 & 2.9 & 0.2 & 0.5 & 1.1 & 3.6 & 78.1 &  &  \\ \cmidrule(lr){2-9}
\multirow{2}{*}{3} & true & 63.5 & 32.2 & 0.0 & 3.2 & 0.1 & 0.7 & 0.2 & \multirow{2}{*}{52} & \multirow{2}{*}{17.55} \\
 & NN & 64.1 & 31.4 & 1.2 & 2.6 & 0.6 & 0.1 & 0.0 &  &  \\ \cmidrule(lr){2-9}
\multirow{2}{*}{4} & true & 1.1 & 21.8 & 0.0 & 42.8 & 34.4 & 0.0 & 0.0 & \multirow{2}{*}{1,316} & \multirow{2}{*}{16.72} \\
 & NN & 1.2 & 22.4 & 0.6 & 42.4 & 32.6 & 0.5 & 0.2 &  &  \\ \cmidrule(lr){2-9}
\multirow{2}{*}{5} & true & 0.8 & 9.6 & 0.1 & 0.6 & 0.0 & 0.1 & 88.7 & \multirow{2}{*}{149,475} & \multirow{2}{*}{24.70} \\
 & NN & 2.9 & 3.0 & 0.0 & 0.1 & 0.9 & 0.3 & 92.7 &  &  \\ \cmidrule(lr){2-9}
\multirow{2}{*}{6} & true & 24.5 & 19.8 & 0.0 & 16.5 & 4.6 & 33.6 & 1.0 & \multirow{2}{*}{1,000} & \multirow{2}{*}{24.61} \\
 & NN & 25.2 & 18.7 & 0.3 & 16.0 & 0.2 & 38.0 & 1.6 &  &  \\ \cmidrule(lr){2-9}
\multirow{2}{*}{7} & true & 1.7 & 49.2 & 48.7 & 0.0 & 0.0 & 0.0 & 0.4 & \multirow{2}{*}{408} & \multirow{2}{*}{21.58} \\
 & NN & 1.7 & 48.3 & 48.6 & 0.4 & 0.0 & 0.1 & 0.9 &  &  \\ \cmidrule(lr){2-9}
\multirow{2}{*}{8} & true & 2.2 & 5.0 & 0.4 & 12.7 & 3.9 & 0.0 & 75.7 & \multirow{2}{*}{519} & \multirow{2}{*}{16.35} \\
 & NN & 2.8 & 4.7 & 1.4 & 12.6 & 3.5 & 2.7 & 72.3 &  &  \\ \cmidrule(lr){2-9}
\multirow{2}{*}{9} & true & 7.3 & 1.3 & 0.2 & 0.0 & 1.7 & 89.0 & 0.6 & \multirow{2}{*}{3,887} & \multirow{2}{*}{20.69} \\
 & NN & 7.6 & 2.6 & 0.6 & 0.3 & 1.2 & 87.5 & 0.2 &  &  \\ \bottomrule
\end{tabular}
}
\label{table:flux_fractions_fermi}
\end{table}
\begin{figure}[htb]
\centering
  \noindent
  \resizebox{\textwidth}{!}{
\includegraphics{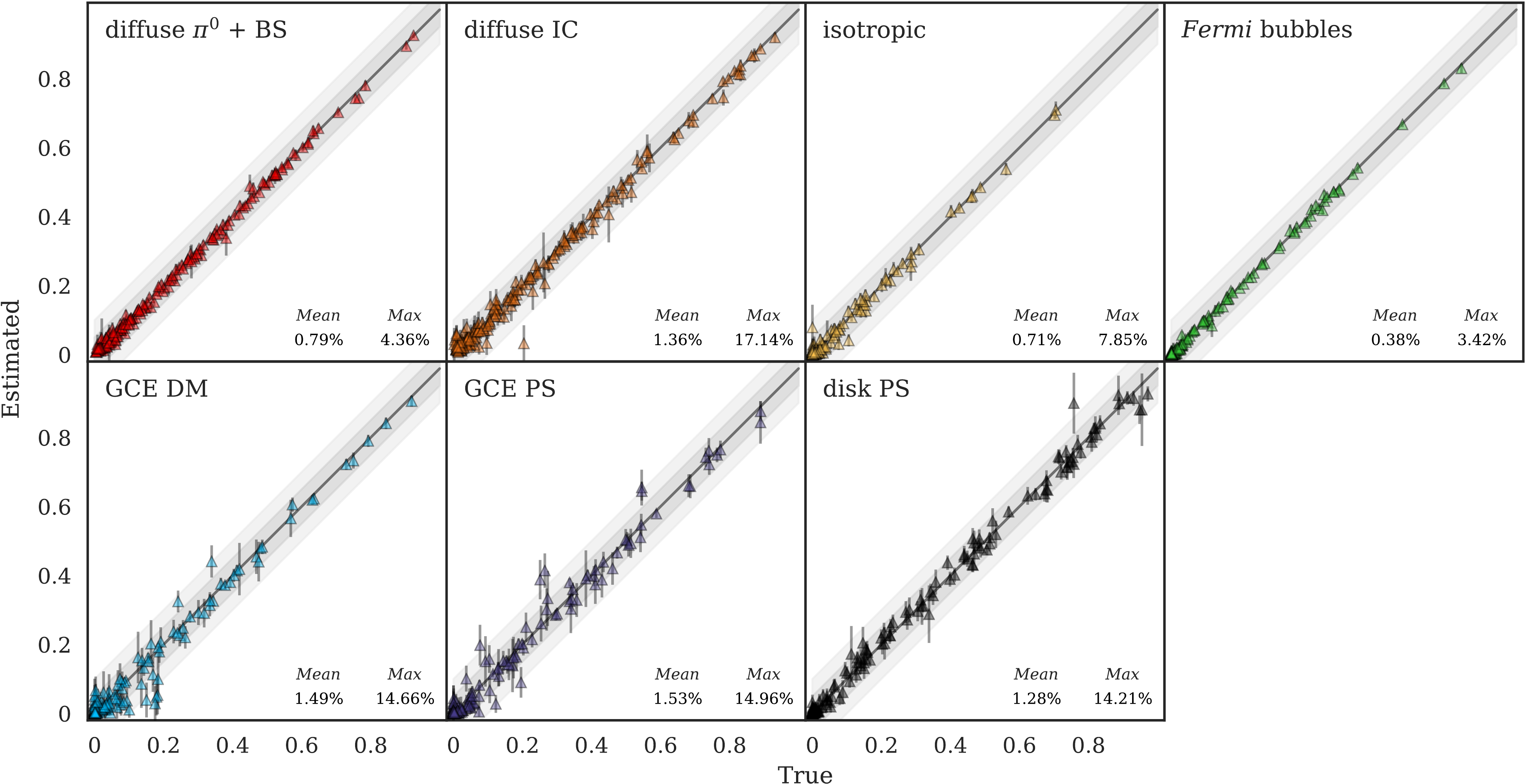} 
}
\caption{True vs. estimated flux fractions for 200 photon-count maps from the test set, for the Bayesian GCNN used in the realistic scenario. The error bars correspond to the $1\sigma$ predictive (aleatoric and epistemic summed in quadrature) errors estimated by the Bayesian GCNN.}
\label{fig:fermi_true_vs_estimated}
\end{figure}
For the training of the NN, we considered ROIs with outer radii between $15$ and $25 ^\circ$, with latitudes $|b| \leq 2 ^\circ$ and known 3FGL sources masked (see Fig.~\ref{fig:fermi_count_map} for the \emph{Fermi} count map within the maximum ROI used herein). We plot the first 9 photons count maps from the test data set (that the NN has not seen during the training) in Fig.~\ref{fig:gce_for_letter_examples}, and the true and estimated flux fractions are listed in Table \ref{table:flux_fractions_fermi}. Note that randomly sampling model parameters from the priors in Table \ref{table:priors} as done for the generation of training and test data yields a large variety among the maps (as already seen in Sec. \ref{sec:add_mat_toy}): whereas the maximum number of photons $s_\text{max} = 30$ for the first depicted map, $s_\text{max} \sim$ 150,000 for map number 5 due to some very bright disk PS, causing the other pixels in the map to appear dark. In comparison, $s_\text{max} = 164$ for the \emph{Fermi} map in our ROI with the 3FGL sources masked. While the NN estimates are mostly accurate, we point out some scenarios that appear to be difficult for the NN: 1) When the photon counts are dominated by extremely strong emission from disk PS, accurately estimating the remaining flux fractions seems to become more challenging (see map 5, where the disk PS template accounts for almost $90 \%$ of the total flux, and the diffuse IC flux fraction is underestimated by the NN by $6.6 \%$). 2) When the GCE is composed of both DM and PSs but one of these contributions is much smaller than the other, it is difficult for the NN to recover it (see map 6, where a GCE DM fraction of $4.6 \%$ is obscured by $33.6 \%$ GCE DM and the NN attributes the entire GCE to GCE DM; also see Sec. \ref{sec:SCDs}). 3) When non-GCE PS emission is very strong, faint GCE DM fluxes may be misattributed to GCE PS (see map 2, where disk PS is largely dominant, and while the NN correctly identifies a GCE, it is predominantly absorbed by the GCE PS template). Comparing the results in Table \ref{table:flux_fractions_fermi} with those for the slightly simplified case in Table \ref{table:flux_fractions_toy} shows that introducing an additional PS template, namely disk-correlated PS, makes it more difficult for the NN to recover the flux fractions, with the GCE templates being most affected. In our numerical experiments, we observed that the loss in accuracy due to the masking of the pixels with bright 3FGL sources in the \emph{Fermi} map and due to the variable ROI is subdominant in comparison to the impact of the additional non-Poissonian disk PS template, and the non-uniform exposure does not seem to affect the NN accuracy at all (the difference in exposure across our small ROI is very small anyway).
\par Now, we consider a larger (and therefore more representative) set of test images consisting of the first 200 maps from our test data set. The true vs. estimated flux fractions are plotted in Fig.~\ref{fig:fermi_true_vs_estimated}. In comparison to the slightly simplified case considered in the proof-of-concept example, the errors in the flux fractions are somewhat larger now, owing to the increased difficulty. Still, the mean absolute error for these 200 maps is $\leq 1.5 \%$ for all the templates. The uncertainty magnitudes reflect the differences in accuracy among the templates and for different maps within a particular template: for the \emph{Fermi} bubbles, for instance, the flux fractions are estimated very accurately and the estimated uncertainties are small (with the error bars mostly hidden behind the markers), whereas the larger errors for the GCE templates are evidenced by larger error bars. For the diffuse IC template, the errors and uncertainties are mostly rather small; however, there is an outlier map for which the diffuse IC flux is significantly underestimated. Still, the NN recognizes that predicting the diffuse IC flux fraction for this specific map is difficult and yields a comparably large uncertainty. This highlights that modeling the uncertainties as heteroscedastic (that is, dependent on the respective input map) enables the NN to assign large uncertainties for difficult maps where the estimated flux fractions might be inaccurate. In general, the higher the flux fraction of a particular template, the more accurate the predictions, which is correctly mirrored by the small uncertainties for large flux fractions. An exception to this are maps that are dominated by very bright disk-correlated PS (note that this case is more of academic interest rather than practically relevant, provided that the bright 3FGL sources are masked in the analysis): in this case, the estimated disk PS flux fractions are scattered around the true values, for which the NN accounts by outputting large uncertainties. We discuss the calibration of the estimated uncertainties below.

\subsection{Calibration of the uncertainties}
\label{sec:calibration}
\begin{figure}[htb]
\centering
  \noindent
  \resizebox{0.5\textwidth}{!}{
\includegraphics{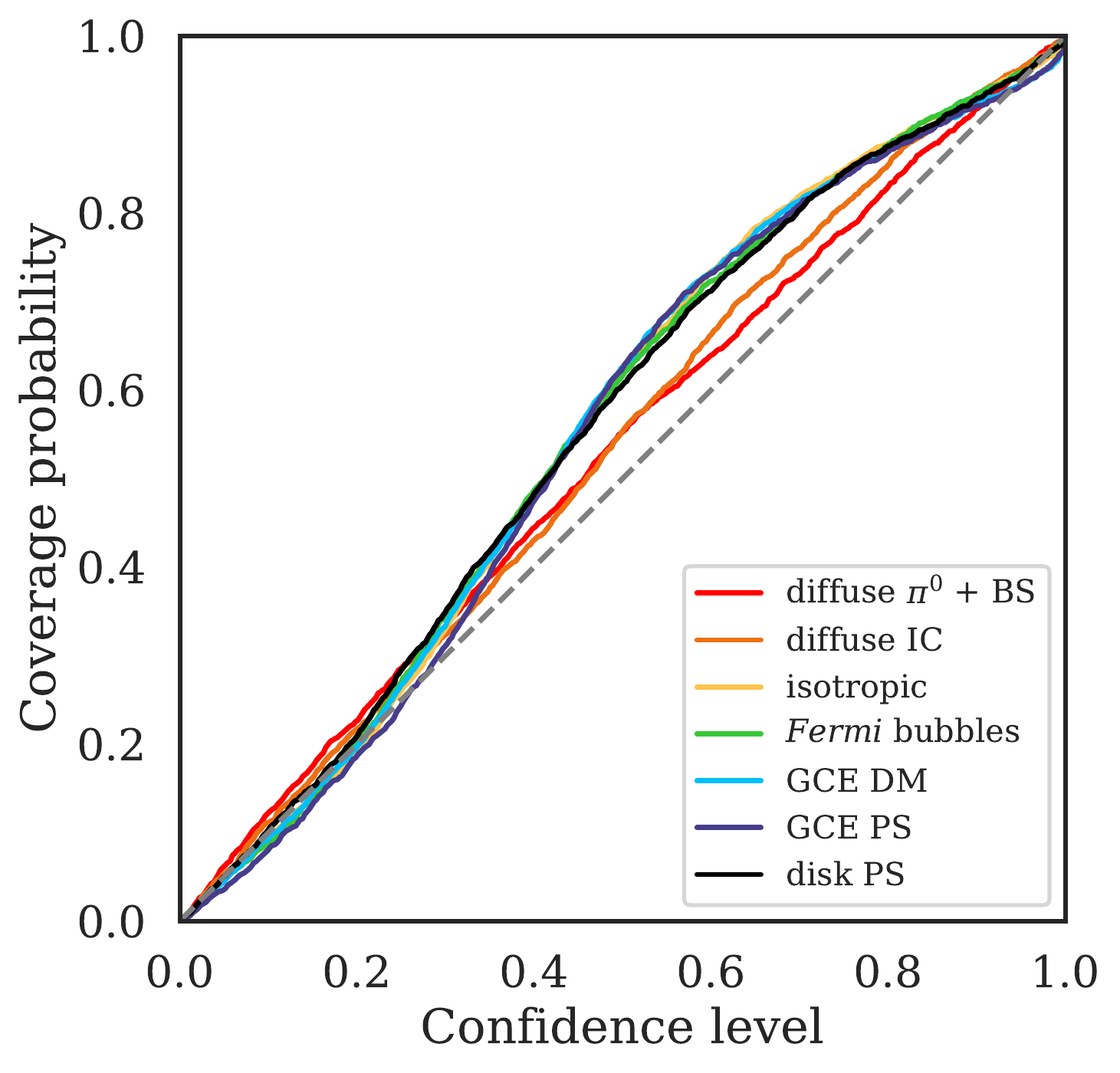} 
}
\caption{Calibration plot for the predictive uncertainties estimated by the Bayesian GCNN, computed from the entire testing data set consisting of 2500 maps. The y-coordinate shows the percentage of maps for which the NN prediction lies within an interval around the true value corresponding to the confidence level specified by the x-coordinate. The dashed gray line indicates perfectly calibrated uncertainties.}
\label{fig:calibration}
\end{figure}
In order to verify that the uncertainties determined by our Bayesian GCNN are sensible, we calculate the coverage probabilities as a function of the confidence level. Coverage probabilities have been used in cosmology for example in \cite{PerreaultLevasseur2017} for assessing the uncertainties of a Bayesian NN for the task of gravitational lensing parameter estimation. For a fixed confidence level $q \in (0, 1)$, the coverage probability is given by the fraction of samples for which the error of the NN falls within the range corresponding to the confidence level $q$; namely, the fraction of samples satisfying
\begin{equation}
    |f^{\bm{\omega}}(\mathbf{x})_t - \mathbf{y}_t| \leq F^{-1}_\mathcal{N}\left(\frac{1+q}{2}\right) \sigma_t^\text{pred},
    \label{eq:coverage_cond}
\end{equation}
where $F^{-1}_\mathcal{N}$ is the inverse CDF (quantile function) of the standard normal distribution, and $\sigma_t^\text{pred}$ is the total predictive uncertainty for template $t$. Note that when the covariance matrix is not assumed to be diagonal (such as in Sec. \ref{subsec:full_covar}), the multi-variate ellipsoid confidence set must be considered on the right hand side of Eq.~\eqref{eq:coverage_cond}. We plot the calibration curves for all the templates in Fig.~\ref{fig:calibration}. For perfectly calibrated uncertainties, the coverage probability equals the respective confidence level $q$ (gray dashed line). For most of the confidence levels, the coverage probabilities are slightly larger than the respective confidence level, implying slightly permissive uncertainties. The best-calibrated uncertainties are those for the diffuse templates. For the GCE DM and PS templates, the coverage probabilities fall below the confidence levels for $q \leq 0.94$, meaning that the number of samples in the testing set for which the true DM and PS flux fractions lie more than $2 \sigma$ away from the estimated values is a bit too large. Splitting up the uncertainties into their aleatoric and epistemic parts reveals that the bulk of the uncertainties is of aleatoric nature, and the epistemic uncertainties are very small. For the \emph{Fermi} map, for instance, the aleatoric uncertainty is more than 5 times larger than the epistemic uncertainty for all templates and ROI radii, and the ratio between aleatoric and epistemic uncertainty is $> 60$ for all the templates when averaged over all ROI radii, for which reason neglecting the epistemic uncertainties would leave the credible regions almost unaffected.
A further improvement in the calibration of the uncertainties can be achieved by applying re-calibration methods such as Platt Scaling \cite{Platt1999} or Temperature Scaling (see \cite{Hortua2019} for the re-calibration of a Bayesian NN for CMB parameter estimation based on Beta Calibration \cite{Kull2017}), which we leave for future work.

\subsection{Full uncertainty covariance matrix}
\label{subsec:full_covar}
\begin{figure*}
\centering
  \noindent
  \resizebox{1\textwidth}{!}{
\includegraphics{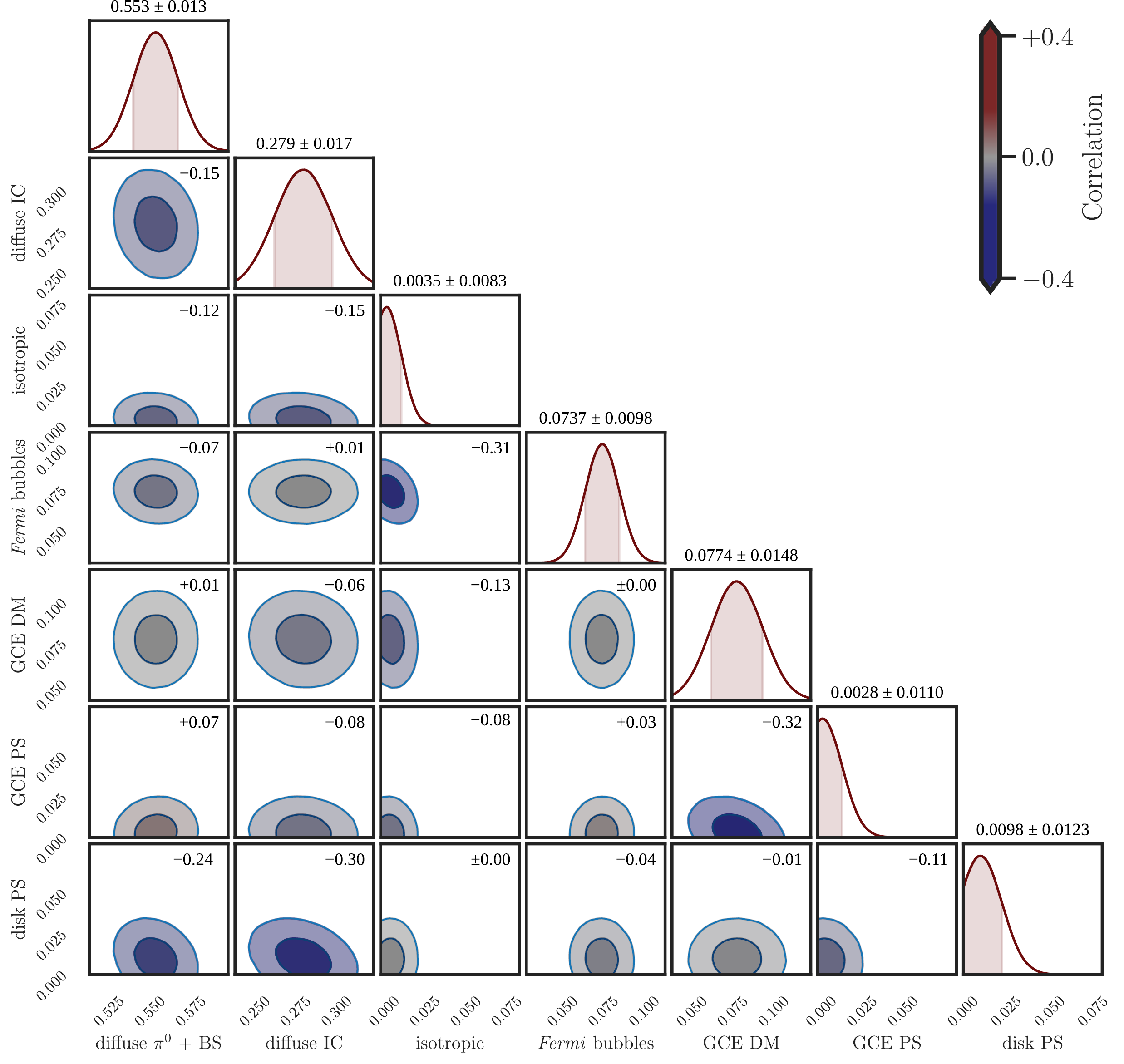}
}
\caption{Corner plot of the Bayesian GCNN estimates for the flux fractions in the \emph{Fermi} map within a radius of $25 ^\circ$ around the GC (with $|b| \leq 2 ^\circ$ and 3FGL sources masked). The shaded regions show the $1\sigma$ and $2\sigma$ credible regions, and their color indicates the Pearson correlation coefficient between the respective two flux fractions, which is also given in the upper right corner of each panel. The color bar corresponds to the more saturated colors of the $1\sigma$ regions. The width of each panel is exactly 0.08, so that the uncertainty contours can be directly compared.}
\label{fig:corner}
\end{figure*}
In Fig.~\ref{fig:fermi_results}, we presented the NN estimates for the flux fractions in the \emph{Fermi} count map together with the predictive uncertainties, treating the flux fractions as independent from each other. Now, we consider the extension to non-diagonal covariance matrices that reveal the correlations between the uncertainties in the flux fractions. For this purpose, we re-train the NN that was applied to the \emph{Fermi} map using the loss function in Eq.~\eqref{eq:loss_full_covar}. Although we make use of the procedure in \cite{Russell2019} to improve the numerical stability, we find that training the NN with this loss function is frequently unstable, for which reason we employ the following procedure: first, we train the NN with the simple $l^2$ loss in Eq.~\eqref{eq:l2loss} for 30,000 mini-batch iterations. After this step, the NN has learned how to estimate the flux fractions but is ignorant about the uncertainties. Then, we freeze all the NN weights except for the weights of the last fully connected layer, which maps to the output flux fractions and the elements of the uncertainty covariance matrix, and re-train the NN for another 30,000 mini-batch iterations using the full loss function defined in Eq.~\eqref{eq:loss_full_covar} in order for the NN to determine the components of the uncertainty covariance matrix as a function of its input. 
\par Fig.~\ref{fig:corner} shows a corner plot for the NN estimates for the \emph{Fermi} map within an outer ROI radius of $25 ^\circ$ and the predictive uncertainty covariances between the flux fractions, created with the aid of the \texttt{ChainConsumer} package \cite{ChainConsumer}. The color of the shaded credible regions corresponds to the Pearson correlation coefficient $\rho_{t_1, t_2} = \bm{\Sigma}^{\text{pred}}_{t_1, t_2} / (\sigma^\text{pred}_{t_1} \sigma^\text{pred}_{t_2}) \in [-1, 1]$ between the flux fractions for two templates $t_1$ and $t_2$, which is also quoted in the upper right corner of each panel. The predictive uncertainty covariance matrix is calculated using Eq.~\eqref{eq:pred_covar}. The flux fraction estimates are similar to the values in Fig.~\ref{fig:fermi_results} for which the NN was trained using the loss function in Eq.~\eqref{eq:aleatoric_llh}, although the flux fractions ascribed to GCE DM and the \emph{Fermi} bubbles are slightly smaller now. Since the total flux fractions sum up to 1, one generally expects negative correlations between the flux fractions and indeed, most of the correlation coefficients are negative or close to zero. The largest negative uncertainty correlation is observed between the two GCE templates ($-0.32)$. Other flux fraction pairs with negative correlations are \emph{Fermi} bubbles -- isotropic ($-0.31$), disk PS -- diffuse IC ($-0.30$), and disk PS -- diffuse $\pi^0 + \text{BS}$ ($-0.24$). The correlation between the disk PS flux and the diffuse templates, as well as between the two diffuse components ($-0.15$), is comprehensible in view of the strong emission around the Galactic Disk for all the three templates (see Fig.~\ref{fig:templates}). Both the \emph{Fermi} bubbles and the isotropic template are spatially uniform (albeit within different regions), explaining the negative correlation between the two. Interestingly, the magnitude of the correlation between the two PS templates (GCE PS and disk PS) is rather small ($-0.11$), suggesting that the distinct spatial templates prevent the NN from confusing the associated fluxes despite their common granular nature (however, cf. Sec.~\ref{sec:injection_test}). In summary: \textbf{the uncertainties between similar templates have the strongest negative correlation, providing further evidence that the NN has understood the underlying physics.} 

\orange{
\subsection{Dependence of the NN accuracy on the number of training samples}
\label{subsec:no_training_maps}
\begin{figure*}
\centering
  \noindent
  \resizebox{0.55\textwidth}{!}{
\includegraphics{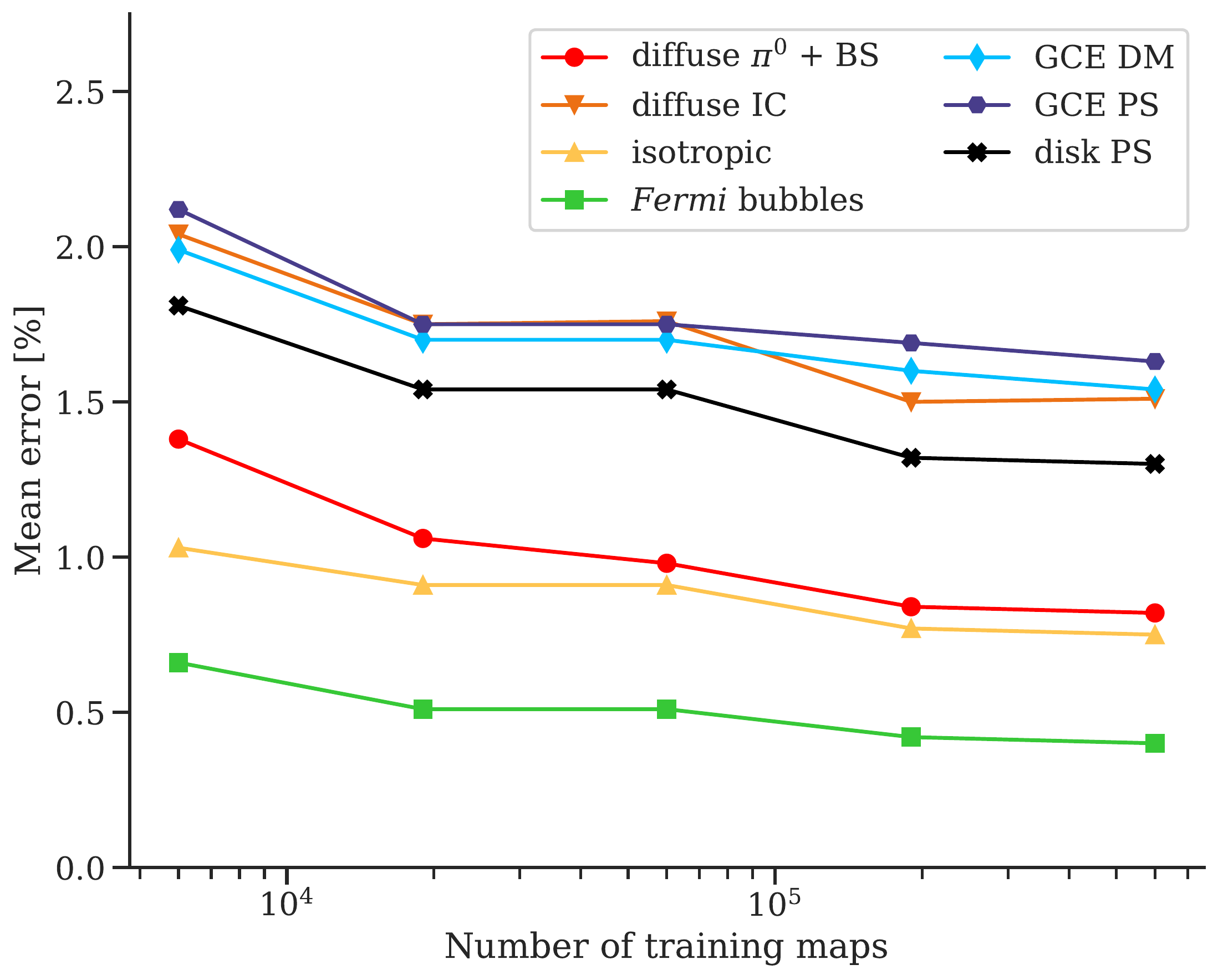}
}
\caption{\orange{Mean error on the validation data set as a function of the number of training samples.}}
\label{fig:no_training_samples}
\end{figure*}
In order to judge whether the number of training samples is sufficient, we examine how the NN accuracy depends on the size of the training data set. For this purpose, we re-train the NN for the realistic scenario with 6,000, 19,000, 60,000, and 190,000 training maps that are randomly drawn from our full training data set with 600,000 maps. The mean error for the flux fraction of each template on the validation data set, which always contains the same 2500 maps, is plotted in Fig.~\ref{fig:no_training_samples} for each training set size. In comparison to 6,000 maps, training on our entire training data set gives an average improvement of roughly $50\%$, however, the difference between the mean flux fraction errors for 190,000 and 600,000 maps is less than $0.1\%$ for all the templates, suggesting that the effect of further increasing the size of our training data set would be rather modest (see also Sec.~\ref{sec:narrow_priors}, where we generate the Poissonian realizations on-the-fly during the training, implying that each training map is unique).

\subsection{Extrapolation to millisecond-pulsar SCDs}
\label{subsec:bartels_SCD}
\begin{figure*}
\centering
  \noindent
  \resizebox{1\textwidth}{!}{
\includegraphics{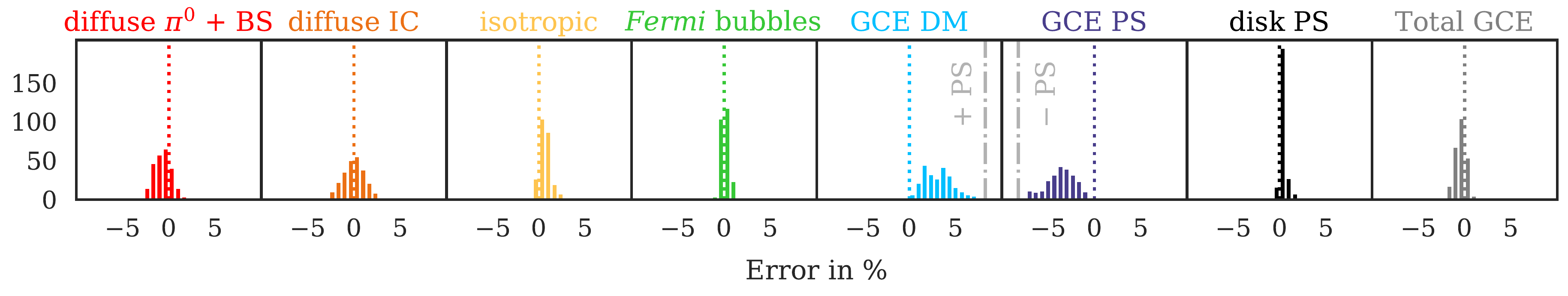}
}
\caption{\orange{Histogram of the NN errors for our largest ROI (outer radius $25^\circ$) for 250 simulated map corresponding to the best-fit parameters for the \emph{Fermi} map as determined by \texttt{NPTfit} within a ROI of outer radius $10^\circ$, but with the GCE PS fluxes drawn from the MSP SCD from \cite{Bartels2018} instead of a broken power law, cut off at fluxes corresponding to less than 1 photon counts (see Fig. \ref{fig:SCDs}).}}
\label{fig:error_hist_bartels}
\end{figure*}
The SCDs that we use for the NN training span the entire relevant flux range, with source count breaks randomly drawn from the interval $[0.05, 60]$ (see Table~\ref{table:priors} for the full specification of the priors ranges). Broken power laws are ubiquitous in the modeling of SCDs and have been shown to give good fits to PS populations in the \emph{Fermi} data \cite[e.g.][]{Zechlin2016}; also, they are widely used in the NPTF literature as they allow analytically integrating the arising integrals over the SCD \cite[e.g.][Appendix C]{Mishra-Sharma2017}. Since no analytic treatment of the SCD is required for the NN, however, we can in principle train and evaluate our NN on maps with PSs that follow arbitrary SCDs. With MSPs being the most likely explanation of a PS-like GCE, it is likely that their SCD would be similar to that seen for MSPs in the Galactic disk and is not exactly described by a singly-broken power law (but important differences between a bulge MSP population and the disk MSPs could arise from the differing star formation histories, resulting in the bulge MSPs to be older and therefore fainter, see \cite{Ploeg2020}). Hence, an interesting test is to evaluate our NN for the realistic scenario, which was trained with singly-broken power law SCDs, on maps with a contribution from GCE PS whose SCD has a different shape. We take a smooth MSP SCD that was obtained using the luminosity function for disk MSPs in \cite{Bartels2018} with the energy spectrum from \cite{Cholis2014}, and which was also used in \cite{Chang2019}. Since we have already extensively investigated the impact of the GCE PS \emph{brightness} in Sec.~\ref{sec:SCDs} and are now focusing on the SCD \emph{shape}, we cut off the SCD at the lower end where the flux corresponds to $1$ detected photon, and scale it such that it accounts for the entire GCE PS flux as determined by \texttt{NPTFit}. This SCD is shown by the green dash-dotted line in Fig.~\ref{fig:SCDs}. For the other templates, we use the same parameters as for the simulated mock data in Fig.~\ref{fig:fermi_results}, given by the \texttt{NPTFit} best-fit parameters within a radius of $10^\circ$. 
\par Fig.~\ref{fig:error_hist_bartels} shows the error histogram over 250 realizations with the MSP SCD for the GCE PS flux (for the same plot in the case of a singly-broken power law SCD see Fig.~\ref{fig:fermi_mock_histogram}). In view of $F^2 \, dN/dF$ peaking at a low flux of $F = 3 \times 10^{-11} \ \text{counts} \ \text{cm}^{-2} \ \text{s}^{-1}$ (corresponding to $\sim 2$ photon counts), it is not surprising that there is substantial misattribution from GCE PS to GCE DM. However, for this case where the GCE almost entirely consists of PSs, the probability of DM absorbing the total GCE is low, even though the NN has not seen such a SCD during the training. We emphasize that the confusion between the two PS templates is small, and the disk PS prediction is very close to zero for almost all the maps.
}

\section{Additional tests for the \emph{Fermi} photon-count map}
\label{sec:add_tests}
In this section, we present the results of three additional tests for the estimation of the flux fractions within the \emph{Fermi} map: first, we replace the thin disk PS template by a thick disk PS template that was previously used literature. Then, we model the diffuse emission with the \emph{Fermi} model \texttt{p6v11} and with the \texttt{GALPROP}-based Models A and F instead of Model~O. Finally, we train NNs to estimate the flux fractions separately for the northern and southern hemispheres. 

\subsection{Thick disk PS}
\label{subsec:fermi_thick_disk}
\begin{figure*}
\centering
  \noindent
  \resizebox{0.5\textwidth}{!}{
\includegraphics{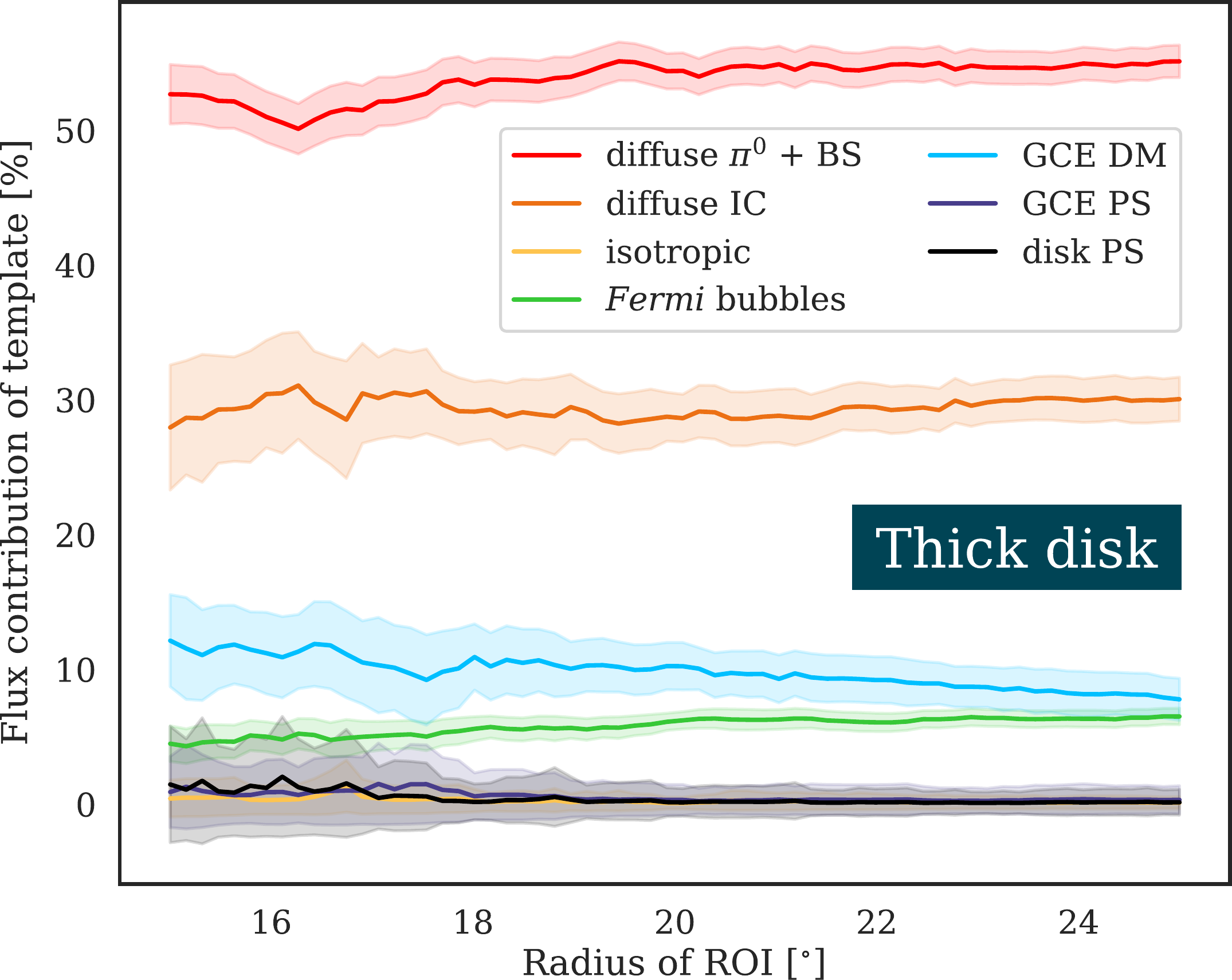}
}
\caption{Prediction of the Bayesian GCNN for the \emph{Fermi}-LAT photon-count map as a function of ROI radius when modeling the disk PS with a thick disk template. The shaded regions show the predictive (aleatoric and epistemic summed in quadrature) 1$\sigma$ uncertainty.}
\label{fig:fermi_results_thick}
\end{figure*}
We train the same NN that was used for the \emph{Fermi} map, but this time with a scale height of $z_s = 1 \ \text{kpc}$ for the disk PS (see Eq.~\eqref{eq:disk_PS}). This parameter choice was made by e.g. \cite{Buschmann2020, Leane2019a}, while \cite{Lee2016} tested both thick and thin disk templates and found the differences in their fits with \texttt{NPTFit} to be small. Figure \ref{fig:fermi_results_thick} shows the estimates of our NN for the \emph{Fermi} map as a function of ROI radius, evaluated for 64 ROI radii (cf. Fig.~\ref{fig:fermi_results} for the thin disk model). With the thick disk PS template, we find the estimated disk PS flux to be consistent with zero for all ROI radii. The diffuse IC template absorbs slightly more flux than when using the thin disk template ($30.0 \%$ vs. $26.9 \%$), and the isotropic flux fraction for small ROIs that the NN predicts with the thin disk template has disappeared. The flux estimates for the \emph{Fermi} bubbles are very consistent and the estimated uncertainties are again small. With regard to the GCE, the NN results confirm the picture presented in the main part of this \emph{Letter}: whereas the GCE PS flux is consistent with zero for all ROI radii, a GCE DM contribution is found that decreases as a function of the ROI radius, as expected. The estimated DM flux fraction is consistent with that in Fig.~\ref{fig:fermi_results}, and the uncertainty becomes small for large ROI radii. For our largest ROI with an outer radius of $25 ^\circ$, we run \texttt{NPTFit} using the same templates and compare the NN predictions with the \texttt{NPTFit} median posterior flux fractions (NN / \texttt{NPTFit}, column-wise in the order of the legend entries in Fig.~\ref{fig:fermi_results_thick} \orange{in \%): $55.1$ / $54.9$, $30.0$ / $29.5$, $0.1$ / $0.3$, $6.5$ / $6.4$, $7.8$ / $0.1$, $0.3$ / $8.2$, $0.1$ / $0.6$.} \textbf{Again, the NN and \texttt{NPTFit} results are strikingly similar except for the attribution of the GCE flux to DM by the NN}.

\subsection{Varying the diffuse model}
\label{subsec:fermi_p6v11}
\begin{figure*}
\centering
  \noindent
  \resizebox{1\textwidth}{!}{
\includegraphics{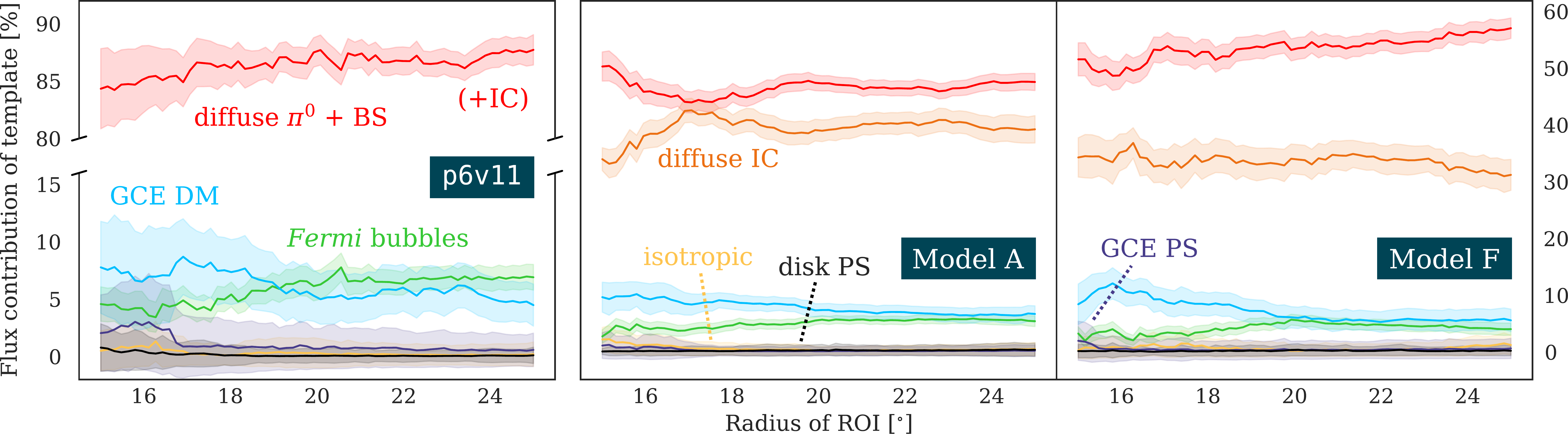}
}
\caption{Prediction of the Bayesian GCNN for the \emph{Fermi}-LAT photon-count map when modeling the Galactic foreground emission with \texttt{p6v11}, Model~A, and Model~F instead of Model~O. The shaded regions show the predictive (aleatoric and epistemic summed in quadrature) 1$\sigma$ uncertainty. Note that since \texttt{p6v11} accounts for both $\pi^0 + \text{BS}$ and IC scattering with a single template, its diffuse flux contribution is very large as compared to the other templates, for which reason the $y$-axis is broken so as to improve the visibility of the other flux contributions. The $y$-axes for Models A and F are identical and labeled on the right hand side of the plot.}
\label{fig:fermi_results_p6v11}
\end{figure*}
It has been pointed out that the model for the diffuse Galactic foreground emission constitutes the largest source of uncertainty when ascertaining the origin of the GCE. Although Model~O has been shown to be superior to the \emph{Fermi} model \texttt{p6v11} \cite{Buschmann2020}, we re-train the NN on maps with \texttt{p6v11} diffuse flux and evaluate it on the \emph{Fermi} map as a consistency check. Additionally, we consider NNs trained on maps with diffuse flux contributions given by Models A and F. In our largest ROI (outer radius $25 ^\circ$, $|b| > 2 ^\circ$), these two diffuse models have been shown to be favored over \texttt{p6v11} in a (purely Poissonian) analysis of the \emph{Fermi} map (see e.g. \cite[Fig. 3]{Buschmann2020}), while Model~O provides the best fit among these four templates for all energies (however, cf. Sec.~\ref{sec:injection_test}).
We take again the thin disk template for the disk-correlated PS as in Fig.~\ref{fig:fermi_results}. The resulting flux fraction estimates are shown in Fig.~\ref{fig:fermi_results_p6v11}. Recall that diffuse emission from pion decay, bremsstrahlung, and IC scattering is described by a single template in \texttt{p6v11}. Therefore, the estimated diffuse flux makes up the vast majority of the total flux, as expected. The NN estimate for the \emph{Fermi} bubbles is similar to the Model~O case, and isotropic emission and flux from disk PS are found to be consistent with zero. Again, the NN identifies a GCE, whose magnitude is somewhat smaller than for Model~O. Also for the \texttt{p6v11} diffuse model, a DM explanation for the GCE is preferred by the NN (although the estimated GCE PS flux fraction rises slightly when using small ROIs). The estimated uncertainties are generally larger when using \texttt{p6v11} instead of Model~O. The similarity between the results from the NN and \texttt{NPTFit} with the exception of the GCE attribution is again remarkable: the estimated flux fractions for the NN / \texttt{NPTFit} for an outer ROI radius of $25 ^\circ$ are \orange{(in $\%$): $87.7$ / $88.5$ (diffuse), $0.2$ / $0.05$ (isotropic), $6.9$ / $6.3$ (\emph{Fermi} bubbles), $4.7$ / $0.04$ (GCE DM), $0.5$ / $4.2$ (GCE PS), $0.07$ / $0.9$ (disk PS)}. 
\par For Models A and F, the estimates for the total diffuse flux fraction are consistent with each other and with \texttt{p6v11}, whereas the preferred diffuse flux with Model~O is lower. However, the attribution of the diffuse flux to the two components differs between Models A and F: the diffuse IC template absorbs roughly $8 \%$ more flux for Model~A than for Model~F within a $25 ^\circ$ ROI, while the flux fraction from pion decay and bremsstrahlung is $\sim~9 \%$ lower. Also for these two diffuse models, the NN identifies contributions from the \emph{Fermi} bubbles and from GCE DM that increase and decrease, respectively, as the ROI is enlarged. The NN estimates within our largest ROI are given by \orange{(Model~A / Model~F, in $\%$): $47.7$ / $57.1$ (diffuse $\pi^0$ + BS), $39.2$ / $31.2$ (diffuse IC), $0.5$ / $1.2$ (isotropic), $5.4$ / $4.0$ (\emph{Fermi} bubbles), $6.7$ / $5.6$ (GCE DM), $0.2$ / $0.6$ (GCE PS), and $0.4$ / $1.8$ (disk PS).} In summary: \textbf{A GCE described by the DM template is identified for all the Galactic foreground models considered in this work.} We confirm the findings of \cite{Buschmann2020} as to the highest GCE flux being found with Model~O and the GCE flux magnitude varying by a factor of up to $\sim~2$ among the diffuse models. The \texttt{NPTFit} best-fit flux fractions for Models A and F can be found in Table \ref{table:results_summary}.
In future work, we will extend our analysis to further diffuse models. 

\subsection{Separately fitting the northern and southern hemisphere}
\label{subsec:hemispheres}
\begin{figure}[htb]
\centering
  \noindent
  \resizebox{1\textwidth}{!}{
\includegraphics{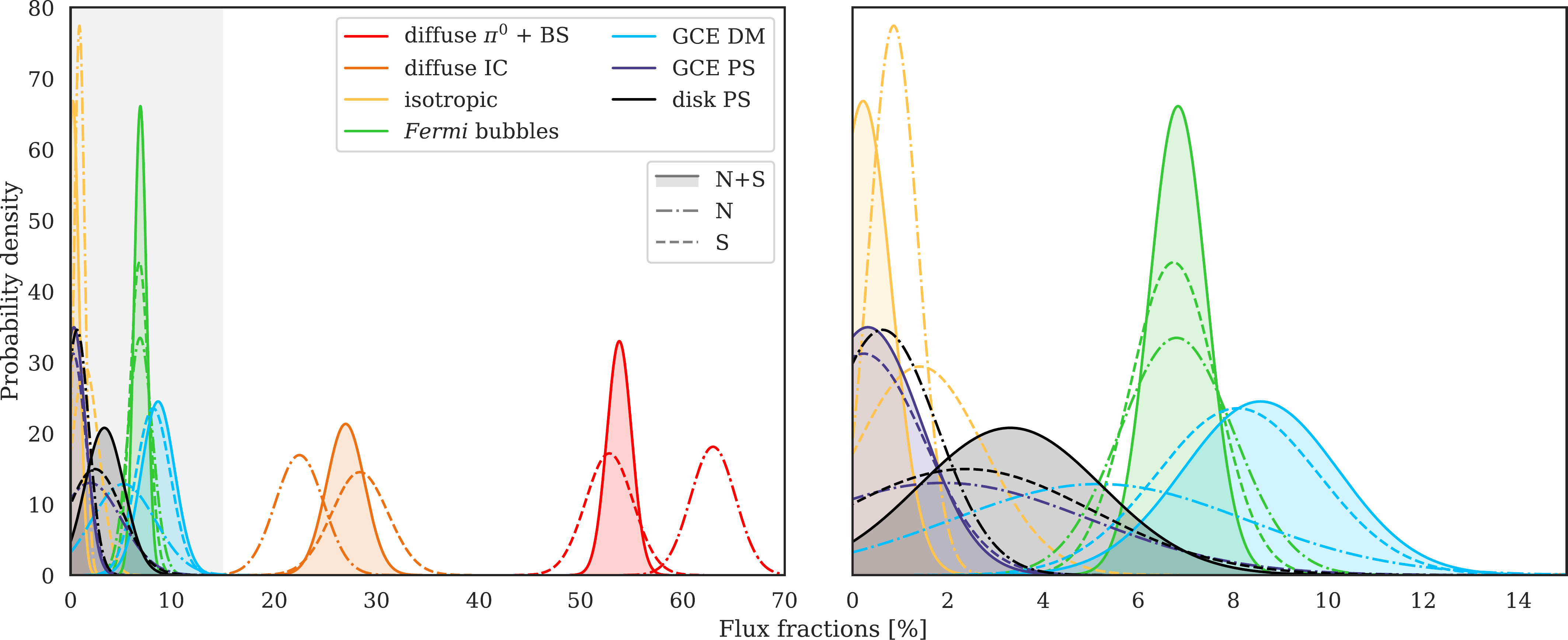} 
}
\caption{Flux fraction distributions as estimated by the Bayesian GCNNs for both hemispheres (N+S) and for the northern (N) and southern (S) hemisphere separately. The right panel shows a zoom into the gray shaded region in the left panel (flux fractions $\leq 0.15$). Since the amount of information decreases when considering each hemisphere individually, most of the distributions become broader than for the joint N+S analysis.}
\label{fig:N_S_flux_fractions}
\end{figure}
As an additional consistency check, we train our Bayesian GCNN individually on the northern and southern hemispheres. In order for the NN to have a substantial amount of pixels at its disposal despite the masking of either hemisphere, we fix the outer ROI radius to be $25 ^\circ$ around the GC for the training data and only estimate the \emph{Fermi} flux fractions within this region. The estimated flux fractions distributions obtained from the data are shown in Fig.~\ref{fig:N_S_flux_fractions}, for the fits in each hemisphere as well as in both hemispheres simultaneously (also for an outer ROI radius of $25 ^\circ$) using the NN from the main part of this \emph{Letter}. The left panel covers the range of flux fractions up to \orange{$70 \%$} such that the flux fractions of all the templates are visible, whereas the right panel is a zoom into the gray shaded region of the left panel (flux fractions \orange{$\leq 15 \%$}). Unsurprisingly, the distributions generally broaden when the fits are performed in a single hemisphere. While the results are similar, the two most important differences are: first, in the northern hemisphere, the NN prefers a larger contribution from neutral pions and bremsstrahlung and slightly less flux from IC scattering. Secondly, the GCE in the northern hemisphere is found to be a bit less pronounced as compared to the other two cases, and the GCE PS distribution is wider. In contrast, \texttt{NPTFit} attributes the majority of the GCE flux to GCE DM within the northern hemisphere when allowing for separate GCE template normalizations in the two hemispheres \cite[Figs. 1~\&~2]{Leane2020}, although these north-south discrepancies may be the consequence of modeling errors. In any case, we showed in Sec. \ref{subsec:GCE_mismodelling} that the NN predictions are comparably insensitive to modest north-south GCE asymmetries (see Fig.~\ref{fig:gce_mismodelling_max_2}), and \textbf{Fig.~\ref{fig:N_S_flux_fractions} confirms that a GCE consistent with smooth emission is detected by the NN in both hemispheres}, in line with our findings in the main part of this \emph{Letter} for the full map.

\orange{
\section{Using narrow priors around the \emph{Fermi} map parameters}
\label{sec:narrow_priors}
\begin{figure}[htb]
\centering
  \noindent
  \resizebox{1\textwidth}{!}{
\includegraphics{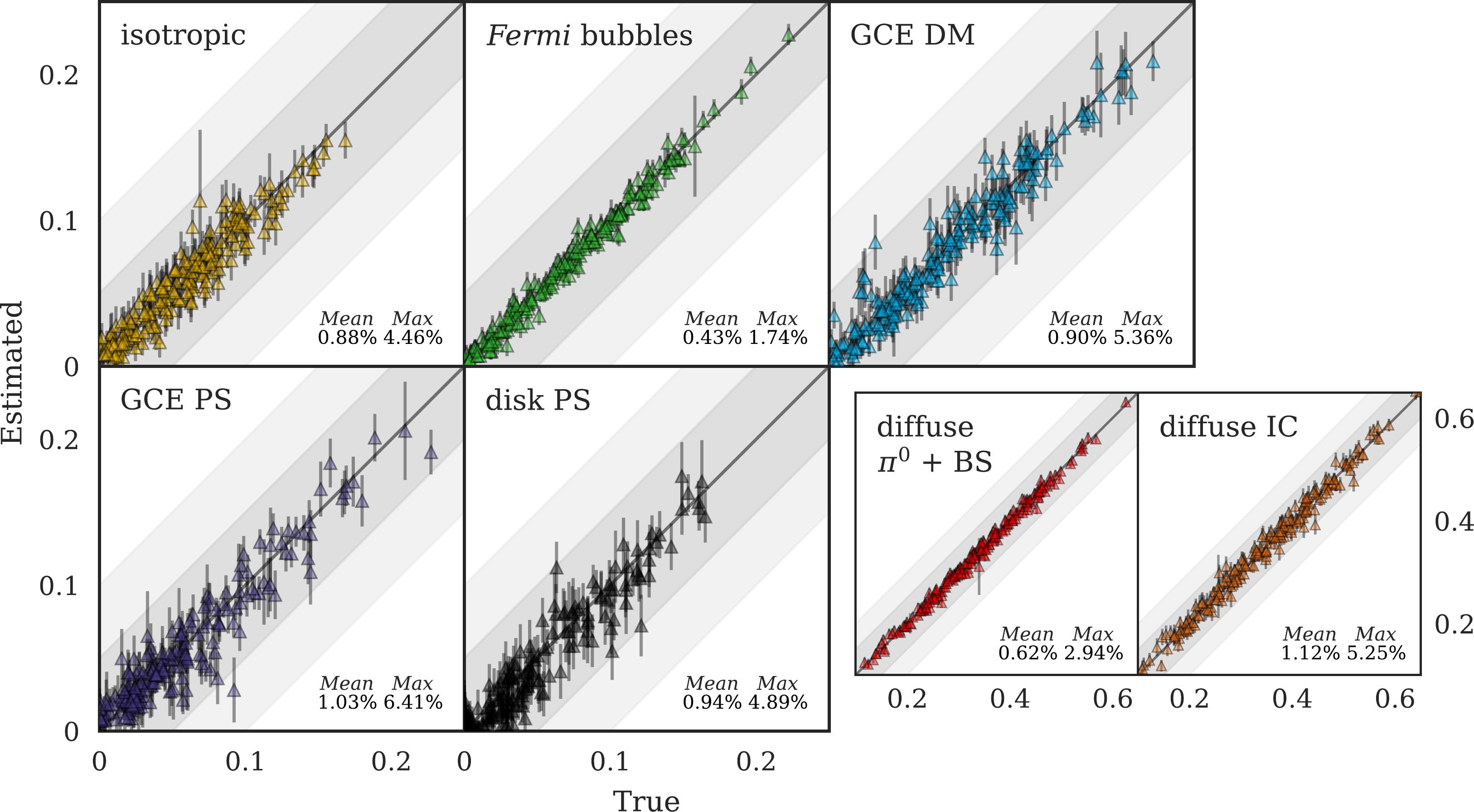} 
}
\caption{\orange{True vs. estimated flux fractions for 200 photon-count test maps when using a narrow prior range around the expected \emph{Fermi} map parameters and a fixed ROI of outer radius $25^\circ$. We zoom into the relevant flux fraction range for each template. The error bars correspond to the $1\sigma$ predictive (aleatoric and epistemic summed in quadrature) errors estimated by the Bayesian GCNN.}}
\label{fig:narrow_priors_results_model_O}
\end{figure}
\begin{figure}[htb]
\centering
  \noindent
  \resizebox{1\textwidth}{!}{
\includegraphics{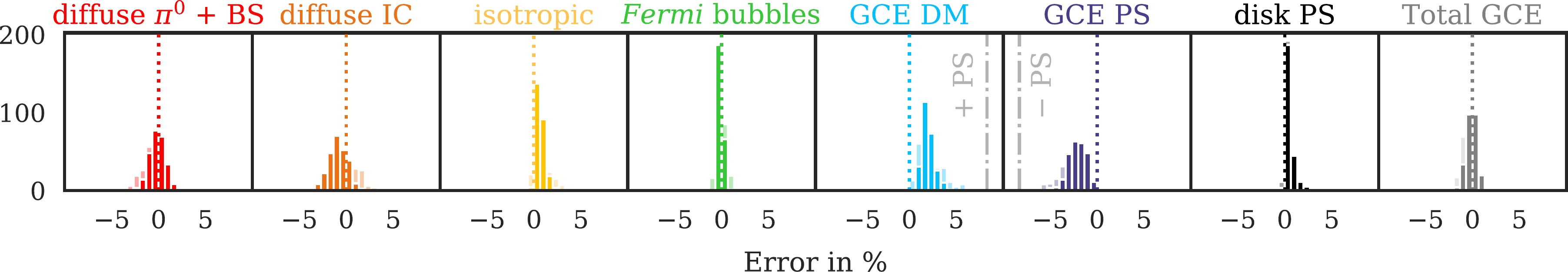} 
}
\caption{\orange{Histogram of the NN errors in our largest ROI for the 250 simulated maps corresponding to the best-fit parameters for the \emph{Fermi} map as determined by \texttt{NPTFit} within a ROI of outer radius $10 ^\circ$, but now for the NN that was trained using a narrow prior range and a fixed ROI of $25^\circ$ in order to maximize the NN performance. The faint bars in the background show the predictions of our default NN with a wider prior range for comparison (see Fig.~\ref{fig:fermi_mock_histogram}). With the narrower priors and the fixed ROI, the misattribution from GCE PS to GCE DM becomes somewhat smaller, and the NN underestimates the GCE PS flux by less than $4\%$ for $96.8\%$ of the maps, as compared to only $85\%$ of the maps with our default NN for the realistic scenario.}}
\label{fig:narrow_priors_hist_best_fit_data}
\end{figure}
\begin{figure}[htb]
\centering
  \noindent
  \resizebox{1\textwidth}{!}{
\includegraphics{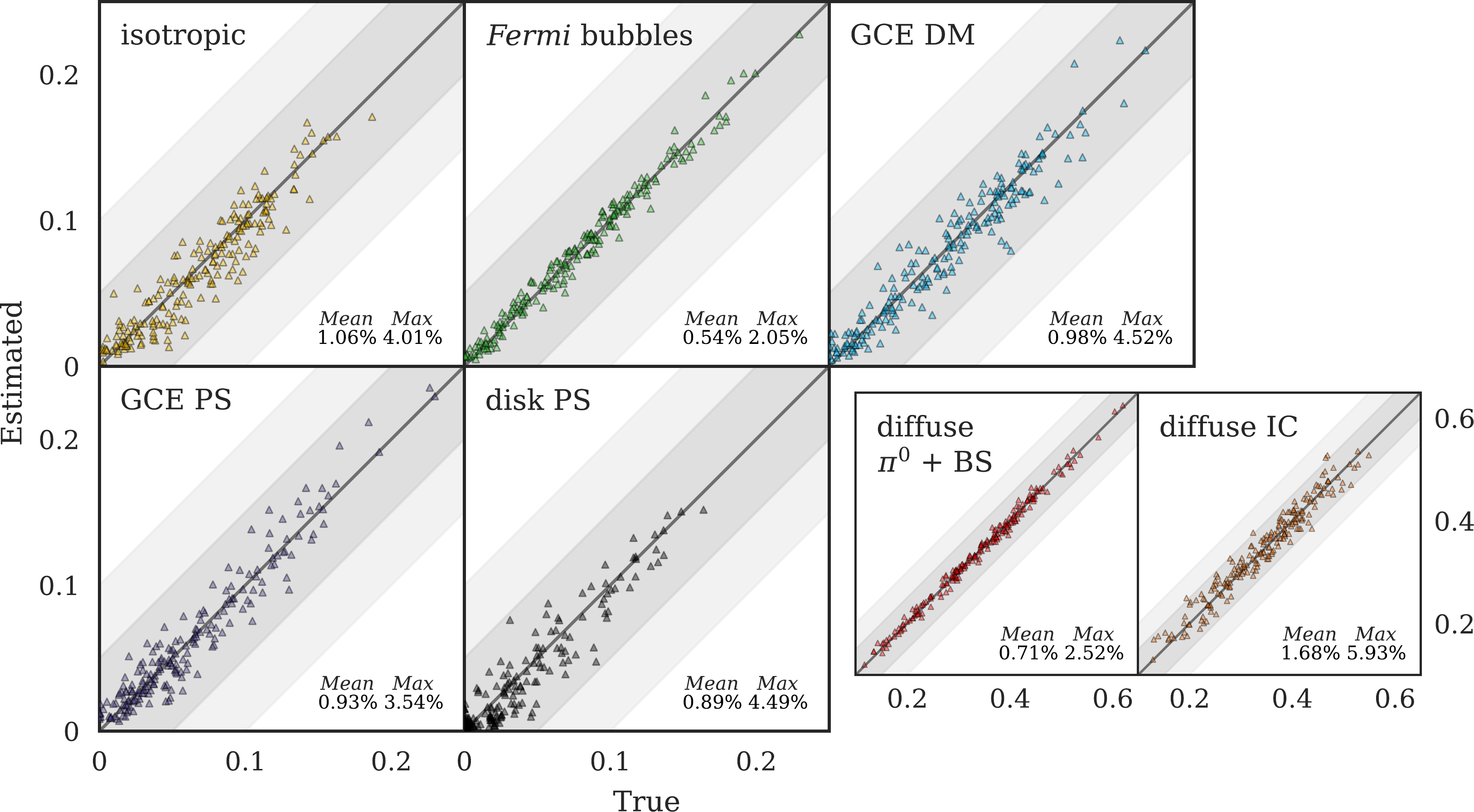} 
}
\caption{\orange{True vs. estimated flux fractions for 200 photon-count test maps when using a narrow prior range around the expected \emph{Fermi} map parameters, a fixed ROI of outer radius $25^\circ$, and random linear combinations of the diffuse Models A, F \& O (for $\pi^0 + \text{BS}$ and IC).}}
\label{fig:narrow_priors_results_model_variants}
\end{figure}
\begin{figure}[htb]
\centering
  \noindent
  \resizebox{1\textwidth}{!}{
\includegraphics{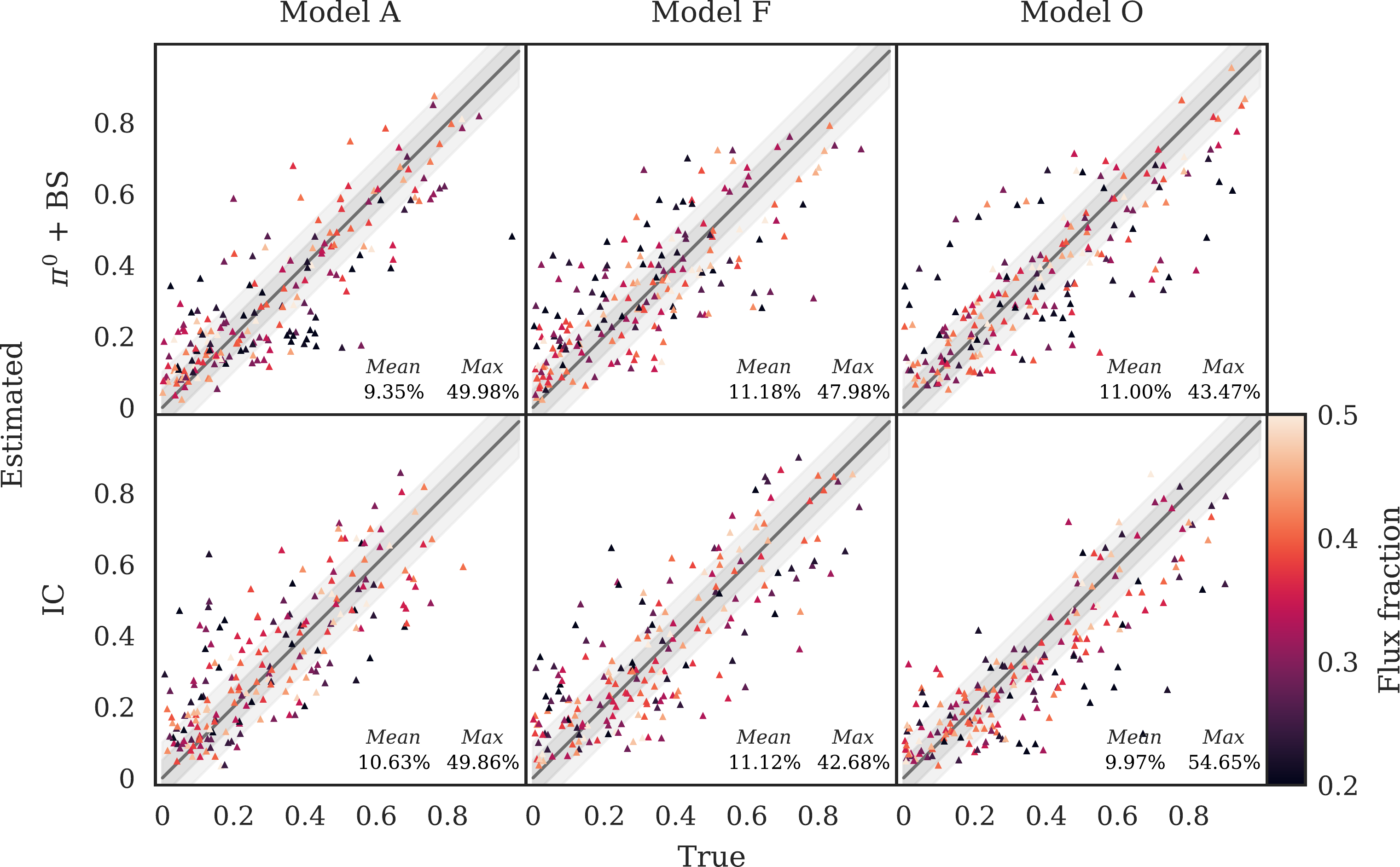} 
}
\caption{\orange{True vs. estimated coefficients for the flux contributions of the three different templates (Models A, F \& O) within the diffuse fluxes from pions \& bremsstrahlung (top) and from IC scattering (bottom), for the same 200 photon-count maps as in Fig.~\ref{fig:narrow_priors_results_model_variants}. The marker color corresponds to the flux fraction of the respective template ($\pi^0 + \text{BS}$ or IC) in the map. Evidently, disentangling the three underlying constituents of the diffuse fluxes is much more challenging than estimating the flux fractions from the different templates.}}
\label{fig:narrow_priors_variant_estimates}
\end{figure}
For training our NN, we have chosen wide prior ranges to enable a thorough analysis of what the NN has learned across diverse maps, thus allowing us to examine the behavior in limit cases such as high fluxes from mismodeled \emph{Fermi} bubbles or a smooth GCE with an unmodeled asymmetry. Two natural questions to ask are whether a tighter prior range around the expected values for the \emph{Fermi} map improves the NN performance, and if the resulting predictions for the \emph{Fermi} map remain consistent with those from our wide default priors.
\par Now, we take linear priors for the template normalizations $A$ of the Poissonian templates such that the flux for each of the diffuse templates lies between $15 - 70\%$, and for the other Poissonian templates between $0 - 20\%$, of the total flux in the \emph{Fermi} map within our ROI with outer radius $25^\circ$. In order to maximize the NN performance, we create the Poissonian template maps directly during the training, implying that \emph{every training map is unique}, and almost 2~million different maps are shown to the NN during the training, which comprises 30,000 batch iterations with batch size $64$. For each of the non-Poissonian templates GCE PS and disk PS, we generate a new catalog of 100,000 template maps, whereby we require that the expected flux in each map be below $20\%$ of the total \emph{Fermi} map flux within our ROI and abort the generation of the respective map otherwise. We randomly put aside 6,667 maps for each non-Poissonian template for testing and use the remaining ones as training data, randomly combining them with the Poissonian template maps at training time. 
We consider two different scenarios: 1) a Bayesian NN with the same templates as in the realistic scenario in the main body (in particular with Model O for the diffuse flux and with the disk PS flux described by the thin disk template), and 2) a non-Bayesian NN with diffuse flux from random mixtures of Models~A, F, and O. For both cases, we fix the outer radius of our ROI to be $25^\circ$, mask the 3FGL sources, and use the \emph{Fermi} exposure map.

\subsection{Diffuse Model O}
\par Fig.~\ref{fig:narrow_priors_results_model_O} shows the NN results on 200 simulated test maps. Note that there is no testing data set as such, but the Poissonian template maps are randomly generated at evaluation time and then combined with the 6,667 template maps for GCE PS and disk PS. Since the flux fractions now span a much tighter range due to the narrower priors, we zoom into the relevant flux range for each template in the figure. For these roughly \emph{Fermi}-like maps, the mean errors are approximately $1\%$ for all the template maps ($<0.5\%$ for the \emph{Fermi} bubbles). The significantly smaller maximum errors as compared to Fig.~\ref{fig:fermi_true_vs_estimated} can be explained by noting that 1) the NN has seen far more training maps within this parameter region, improving the accuracy for (approximately) \emph{Fermi}-like maps, 2) there are no ``unphysical'' testing maps with e.g. a $50\%/50\%$ DM/PS GCE that accounts for most of the total flux in the map, where the NN can easily misattribute $\mathcal{O}(10\%)$ of the GCE flux from one component to the other, and 3) the ROI is fixed to 25$^\circ$ instead of varying between $15 - 25^\circ$, which makes the task of the NN easier.
\par Evaluating this NN on the 250 simulated maps corresponding to the \texttt{NPTFit} best-fit parameters when fitting the \emph{Fermi} map within a $10^\circ$ ROI produces the error histogram shown in Fig.~\ref{fig:narrow_priors_hist_best_fit_data}. The histogram with our default NN for the realistic scenario that was trained with a wider prior range and a variable ROI is indicated by the faint bars in the background for comparison (identical to that in Fig.~\ref{fig:fermi_mock_histogram}; see also Fig.~\ref{fig:fermi_results} for the results of our default NN on these maps as a function of the ROI radius). Whereas the error distributions for the non-GCE templates are quite similar for the two NNs, the tails of the error distributions for the GCE templates drop off more quickly now, and the amount of misattribution from GCE PS to DM is somewhat smaller. For example, the NN underestimates the GCE PS flux fraction by less than $4\%$ for $96.8\%$ of the maps (recall that the true GCE PS magnitude in the maps is $8.3\%$), compared with $85\%$ of the maps for our default NN, and the maximum predicted GCE DM flux is only $4.6\%$ compared with $7.5\%$ for the default NN. This shows that although the degeneracy between the two GCE templates is ultimately physical, there is still potential to further improve the accuracy of our NN, and narrowing down the task of the NN (e.g. by fixing the ROI and considering a smaller cube in parameter space) can potentially enhance the NN accuracy.

\par Finally, we report the results of the NN with narrow priors for the $\emph{Fermi}$ map (in $\%$): $54.4$ (diffuse $\pi^0 + \text{BS}$), $26.3$ (diffuse IC), $1.9$ (isotropic), $6.1$ (\emph{Fermi} bubbles), $6.4$ (GCE DM), $1.1$ (GCE PS), $3.8$ (disk PS). The total GCE flux magnitude is slightly lower than with our default NN ($7.5\%$ vs. $8.9\%$), and the NN finds a $\sim 1\%$ contribution from GCE PS, but the preference for a predominantly smooth GCE persists ($6.4\%$ GCE DM). Further, the flux fractions of all templates are consistent with the procedure pursued in the main body, within the quoted errors, with the marginal exception of the isotropic template.

\subsection{Mixtures of diffuse Models A, F \& O}
An interesting avenue for alleviating the sensitivity of the NN results with respect to the chosen diffuse model is to train the NN simultaneously on multiple diffuse templates. We take the NN with the training procedure described above, but now we draw $2 \times 3$ coefficients $\alpha_{c, v}$ from a flat Dirichlet distribution (that is, a uniform distribution over the standard $2$-simplex) for each map, which determine the contributions of diffuse Model $v \in \{\text{A, F, O}\}$, for $c \in \{\pi^0 + \text{BS}, \text{IC}\}$. Note that we draw three different coefficients for each of the two diffuse templates, implying that their compositions are independent. The composite templates are then given as $T_\text{dif}^{c, \text{mix}} = \sum_{v \in \{\text{A, F, O}\}} \alpha_{c, v} T_\text{dif}^{c, v}$. Thus, the two diffuse templates are random mixtures of the three diffuse Models A, F, and O for each training map. The template normalizations are drawn as usual. We append the 6 coefficients $\{\alpha_{c, v}\}$ as additional outputs that the NN is trained to estimate using an $l^2$ loss function, and we use a Softmax activation function to enforce that they sum up to unity for $\pi^0 + \text{BS}$ and IC. For simplicity, we do not estimate uncertainties for the flux fractions in this case. 
\par Fig.~\ref{fig:narrow_priors_results_model_variants} shows the true vs. estimated flux fractions for this NN, for 200 test maps. As expected, the errors for the two diffuse templates are a bit larger than in Fig.~\ref{fig:narrow_priors_results_model_O} for diffuse Model O only, because the NN now needs to associate different spatial distributions of the photon counts with the diffuse templates, but the increase is modest. Interestingly, the mean errors for the PS templates are slightly smaller now, suggesting that diverse (Poissonian) diffuse flux contributions do not have a detrimental effect on the estimation of the non-Poissonian flux fractions. Clearly, estimating the coefficients of the three constituents of the diffuse templates is a more difficult task than simply learning how to estimate the flux fractions in the presence of multiple diffuse models, in particular because the three diffuse templates are very similar. This is reflected in the NN estimates of the coefficients, plotted in Fig.~\ref{fig:narrow_priors_variant_estimates}, which have a mean error of $\sim 10\%$ for diffuse flux from pion decay \& bremsstrahlung and from IC. Still, it is worthwhile to apply this NN to the \emph{Fermi} map in order to simultaneously obtain estimates for the flux fractions and the composition of the diffuse flux in the \emph{Fermi} map. The flux fraction estimates are (in $\%$): $53.7$ (diffuse $\pi^0 + \text{BS}$), $29.9$ (diffuse IC), $1.9$ (isotropic), $6.4$ (\emph{Fermi} bubbles), $5.6$ (GCE DM), $1.7$ (GCE PS), $0.9$ (disk PS), and the estimated diffuse flux composition given by the 6 coefficients $\{\alpha_{c, v}\}$ is
\begin{itemize}
    \item{\makebox[3cm][l]{$\pi^0 + \text{BS}$:} $7\%$ (A) + $22\%$ (F) + $72\%$ (O),}
    \item{\makebox[3cm][l]{IC:} $33\%$ (A) + $2\%$ (F) + $65\%$ (O).}
\end{itemize}
With these additional degrees of freedom for the diffuse model, the NN assigns a flux fraction of $1.7\%$ to GCE PS, but GCE DM is still the dominating GCE template with $5.6\%$ of the flux. As to the diffuse flux, \textbf{the NN finds the recent Model O to provide the best fit to the \emph{Fermi} map}, with moderate contributions of Model F for $\pi^0 + \text{BS}$ flux and of Model A for IC flux. We will explore further options for introducing degrees of freedom to the modeling of the inner Galaxy, e.g. by utilizing spherical harmonics \cite{Buschmann2020} or Gaussian processes \cite{Mishra-Sharma2020}, in future work.
}

\section{Recovering artificially injected flux from the \emph{Fermi} map}
\label{sec:injection_test}
\begin{figure}[htb]
\centering
  \noindent
  \resizebox{1\textwidth}{!}{
\includegraphics{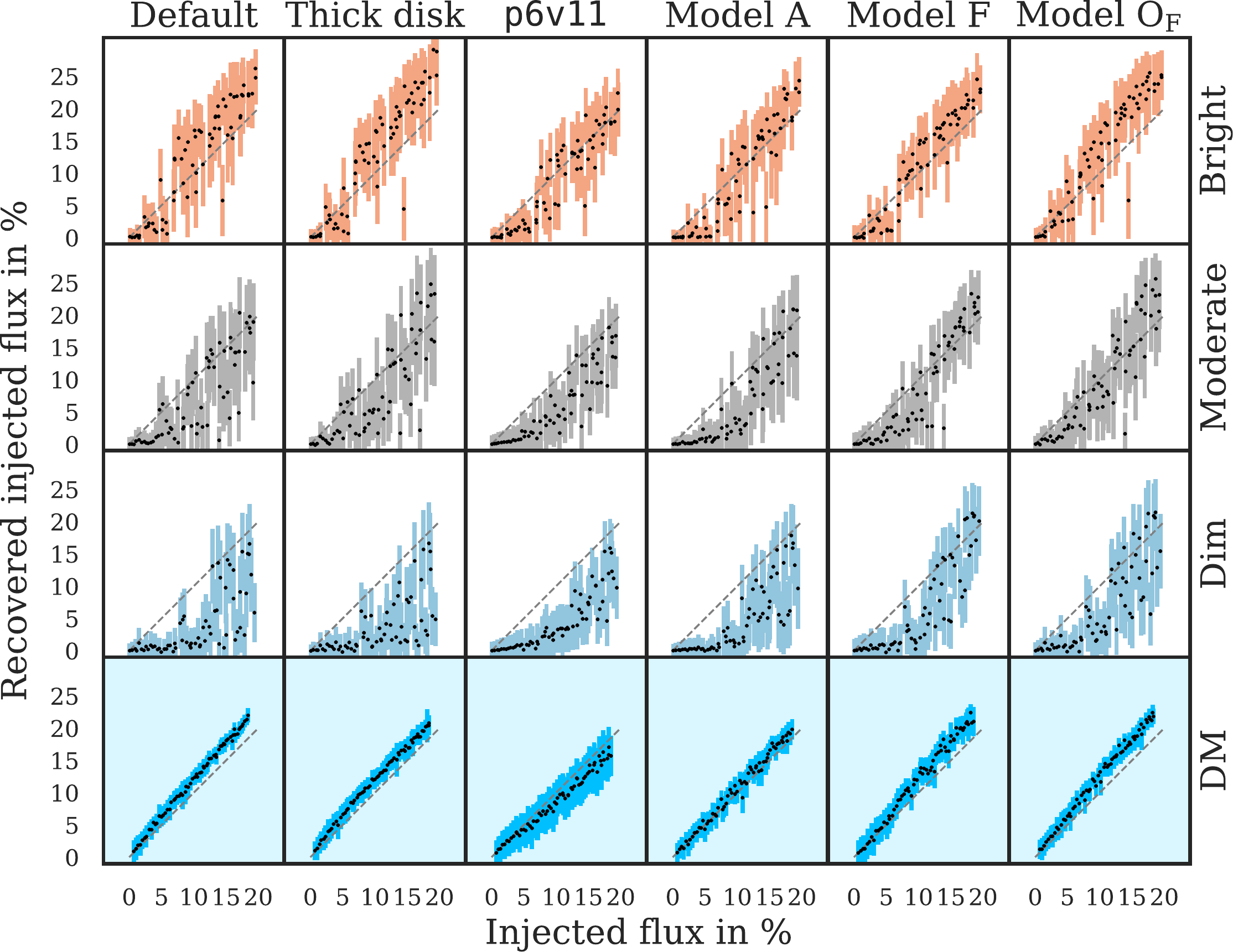} 
}
\caption{Injection of artificial flux from GCE PS (upper three rows) and GCE DM (bottom row) into the \emph{Fermi} map. The $y$-axis shows how much of the artificially injected fluxes as given on the $x$-axis (post-injection percentage) is recovered by the NN. In the ideal case, the markers would trace the dashed gray line ($y = x$).}
\label{fig:injection_results}
\end{figure}
An effective way of scrutinizing whether the estimates for the GCE may be an artifact caused by some form of mismodeling is to consider the recovery of artificially injected flux into the \emph{Fermi} photon-count map. In \cite{Leane2019a}, it was demonstrated that a GCE DM flux of $6.7 \%$ injected into the \emph{Fermi} map is not recognized by \texttt{NPTFit} within a $30 ^\circ$ ROI when using the \texttt{p6v11} diffuse model, and is absorbed by the GCE PS template instead, while for an injected flux fraction of $15.2 \%$, around half of the DM flux is recovered. In \cite[Fig. 12]{Buschmann2020}, it was shown that the recovered flux fractions are much more consistent with expectation when resorting to Model~O instead of \texttt{p6v11}, and a spherical-harmonic marginalization procedure leads to major improvements (also for $\texttt{p6v11}$). For the NPTF, the motivation of this injection experiment is to examine whether a real DM flux contribution might be missed due to over-subtraction from other templates (mainly the Galactic foregrounds) in view of its general preference for a GCE with PS characteristics (however, cf. \cite{Leane2020}). In contrast, since our NN favors a smooth GCE in all our experiments, the converse question as to whether the NN is able to recover synthetic GCE PS flux from the \emph{Fermi} map is more crucial in view of the worry that the NN might base its judgment on a non-GCE flux component that biases the NN estimates towards a smooth GCE. 
\par For this experiment, \orange{we use again our default NN for the realistic scenario, and} we consider the three central SCDs from Section \ref{sec:SCDs} (bright, moderate, and dim). The very bright SCD is irrelevant here since many of the GCE PSs would be expected to be resolved in that case. Reliably distinguishing GCE PSs modeled by the very dim SCD (with source count break at $S_1^\text{gce} = 1.31$) from GCE DM is currently an unfeasible undertaking with our NN, even if these PSs compose the entire GCE and all the templates provide a perfect description of the data (see Sec.~\ref{sec:SCDs}). We generate $64$ realizations for each of the SCDs, linearly varying the template normalization $A_\text{gce}^\text{PS}$ such that the injected flux ranges from $0$ to $\sim~20 \%$ (post-injection). Additionally, we draw 64 Poissonian samples according to the GCE DM template that span the same flux range. We perform this experiment for our default setup, for the case of a thick disk instead of a thin disk (Sec.~\ref{subsec:fermi_thick_disk}), and for the diffuse models $\texttt{p6v11}$, Model~A, and Model~F (Sec.~\ref{subsec:fermi_p6v11}); moreover, we consider a hybrid case (Model~O$_\text{F}$) where the diffuse flux from pion decay \& bremsstrahlung is described by Model~O, whereas the diffuse IC template is taken from Model~F. We consider a fixed ROI with an outer radius of $25 ^\circ$ around the GC and latitudes $|b| > 2 ^\circ$ for this experiment.
\par Fig.~\ref{fig:injection_results} shows how much of the flux injected into the \emph{Fermi} map is recovered by the NN and the associated predictive uncertainties ($1\sigma)$, with the true injected flux fractions (post-injection) for the respective template on the $x$-axis and the recovered injected flux fractions on the $y$-axis. They are calculated as
\begin{equation}
    f^{\bm{\omega}}_{\text{recovered}}\left(\mathbf{x}_{\it{Fermi}}^{\text{post}}\right)_t = f^{\bm{\omega}}\left(\mathbf{x}_{\it{Fermi}}^{\text{post}}\right)_t - \frac{F^{\text{pre}}_{\text{tot}}}{F^{\text{post}}_{\text{tot}}} f^{\bm{\omega}}\left(\mathbf{x}_{\it{Fermi}}^{\text{pre}}\right)_t,
\end{equation}
where $f^{\bm{\omega}}(\mathbf{x}_{\it{Fermi}}^{\text{post}})_t$ is the NN prediction for the flux fraction of template $t \in \{\text{GCE PS}, \text{GCE DM}\}$ in the \emph{Fermi} map containing artificially injected flux from template $t$, $F^{\text{pre}}_{\text{tot}}$ and $F^{\text{post}}_{\text{tot}}$ are the total fluxes in the \emph{Fermi} map pre- and post-injection, respectively, and $f^{\bm{\omega}}(\mathbf{x}_{\it{Fermi}}^{\text{pre}})_t$ is the NN prediction for the flux fraction of template $t$ in the original \emph{Fermi} map (which is close to zero for GCE PS and $\sim 5 - 10 \%$ for GCE DM). Thus, if the NN estimates for the \emph{Fermi} map were exactly correct and the injected flux were perfectly recovered by the NN without affecting the fidelity of the NN predictions for the fluxes present in the \emph{Fermi} map, the markers would lie on the gray dashed line $y = x$. The upper three rows belong to the injection of GCE PS flux for the three different SCDs (with the colors of the error bars matching those of the respective SCDs in Fig.~\ref{fig:SCDs}), and the bottom row contains the results for GCE DM injection. For the case of DM injection, even small injected fluxes are detected by the NN, and the recovered DM flux increases very monotonically as a function of the injected DM flux. For most of the diffuse models, the NN tends to slightly overestimate the injected DM flux; not so, however, for \texttt{p6v11} where the recovered DM flux is a bit too low (also, the uncertainties are larger) -- reminiscent of the findings in \cite{Leane2019a, Buschmann2020} for \texttt{NPTFit}. Regardless of whether or not the GCE actually has a smooth origin, the case of artificial GCE PS injection is more challenging for the NN as synthetic PS flux will entail a mixed DM / PS GCE \emph{as deemed by the NN}, which can cause confusion between the two GCE templates (see Sec.~\ref{sec:SCDs}). Therefore, very small injected PS fluxes ($\lesssim 2 - 3 \%$) are not identified by the NN for any of the SCDs. As the injected PS flux increases, it is eventually detected by the NN, which occurs (unsurprisingly) at a smaller injected flux for brighter PSs. For the bright SCD (and less so for the moderate one), the DM flux ascertained by the NN in the \emph{Fermi} map starts to be absorbed by the GCE PS template as the injected PS flux grows, resulting in the recovered GCE PS flux to be too high. For dim GCE PSs, injected fluxes $\lesssim 5 \%$ are generally not detected by the NN (although a small fraction is often found in the case of the \texttt{p6v11} diffuse model). Remarkably, the recovered fluxes increase much more monotonically for Model~F than for our default case with Model~O, the reason for which will become apparent when considering how the predicted fluxes of the other templates change due to the synthetic PS flux (discussed below). 
\par Interestingly, the predictive uncertainties grow as more PS flux is injected, whereas they remain approximately constant in the case of DM injection. This is in line with the expected behavior for a smooth GCE in the \emph{Fermi} data that is roughly consistent with our generalized NFW template: when additional DM flux is injected, the total DM flux in the map simply increases and the GCE is still made up of $100 \%$ DM irrespective of the synthetic flux, for which reason the ``difficulty'' of the estimation is largely independent of the injection, and the estimated uncertainties are hence not affected. In contrast, injecting GCE PSs into the \emph{Fermi} map in the case of a GCE consisting of $100 \%$ DM flux in the data gives rise to a GCE with contributions from both DM and PS, in which case disentangling the individual constituents is challenging (see Fig.~\ref{fig:SCD_test_with_DM}), which would explain the increasing uncertainties.
\par
\begin{figure}[htb]
\centering
  \noindent
  \resizebox{1\textwidth}{!}{
\includegraphics{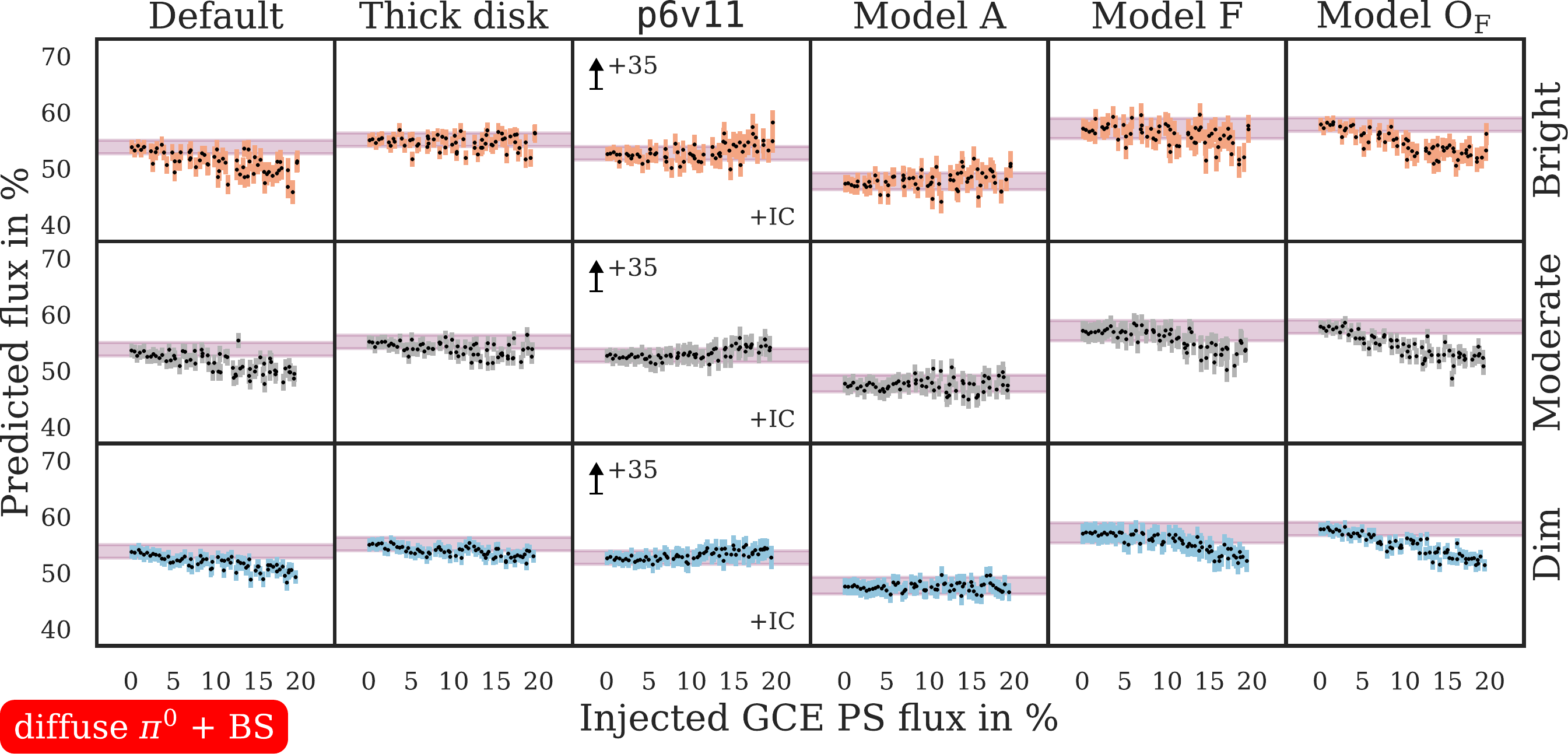}
}
\caption{Injection of artificial GCE PS flux into the \emph{Fermi} map: flux predictions for the pion decay \& bremsstrahlung template. The NN estimates on the $y$-axis are rescaled so as to account for the additional GCE PS flux in the map (whose post-injection flux fraction is given on the $x$-axis). In the ideal case of the estimates not being affected by the additional GCE PS flux, the markers would remain constant. Since the \texttt{p6v11} diffuse model describes the entire diffuse flux with a single template and the resulting NN estimates are consequently high, we subtract a typical IC contribution of $35 \%$ in order for the NN estimates to have a similar magnitude to the other experiments in the figure. The purple bars show the $1\sigma$ credible intervals for $\pi^0$ + BS in the \emph{Fermi} map without a synthetic flux contribution.}
\label{fig:injection_results_dif_O_pibs}
\end{figure}

\begin{figure}[htb]
\centering
  \noindent
  \resizebox{1\textwidth}{!}{
\includegraphics{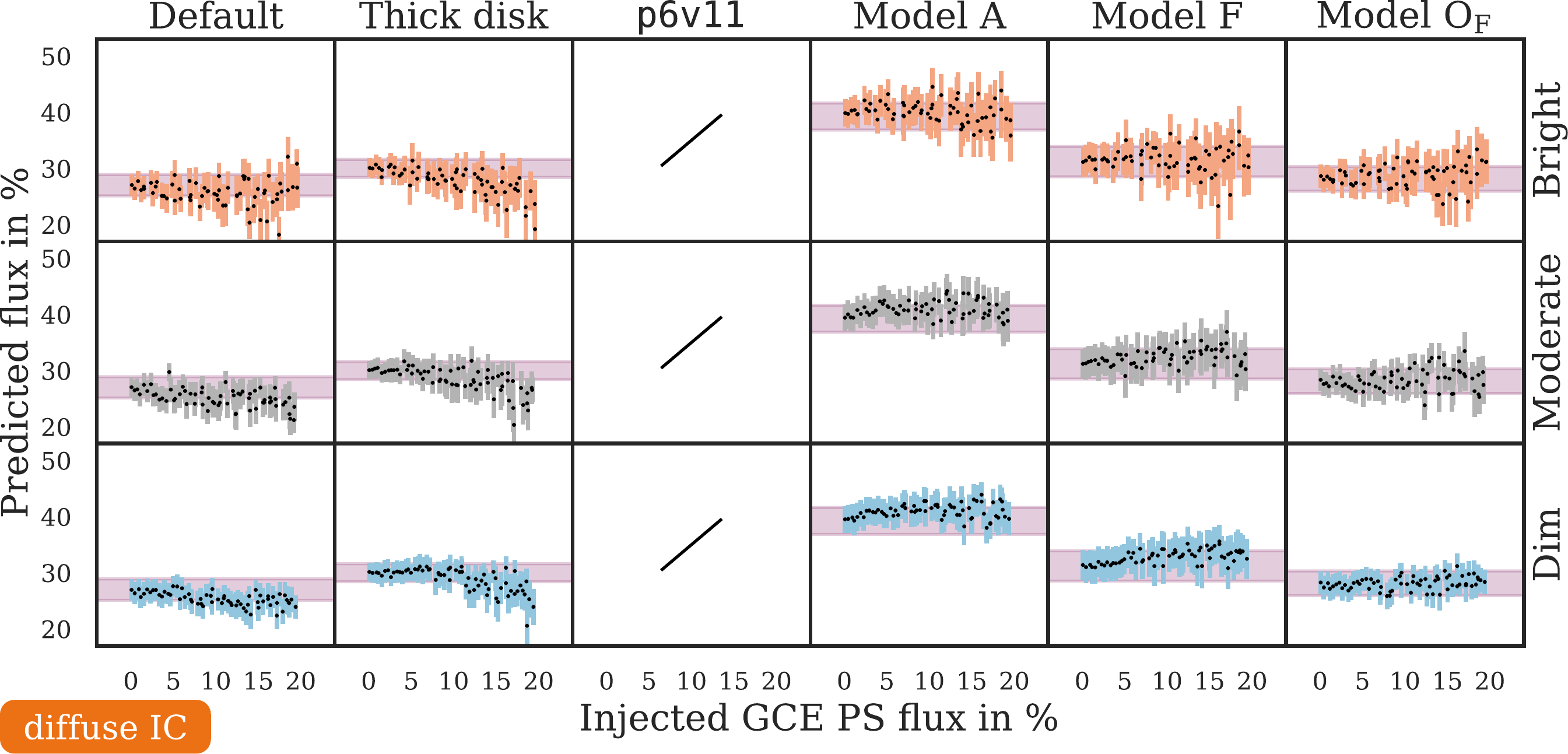}
}
\caption{Same as Fig.~\ref{fig:injection_results_dif_O_pibs}, but for the diffuse IC template. Model \texttt{p6v11} has no separate diffuse IC component, for which reason it is not shown.}
\label{fig:injection_results_dif_O_ic}
\end{figure}

\begin{figure}[htb]
\centering
  \noindent
  \resizebox{1\textwidth}{!}{
\includegraphics{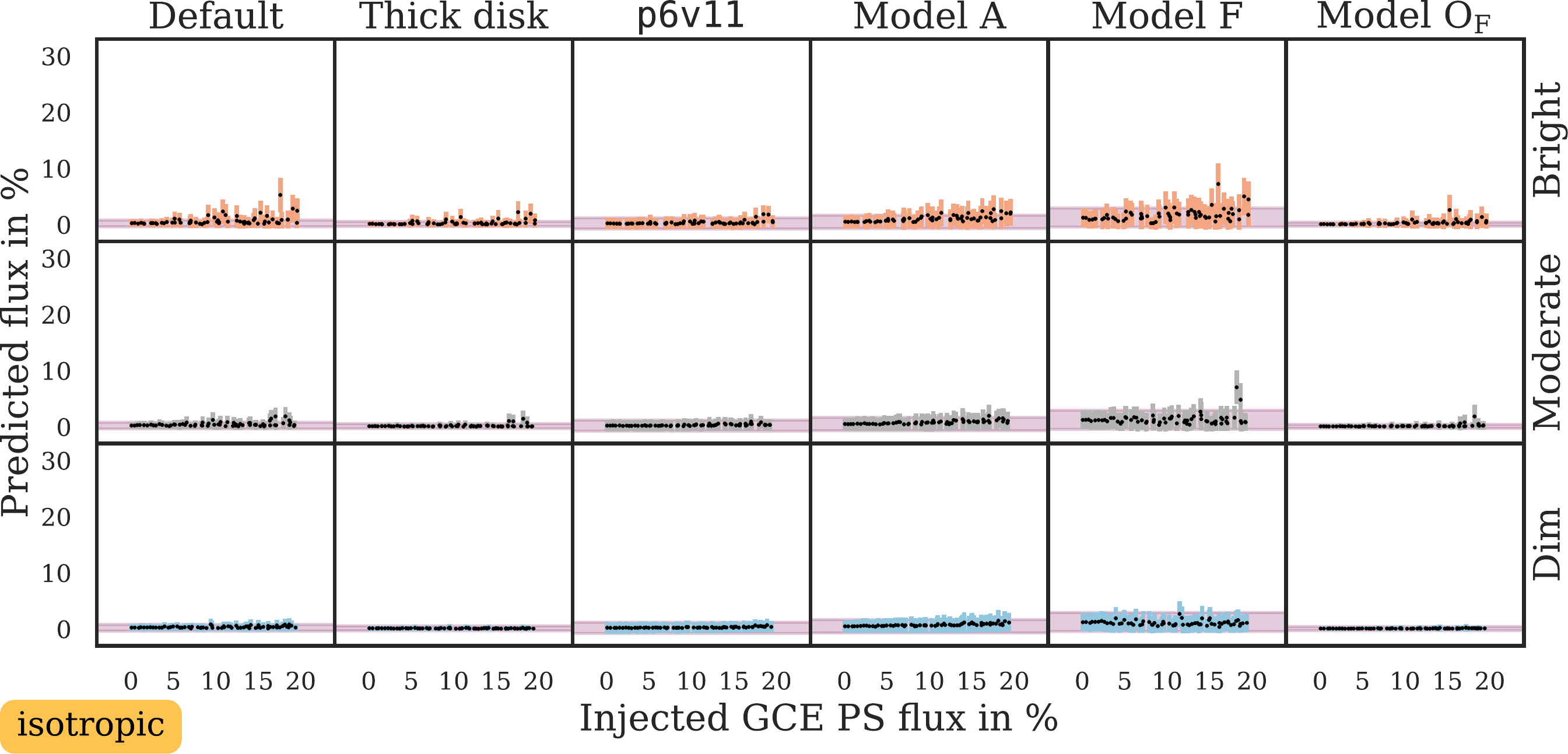}
}
\caption{Same as Fig.~\ref{fig:injection_results_dif_O_pibs}, but for the isotropic template.}
\label{fig:injection_results_iso}
\end{figure}

\begin{figure}[htb]
\centering
  \noindent
  \resizebox{1\textwidth}{!}{
\includegraphics{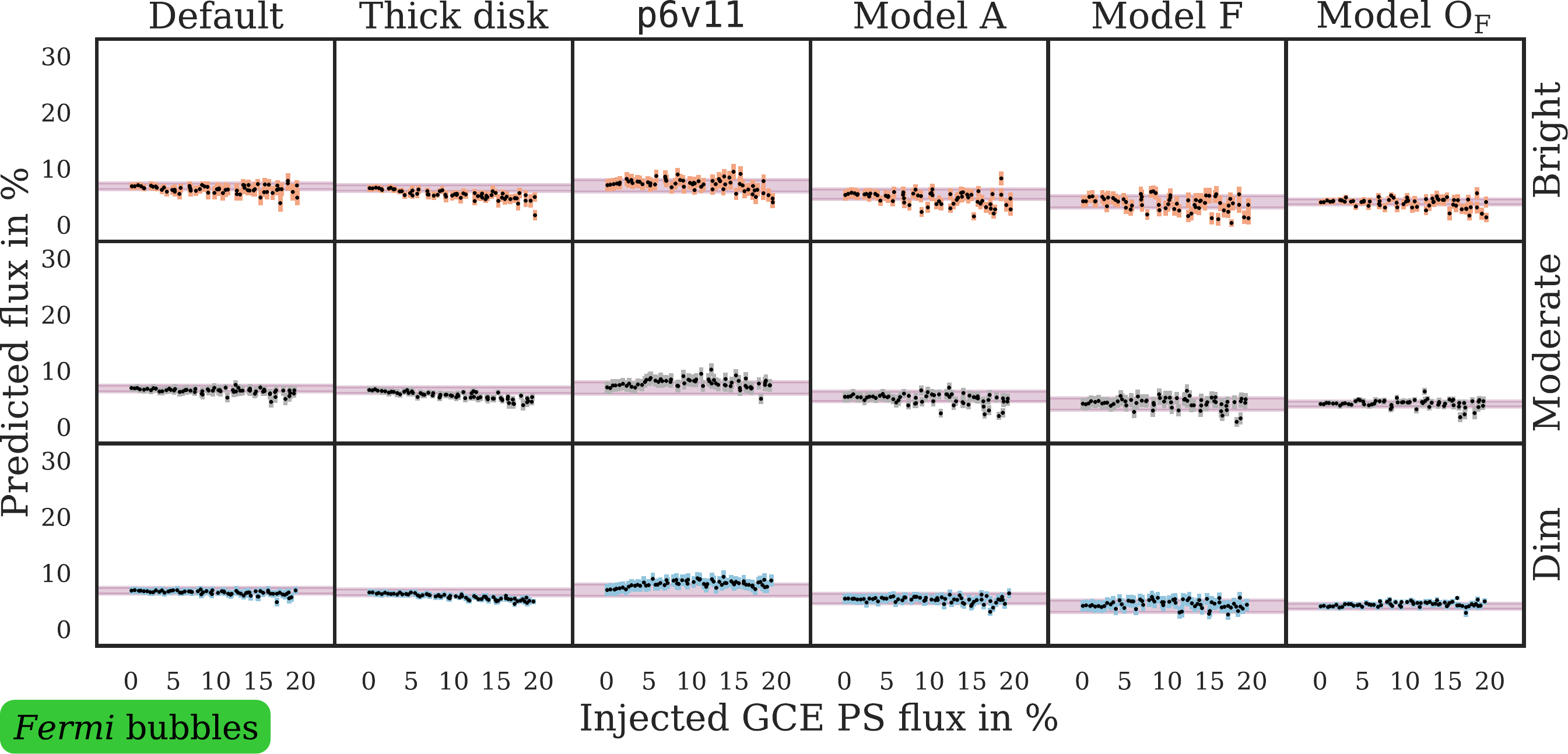}
}
\caption{Same as Fig.~\ref{fig:injection_results_dif_O_pibs}, but for the \emph{Fermi} bubble template.}
\label{fig:injection_results_bub}
\end{figure}

\begin{figure}[htb]
\centering
  \noindent
  \resizebox{1\textwidth}{!}{
\includegraphics{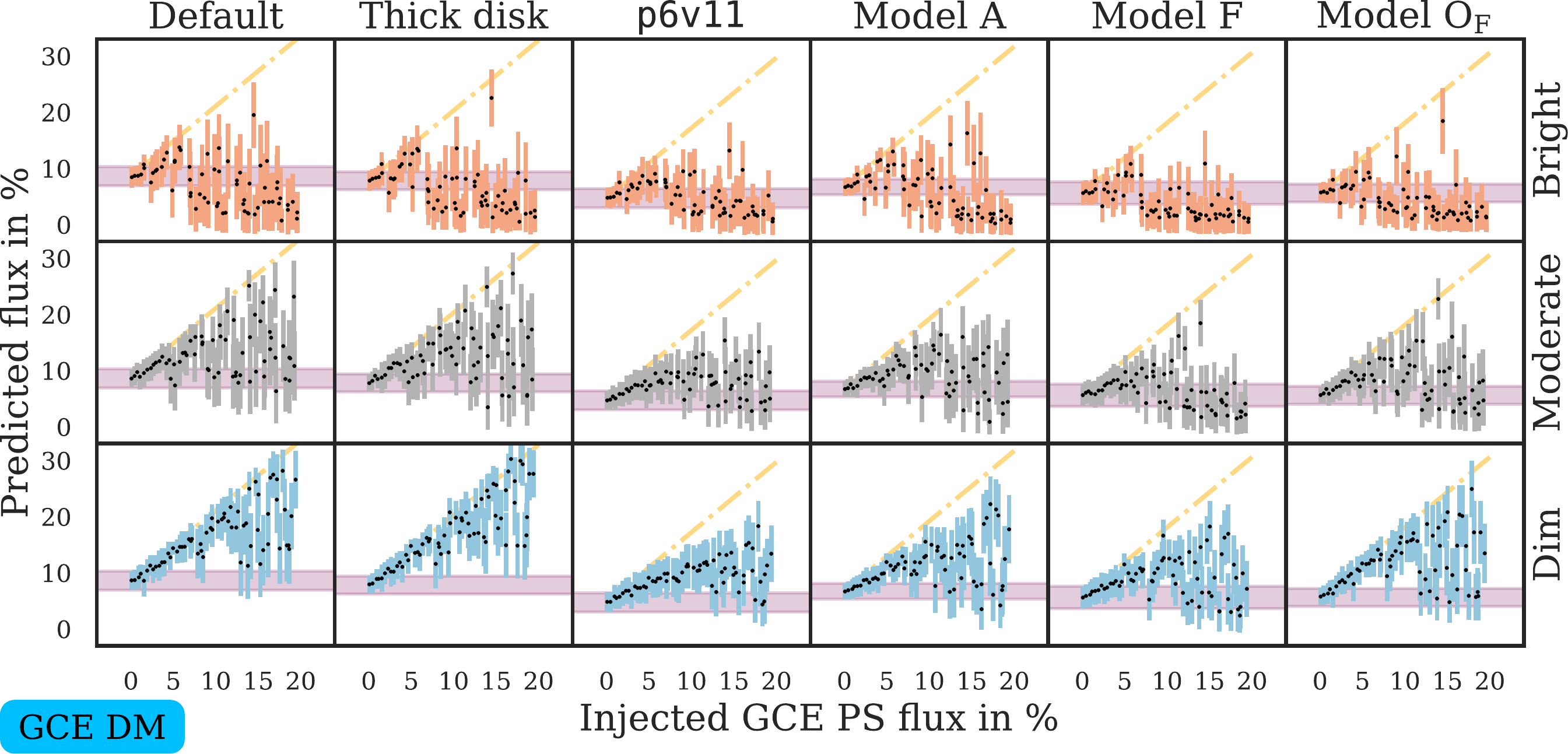}
}
\caption{Same as Fig.~\ref{fig:injection_results_dif_O_pibs}, but for the GCE DM template. The golden dash-dotted lines represent the trend that would be following if the entire injected GCE PS flux (in addition to the DM flux identified in the \emph{Fermi} map) were attributed to GCE DM.}
\label{fig:injection_results_gce}
\end{figure}

\begin{figure}[htb]
\centering
  \noindent
  \resizebox{1\textwidth}{!}{
\includegraphics{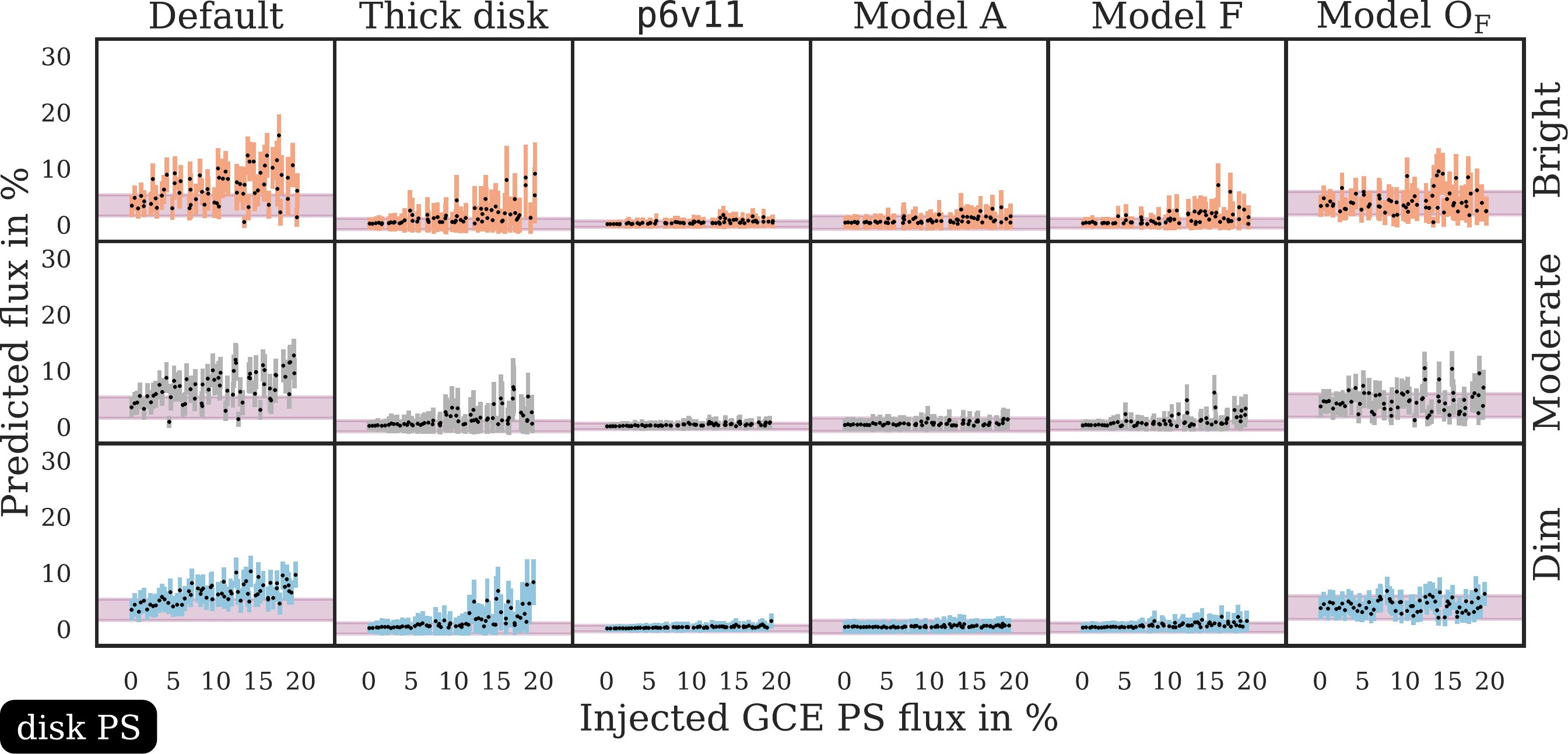}
}
\caption{Same as Fig.~\ref{fig:injection_results_dif_O_pibs}, but for the disk PS template.}
\label{fig:injection_results_disk_PS}
\end{figure}

 In Figs.~\ref{fig:injection_results_dif_O_pibs}~--~\ref{fig:injection_results_disk_PS}, we plot the NN estimates for the other templates (for which no flux is artificially injected) for the GCE PS experiment against the injected GCE PS flux fraction (post-injection). The NN predictions on the $y$-axis are rescaled such that they remain constant if the NN detects the same \emph{total} flux (rather than the same flux fraction) of the respective template as in the original \emph{Fermi} map, despite the additional GCE PS flux. That is, the variable plotted on the $y$-axis is
\begin{equation}
    f^{\bm{\omega}}_{\text{rescaled}}\left(\mathbf{x}_{\it{Fermi}}^{\text{post}}\right)_t = f^{\bm{\omega}}\left(\mathbf{x}_{\it{Fermi}}^{\text{post}}\right)_t \frac{F^{\text{post}}_{\text{tot}}}{F^{\text{pre}}_{\text{tot}}}.
\end{equation}
For most of our template choices, the injected GCE PS flux leads to a decrease in the estimated flux of diffuse pion \& bremsstrahlung, with Models A and \texttt{p6v11} being affected the least. The effect of the PS injection on the diffuse IC template is generally smaller and the estimates stay approximately constant (with somewhat larger uncertainties than for the diffuse $\pi^0$ + BS template), with the exception of the thick disk case, where the estimated diffuse IC flux is negatively correlated with the injected GCE PS flux. The NN estimates for the isotropic template do not change much, although there are a few outliers particularly for the bright SCD, accompanied by larger uncertainties. The estimates for the \emph{Fermi} bubbles remain roughly constant for the default case, decrease monotonically with the thick disk template, increase slightly for dim GCE PSs in the case of model \texttt{p6v11}, and are scattered around the fiducial value with a small negative bias for diffuse Models A and F when the SCD is (moderately) bright. Figure \ref{fig:injection_results_gce} reveals how much of the injected GCE PS flux is misattributed to GCE DM, which is expected to account for the bulk of the misattributed flux as the two templates share the same spatial distribution (albeit not the same statistics). The golden dash-dotted lines designate the predicted DM fluxes that would arise if the NN attributed the entire injected PS flux to DM, in addition to the DM flux detected in the \emph{Fermi} map. For bright GCE PSs, the estimated DM flux rises initially but quickly settles down at very small values as the injected PS flux fraction increases, indicating that the GCE PS template absorbs the flux from both the synthetic PSs and DM. For moderately bright GCE PSs, there are large fluctuations in the DM estimates with Model~O (default and thick disk cases), which are somewhat attenuated when modeling the IC flux with Model~F instead of Model~O (see Model~O$_\text{F}$ in the rightmost column). For dim PSs, the entire injected PS flux is often misattributed to DM in the case of Model~O. Contrarily, the recovery of the GCE PS flux works significantly better when employing Model~F or \texttt{p6v11}: even in the case of dim PSs, the DM template almost never absorbs the entire artificial PS flux for injected flux fractions $\gtrsim 5 \%$, and the DM estimates are scattered between the fiducial DM value and higher values that evidence a partial (but hardly ever complete) misattribution of the PS flux to DM -- as would be expected in the case of a mixed DM / PS GCE. Considering the cross-talk between the injected GCE PSs and the disk PS template depicted in Fig.~\ref{fig:injection_results_disk_PS} exposes another contribution to the ``jumping'' GCE PS predictions in Fig.~\ref{fig:injection_results} for Model~O: for all the scenarios in which the diffuse flux is described by Model~O, the disk PSs estimates fluctuate significantly and reach values $\gtrsim 10 \%$. This is in stark contrast to Model~A, \texttt{p6v11}, and Model~F (to a lesser extent), for which the disk PS estimates barely vary as the GCE PS flux increases and exhibit much smaller uncertainties.  
\par We summarize our main conclusions from this injection experiment:
\begin{itemize}
    \item The NN generally recovers artificially injected GCE DM flux from the \emph{Fermi} map. For some template choices, the recovered DM flux exceeds the injected flux by a few per cent. The uncertainties remain approximately constant for the case of DM injection.
    \item Artificially injected GCE PS flux may be misattributed to GCE DM and, for some template choices, to disk PS. The probability of this occurring is negatively correlated to the brightness of the synthetic GCE PSs. Therefore, we cannot currently exclude the possibility of there being faint PSs in the GC that are overlooked by our NN.  
    \item The NN recovers the injected GCE PS flux more faithfully with Model~F and \texttt{p6v11} than when using Model~O.
    For Model~F, Model~A, and \texttt{p6v11}, the GCE DM template almost never absorbs GCE PS flux fractions $\gtrsim 5 - 7\%$ entirely, even for dim GCE PSs, and for either of the latter two diffuse models, the misattribution of the GCE PS flux to disk PS is very small.
    \item The NN mostly reacts to injected GCE PSs with increased uncertainties for the GCE templates as compared to those for the original \emph{Fermi} map, becoming aware of the increased difficulty due to the presence of both (seemingly) smooth and point-like emission from the GC.
\end{itemize}
We emphasize again that the current version of our NN prefers a smooth GCE for all template choices that we investigated.


